\documentclass[10pt]{revtex4-1}
\usepackage[utf8]{inputenc}
\usepackage[T1]{fontenc}
\usepackage{graphicx}
\usepackage{nicefrac}
\usepackage{amsfonts}
\usepackage{amssymb}
\usepackage{amsmath} 
\usepackage{subfigure}
\usepackage{multirow} 
\usepackage{tabularx} 
\usepackage{array}
\usepackage{units}
\usepackage{tensor} 
\usepackage{braket}
\usepackage{bm}
\usepackage{hyperref}
\usepackage[resetlabels, labeled]{multibib}
\usepackage{xcolor}
\usepackage{bbm}

\newcommand{\indicator}[1]{\mathbbm{1}_{\mathcal{#1}}} 

\newcommand\nHam{6561}  
\newcommand\epsilonValue{0.01}  
\newcommand\nExp{100}   
\newcommand\sshOneTestNHam{1005}
\newcommand\sshOneTestFracHam{15.32}
\newcommand\sshOneTrainPlusValNHam{5556}
\newcommand\sshOneTrainNHam{556}
\newcommand\sshOneTrainFracHam{8.47}
\newcommand\sshOneValNHam{5000}
\newcommand\sshOneValFracHam{76.21}
\newcommand\sshOneTrainPlusValWindZeroFracHam{50.79}
\newcommand\sshOneTrainPlusValWindOneFracHam{49.21}
\newcommand\sshOneTestWindZeroFracHam{44.79}
\newcommand\sshOneTestWindOneFracHam{55.21}
\newcommand\sshOneHamTestBaseline{0.5521}
\newcommand\nEst{25}   
\newcommand\sshTwoTestNHam{1040}
\newcommand\sshTwoTestFracHam{15.85}
\newcommand\sshTwoTrainPlusValNHam{5521}
\newcommand\sshTwoTrainNHam{2761}
\newcommand\sshTwoTrainFracHam{42.08}
\newcommand\sshTwoValNHam{2760}
\newcommand\sshTwoValFracHam{42.07}
\newcommand\sshTwoTrainPlusValWindMinusOneFracHam{17.90}
\newcommand\sshTwoTrainPlusValWindZeroFracHam{32.51}
\newcommand\sshTwoTrainPlusValWindOneFracHam{32.26}
\newcommand\sshTwoTrainPlusValWindTwoFracHam{17.33}
\newcommand\sshTwoTestWindMinusOneFracHam{36.32}
\newcommand\sshTwoTestWindZeroFracHam{11.06}
\newcommand\sshTwoTestWindOneFracHam{12.69}
\newcommand\sshTwoTestWindTwoFracHam{39.93}
\newcommand\sshTwoHamTestBaseline{0.3993}

\newcommand\sitesExpOne{(0, 1, 3, 50, 51, 53)}
\newcommand\sitesExpTwo{(0, 1, 2, 3, 4, 5, 46, 48, 49, 50, 51, 53, 94, 95, 96, 97, 98, 99)}

\newcommand\sitesSOne{(0, 1, 3, 50)}
\newcommand\sitesEOne{(1, 2, 36, 49)}
\newcommand\sitesOOne{(0, 18, 28, 30)}
\newcommand\sitesSTwo{(0, 1, 2, 3, 4, 5, 6, 7, 46, 48, 49, 50)}
\newcommand\sitesETwo{(0, 1, 2, 3, 4, 5, 6, 7, 47, 48, 49, 50)}
\newcommand\sitesOTwo{(0, 1, 2, 3, 4, 5, 43, 44, 45, 46, 47, 48)}
\newcommand\xalpha{$X_\alpha$}
\newcommand\xSalpha{$X_{\mathcal{S}_\alpha}$}
\newcommand\xcalpha{$\hat{X}^c_\alpha$}
\newcommand\xcEalpha{$\hat{X}^c_{\mathcal{E}_\alpha}$}
\newcommand\xcSEalpha{$\hat{X}^c_{\mathcal{S}_\alpha,\mathcal{E}_\alpha}$}
\newcommand\xsalpha{$\hat{X}^s_\alpha$}
\newcommand\xsOalpha{$\hat{X}^s_{\mathcal{O}_\alpha}$}
\newcommand\xsSOalpha{$\hat{X}^s_{\mathcal{S}_\alpha,\mathcal{O}_\alpha}$}

\newcommand{\tableRow}[1]{\\[#1 cm]}
\newcommand{\tableRowHeader}[1]{\\[#1 cm]}
\newcommand\tableRowHeaderEnd{\tableRowHeader{0.3}}
\newcommand\tableRowEnd{\tableRow{0.15}}
\newcommand\SSHSys{SSH system}         
\newcommand\Feat{Features}              
\newcommand\ValEig{Val. eigenvectors}   
\newcommand\TestEig{Test eigenvectors}  
\newcommand\ValHam{Val. Hamiltonians}   
\newcommand\TestHam{Test Hamiltonians}  
\newcommand\xOne{$X_1$}
\newcommand\xOneEigTrain{0.9814}
\newcommand\xOneEigVal{0.9639}
\newcommand\xOneEigTest{0.7897}
\newcommand\xOneHamTrain{1.000}
\newcommand\xOneHamVal{1.000}
\newcommand\xOneHamTest{0.9919}
\newcommand\xSOne{$X_{\mathcal{S}_1}$}

\newcommand\xSOneEigVal{0.9444}
\newcommand\xSOneEigTest{0.7763}

\newcommand\xSOneHamVal{0.9853}

\newcommand\xcOne{$\hat{X}^c_1$}

\newcommand\xcOneEigVal{0.9521}
\newcommand\xcOneEigTest{0.7374}

\newcommand\xcOneHamVal{0.9976}
\newcommand\xcOneHamTest{0.9916}
\newcommand\xcEOne{$\hat{X}^c_{\mathcal{E}_1}$}

\newcommand\xcEOneEigVal{0.8280}
\newcommand\xcEOneEigTest{0.6067}

\newcommand\xcEOneHamVal{0.9979}
\newcommand\xcEOneHamTest{0.9191}
\newcommand\xcSEOne{$\hat{X}^c_{\mathcal{S}_1,\mathcal{E}_1}$}

\newcommand\xcSEOneEigVal{0.9444}
\newcommand\xcSEOneEigTest{0.8176}

\newcommand\xcSEOneHamVal{0.9853}
\newcommand\xcSEOneHamTest{0.9934}
\newcommand\xsOne{$\hat{X}^s_1$}

\newcommand\xsOneEigVal{0.9533}
\newcommand\xsOneEigTest{0.7314}

\newcommand\xsOneHamVal{0.9906}
\newcommand\xsOneHamTest{0.9856}
\newcommand\xsOOne{$\hat{X}^s_{\mathcal{O}_1}$}

\newcommand\xsOOneEigVal{0.6942}
\newcommand\xsOOneEigTest{0.5420}

\newcommand\xsOOneHamVal{0.7088}
\newcommand\xsOOneHamTest{0.4798}
\newcommand\xsSOOne{$\hat{X}^s_{\mathcal{S}_1,\mathcal{O}_1}$}

\newcommand\xsSOOneEigVal{0.9456}
\newcommand\xsSOOneEigTest{0.7818}

\newcommand\xsSOOneHamVal{0.9854}
\newcommand\xsSOOneHamTest{0.9399}
\newcommand\xTwo{$X_2$}
\newcommand\xTwoEigTrain{0.9997}
\newcommand\xTwoEigVal{0.9709}
\newcommand\xTwoEigTest{0.6634}
\newcommand\xTwoHamTrain{1.000}
\newcommand\xTwoHamVal{0.9972}
\newcommand\xTwoHamTest{0.8797}
\newcommand\xSTwo{$X_{\mathcal{S}_2}$}

\newcommand\xSTwoEigVal{0.9590}
\newcommand\xSTwoEigTest{0.6168}

\newcommand\xSTwoHamVal{0.9961}

\newcommand\xcTwo{$\hat{X}^c_2$}

\newcommand\xcTwoEigVal{0.9740}
\newcommand\xcTwoEigTest{0.6895}

\newcommand\xcTwoHamVal{0.9976}
\newcommand\xcTwoHamTest{0.8862}
\newcommand\xcETwo{$\hat{X}^c_{\mathcal{E}_2}$}

\newcommand\xcETwoEigVal{0.8990}
\newcommand\xcETwoEigTest{0.5357}

\newcommand\xcETwoHamVal{0.9955}
\newcommand\xcETwoHamTest{0.7897}
\newcommand\xcSETwo{$\hat{X}^c_{\mathcal{S}_2,\mathcal{E}_2}$}

\newcommand\xcSETwoEigVal{0.8999}
\newcommand\xcSETwoEigTest{0.5030}

\newcommand\xcSETwoHamVal{0.9956}
\newcommand\xcSETwoHamTest{0.7671}
\newcommand\xsTwo{$\hat{X}^s_2$}

\newcommand\xsTwoEigVal{0.9735}
\newcommand\xsTwoEigTest{0.6878}

\newcommand\xsTwoHamVal{0.9971}
\newcommand\xsTwoHamTest{0.8899}
\newcommand\xsOTwo{$\hat{X}^s_{\mathcal{O}_2}$}

\newcommand\xsOTwoEigVal{0.9042}
\newcommand\xsOTwoEigTest{0.5527}

\newcommand\xsOTwoHamVal{0.9918}
\newcommand\xsOTwoHamTest{0.7427}
\newcommand\xsSOTwo{$\hat{X}^s_{\mathcal{S}_2,\mathcal{O}_2}$}

\newcommand\xsSOTwoEigVal{0.8204}
\newcommand\xsSOTwoEigTest{0.4127}

\newcommand\xsSOTwoHamVal{0.9942}
\newcommand\xsSOTwoHamTest{0.5903}

\newcommand\SSHOnePhaseDiagram{./ssh1_phase_diagram.png}
\newcommand\SSHTwoPhaseDiagram{./ssh2_phase_diagram.png}
\newcommand\SSHOneTrainValTestSplit{./ssh1_scatter_train_val_test_experiment_0.png}
\newcommand\SSHOneWindingTrain{./ssh1_scatter_winding_train_experiment_0.png}
\newcommand\SSHOnePredictionGrid{./ssh1_pcolormesh_prediction_grid_experiment_0.png}
\newcommand\SSHTwoTrainValTestSplit{./ssh2_scatter_train_val_test_experiment_0.png}
\newcommand\SSHTwoWindingTrain{./ssh2_scatter_winding_train_experiment_0.png}
\newcommand\SSHTwoPredictionGrid{./ssh2_pcolormesh_prediction_grid_experiment_0.png}
\newcommand\SSHOneWindingZero{./ssh1_imshow_winding_grid_winding_0_sim.png}
\newcommand\SSHOneWindingOne{./ssh1_imshow_winding_grid_winding_1_sim.png}
\newcommand\SSHOneHeatmap{./ssh1_merge_imshow_winding_grids_second_sim.png}
\newcommand\SSHTwoWindingMinusOne{./ssh2_imshow_winding_grid_winding_-1_sim.png}
\newcommand\SSHTwoWindingZero{./ssh2_imshow_winding_grid_winding_0_sim.png}
\newcommand\SSHTwoWindingOne{./ssh2_imshow_winding_grid_winding_1_sim.png}
\newcommand\SSHTwoWindingTwo{./ssh2_imshow_winding_grid_winding_2_sim.png}
\newcommand\SSHTwoHeatmap{./ssh2_merge_imshow_winding_grids_sim.png}
\newcommand\SSHOneEntropySignature{./ssh1_plot_feature_importances.png}
\newcommand\SSHTwoEntropySignature{./ssh2_plot_feature_importances.png}

\newcommand\SSHModel{./supp_ssh_model.pdf}
\newcommand\WindingNumber{./supp_winding.png}
\newcommand\SSHOneCompressedRealSpacePhaseDiagram{./supp_ssh1_real_merge_imshow_winding_grids_second_sim.png}
\newcommand\SSHOneCompressedDCTPhaseDiagram{./supp_ssh1_dct_merge_imshow_winding_grids_second_sim.png}
\newcommand\SSHOneCompressedDSTPhaseDiagram{./supp_ssh1_dst_merge_imshow_winding_grids_second_sim.png}
\newcommand\SSHTwoCompressedRealSpacePhaseDiagram{./supp_ssh2_real_merge_imshow_winding_grids_sim.png}
\newcommand\SSHTwoCompressedDCTPhaseDiagram{./supp_ssh2_dct_merge_imshow_winding_grids_sim.png}
\newcommand\SSHTwoCompressedDSTPhaseDiagram{./supp_ssh2_dst_merge_imshow_winding_grids_sim.png}
\newcommand\SSHOneEntropySignatureOneFourZero{./supp_ssh1_140_plot_feature_importances.png}
\newcommand\SSHOneEntropySignatureOneEightZero{./supp_ssh1_180_plot_feature_importances.png}
\newcommand\SSHOneEntropySignatureTwoTwoZero{./supp_ssh1_220_plot_feature_importances.png}
\newcommand\SSHTwoEntropySignatureOneFourZero{./supp_ssh2_140_plot_feature_importances.png}
\newcommand\SSHTwoEntropySignatureOneEightZero{./supp_ssh2_180_plot_feature_importances.png}
\newcommand\SSHTwoEntropySignatureTwoTwoZero{./supp_ssh2_220_plot_feature_importances.png}
\newcommand\SSHOneCumulativeEntropyOneFourZero{./supp_ssh1_140_plot_cumulative_feature_importances.png}
\newcommand\SSHOneCumulativeEntropyOneEightZero{./supp_ssh1_180_plot_cumulative_feature_importances.png}
\newcommand\SSHOneCumulativeEntropyTwoTwoZero{./supp_ssh1_220_plot_cumulative_feature_importances.png}
\newcommand\SSHTwoCumulativeEntropyOneFourZero{./supp_ssh2_140_plot_cumulative_feature_importances.png}
\newcommand\SSHTwoCumulativeEntropyOneEightZero{./supp_ssh2_180_plot_cumulative_feature_importances.png}
\newcommand\SSHTwoCumulativeEntropyTwoTwoZero{./supp_ssh2_220_plot_cumulative_feature_importances.png}

\begin{document}

\nocite{*}

\title{Machine learning topological phases in real space}

\author{N. L. Holanda}
\email{linneuholanda@gmail.com, linneu@cbpf.br}
\affiliation{Integrated Quantum Materials, Cavendish Laboratory, University of Cambridge, J. J. Thomson Avenue, Cambridge, CB3 0HE, United Kingdom}
\affiliation{Centro Brasileiro de Pesquisas F\'isicas, \\Rua Dr. Xavier Sigaud, 150 - Urca, 22290-180,  Rio de Janeiro, RJ, Brazil}

\author{M. A. S. Griffith}
\email{griffithphys@gmail.com}
\affiliation{Centro Brasileiro de Pesquisas F\'isicas, \\Rua Dr. Xavier Sigaud, 150 - Urca, 22290-180,  Rio de Janeiro, RJ, Brazil}
\affiliation{Departamento de Ciências Naturais, Universidade Federal de S\~ao Jo\~ao Del Rei, Praça Dom Helv\'ecio 74, 36301-160, S\~ao Jo\~ao Del Rei, MG, Brazil}

\date{\today }

\begin{abstract}
We develop a supervised machine learning algorithm that is able to learn topological phases of finite condensed matter systems from bulk data in real lattice space. The algorithm employs diagonalization in real space together with any supervised learning algorithm to learn topological phases through an eigenvector ensembling procedure. We combine our algorithm with decision trees and random forests to successfully recover topological phase diagrams of Su-Schrieffer-Heeger (SSH) models from bulk lattice data in real space and show how the Shannon information entropy of ensembles of lattice eigenvectors can be used to retrieve a signal detailing how topological information is distributed in the bulk. We further use insights obtained from these information entropy signatures to engineer global topological features from real space lattice data that still carry most of the topological information in the lattice, while greatly diminishing the size of feature space, thus effectively amounting to a topological lattice compression. Finally, we explore the theoretical possibility of interpreting the information entropy topological signatures in terms of emergent information entropy wave functions, which lead us to Heisenberg and Hirschman uncertainty relations for topological phase transitions. The discovery of Shannon information entropy signals associated with topological phase transitions from the analysis of data from several thousand SSH systems illustrates how model explainability in machine learning can advance the research of exotic quantum materials with properties that may power future technological applications such as qubit engineering for quantum computing. 
\end{abstract}

\maketitle

\section{Introduction}
\label{introduction}
The quest for innovative materials that harness exotic quantum properties has lured physicists into the realm of topological insulators and topological states of matter \cite{RevModPhys.82.3045}. These materials feature previously unthought-of traits like bulk insulation coupled with metallic conductance at the surface and the splitting of currents according to spin orientation. Adding to that, these properties are protected by non-trivial topology that renders them robust to many sources of perturbation like thermal noise. Such characteristics make them promising candidates to being the cornerstone of 21st century technologies like spintronics and quantum computing.

These new topological states of matter have been studied in several contexts in condensed matter physics including superconductors \cite {CONTINENTINO2017A1}$^-$\cite{ryu2010topological}, ultracold atoms \cite{atala2013direct}$^-$\cite{meier2016observation}, photonic crystals \cite{hafezi2013imaging}$^-$\cite{PhysRevX.5.031011}, photonic quantum walks \cite{kitagawa2012observation}$^-$\cite{PhysRevX.7.031023} and Weyl semimetals \cite{soluyanov2015type,PhysRevX.5.031013}. Among these, the Su-Schrieffer-Heeger (SSH) model \cite{PhysRevLett.42.1698} has attracted particular theoretical interest due to its simplicity and generality.

The SSH model is the simplest tight-binding model that exhibits a topological phase transition. As such, it can be viewed as the \emph{Drosophila} of the field, providing a simple framework for testing new techniques. The model can be expressed in terms of creation and annihilation operators by the Hamiltonian
\begin{equation}
\label{SSH_ham}
\mathbf{H}(\mathbf{t})=\mathbf{c}^{\dagger}H(\mathbf{t})\mathbf{c}
\end{equation}
and describes e.g. the hopping of electrons along a one-dimensional chain comprising two atoms per unit cell (a brief discussion of the SSH model and its topological properties can be found in the section \textbf{The SSH model} in the Supplementary Material). The SSH model has found several interesting applications in the modelling of diverse systems with non-trivial topology like optical lattices \cite{maffei2018topological}, polymeric materials \cite{RevModPhys.73.681} and topological mechanisms \cite{kane2014topological,Chen13004}.

Many recent papers have explored the possibility of treating the general problem of determining phase transition boundaries of physical systems as machine learning tasks \cite{carrasquilla2017machine}$^-$\cite{rodriguez2018identifying}. In the particular case of topological phase transitions, the usual approach for supervised learning is to generate a data set $\big(H_1(k), W_1\big)$, ..., $\big(H_n(k), W_n\big)$ whose inputs are representations of Hamiltonians in wavevector space $H_i(k)$ and targets are their corresponding topological invariants $W_i$ (for the SSH model the topological invariant is the winding number). Our paper extends this task to the case of learning topological phase diagrams from input data in real space. Strikingly, we find that information localized on a few lattice sites in the bulk is sufficient to predict with high accuracy which topological phase a particular Hamiltonian belongs to.

The main motivation for developing a data-driven approach based on real space is that the canonical method of choice for the analysis of topological systems, i.e. wavevector space computations of topological invariants, is often only feasible for systems with translational symmetry, which many physical systems of current interest (e.g. disordered systems in condensed matter) do not have. Furthermore, it is not always granted that the topological invariants of a physical system being investigated are known in advance, as is presently the case for many gapless insulators. In such cases, being able to engineer topological features that encode the topological states of a system while at the same time reducing the system's complexity may provide an alternative strategy. Moreover, since real space and wavevector space eigenvectors are related by Fourier transforms, the latter are essentially delocalized and therefore so is any information recovered from them. Constructing theoretical methods to trace the distribution of information in topological systems may be an essential prerequisite to discovering new topological invariants and features. The data-driven approach designed in this article addresses these issues. 

To investigate topological phases of matter in real space we have designed a novel supervised learning algorithm (here called eigenvector ensembling algorithm) tailored for the task of learning phase transition boundaries from local features. The algorithm is based on eigenvector decomposition and eigenvector ensembling and therefore will require minimal changes to be applicable to a broader class of data-driven physics problems. We describe the algorithm in detail in section \ref{the_eigenvector_ensembling_algorithm} and demonstrate its effectiveness by combining it with decision trees and random forests to recover the topological phase diagrams of SSH systems from local coordinates of eigenstates in real space. This is performed in section \ref{numerical_experiments}.

The advantage of using decision tree-based algorithms to learn topological phases from local eigenvector data is that their use of entropy-based cost functions (such as Shannon information entropy or Gini impurity) furnishes them with an intrinsinc model explanaibility tool that summarizes how important each feature was to learning the desired patterns in the data. This makes it much easier to trace the localization of relevant information along the features of a data set. Here we use the Shannon information entropy of ensembles of real space eigenvectors to recover a signal quantifying the amount of topological information available from each lattice site. This is a highly non-trivial proposition since the topological phase of a system is a global property of the system as a whole emerging from complex interactions between its components, and therefore even defining a local topological signal is a daunting theoretical task. To our knowledge this is the first time that a signal describing the localization of topological information in the bulk of topological condensed matter systems is presented in the literature. These topological signals, here called information entropy signatures, are the subject of section \ref{information_entropy_signatures}. 

In possession of the information entropy signatures, we demonstrate how the symmetries in real and wavevector space eigenvectors can be manipulated with signal processing tools commonly employed in audio and image processing to execute two standard unsupervised learning tasks, namely dimensionality reduction and feature engineering. The topological features resulting from this analysis, here denominated Discrete Cosine Transform and Discrete Sine Transform topological features, are discussed in section \ref{topological_feature_engineering}.    

The information entropy signatures are finally explored as theoretical constructs in terms of information entropy mass functions along the lattices. By taking their continuum limit and admitting that they are quantum in nature, we show how the information entropy signatures can be naturally understood in terms of emergent information entropy wave functions along the lattices. The theoretical formulation of the results obtained with machine learning in terms of emergent quantum mechanical wave functions allows us to establish Heisenberg and Hirschman uncertainty principles for the localizability of information entropy in topological phase transitions. The emergent information entropy wave functions are the theme of section \ref{emergent_information_entropy_wave_functions}.      

The  discovery of information entropy signatures of topological phase transitions and their description as a measurable emergent phenomenon originating from the information entropy of ensembles of topological systems proximal to phase transition boundaries provide clear illustrations of how model explainability in machine learning can guide new discoveries in condensed matter and quantum materials physics, since the existence of these signals was established by analyzing data from several thousand SSH systems which, taken individually, could not have provided any concrete hint of their existence.

As of yet model explainability \cite{gilpin2018explaining}-\cite{roscher2020explainable} is one of the topics at the edge of machine learning research that has been little explored by the physics community working at the interface between the two disciplines. This raises important questions as to whether machine learning can in fact help to advance theoretical investigation in physics, since the majority of physics papers published on the subject are proofs of concept aimed at showing that modern machine learning techniques are capable of recognizing the relevant patterns in data from physical systems whose properties were known in advance. By proposing new concepts from the data analysis of physical systems of contemporary interest and knitting together ideas from topological phase transitions and information theory by dint of model explainability, we expect to draw the physics community's attention to this essential machine learning tool. 
\begin{figure}
\centering
\subfigure[]{\label{ssh1}\includegraphics[width=.45\textwidth]{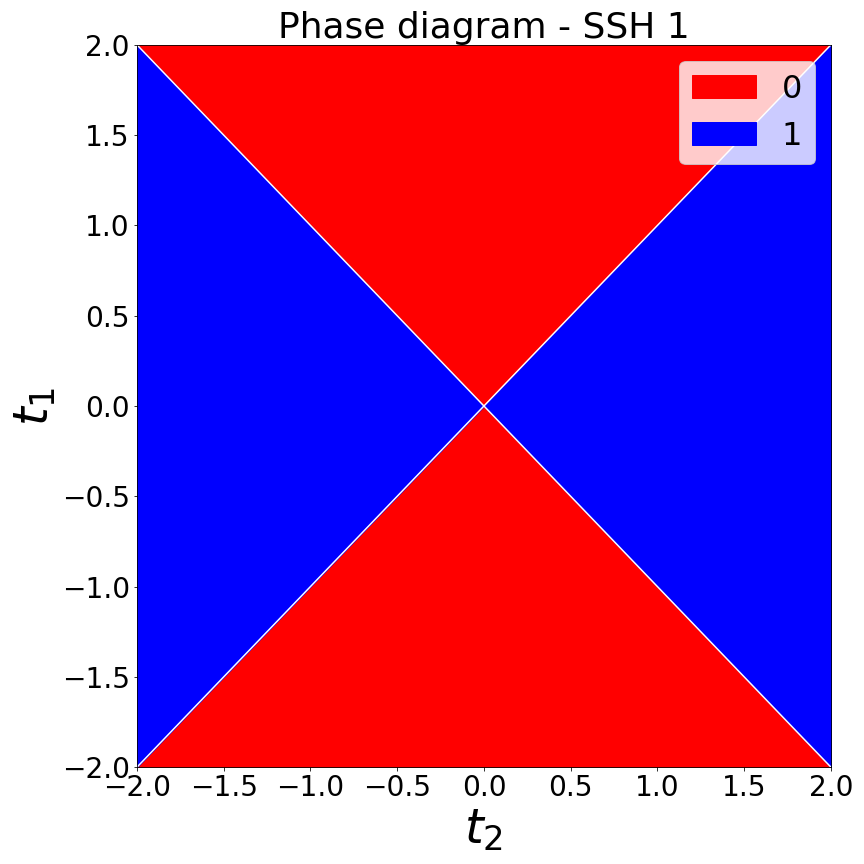}}\quad
\subfigure[]{\label{ssh2}\includegraphics[width=.45\textwidth]{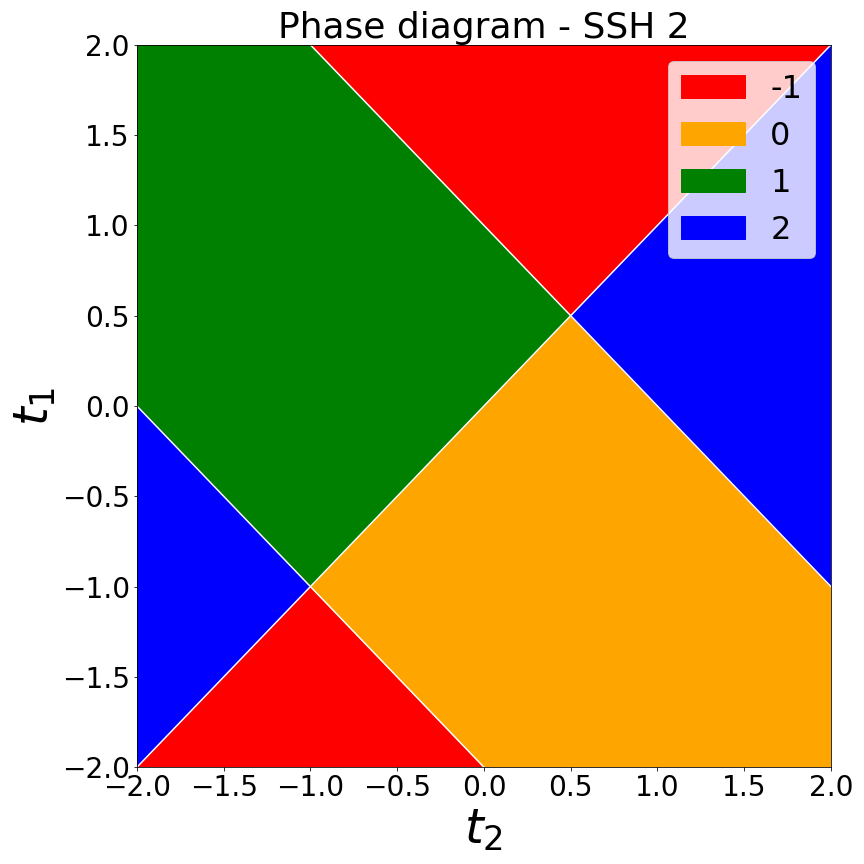}}
\caption{Phase diagrams in parameter space. a) SSH model with first-neighbor hoppings $t_1$ and $t_2$. The (red) regions with winding number $W$ = 0 are trivial, while the (blue) regions with winding number $W$ = 1 are topologically non-trivial. b) SSH model with first ($t_1$ and $t_2$) and second ($T_1$ and $T_2$) nearest-neighbor hoppings. In this article we set $t_1$ = $t_2$ = 1 and renamed the variables $T_1$ $\rightarrow$ $t_1$, $T_2$ $\rightarrow$ $t_2$ for convenience. The (orange) region with winding number $W$ = 0 is trivial while the others with winding numbers $W$ = -1, $W$ = 1 and $W$ = 2 (red, green and blue respectively) are topologically non-trivial.}
\label{fig:phasediagrams}
\end{figure}

\section{The eigenvector ensembling algorithm}
\label{the_eigenvector_ensembling_algorithm}
The eigenvector ensembling algorithm consists of five steps: 1) Generating Hamiltonians in real space and their corresponding winding numbers; 2) Creating training, validation and test sets; 3) Training on real space eigenvectors of Hamiltonians in the training set; 4) Eigenvector ensembling and 5) Bootstrapping. We describe here in detail each of these steps as they were implemented in this work. For a comprehensive introduction to the concepts referenced in the steps below, we recommend the book \cite{friedman2001elements}.    
\vspace{.3cm}
\begin{enumerate}
\item[1)] \textbf{Generating Hamiltonians and winding numbers:} we start generating a number of paremeterized Hamiltonians $H(\mathbf{t})$ in real space and their corresponding winding numbers $W(\mathbf{t})$, where $\mathbf{t} = (t_1, t_2,...,t_h)$ is a vector of $h$ hopping parameters (in the simplest case of the SSH model $h$ = 2). These Hamiltonians are $N\times N$ matrices, where $N$ is twice the number of unit cells in the chain.
\item[2)] \textbf{Creating training, validation and test sets:} we split our set of parameterized Hamiltonians and winding numbers into training, validation and test sets, as is usualy done in machine learning. More explicitly, assume our hopping parameters vector $\mathbf{t}$ takes on the values $\mathbf{t}_1, \mathbf{t}_2, ..., \mathbf{t}_n$ corresponding to the Hamiltonian-winding number pairs ($H_1$, $W_1$), ..., ($H_n$, $W_n$). We partition the set \{($H_i$, $W_i$)$\mid$ $i=1,...,n$\} in three disjoint subsets: the training set, the validation set and the test set.
\item[3)] \textbf{Training on eigenvectors in real space:} since each Hamiltonian $H_i$ is represented by an $N\times N$ matrix, each one will provide $N$ eigenvectors $\mathbf{v}_i^{(1)}, \mathbf{v}_i^{(2)},...,\mathbf{v}_i^{(N)}$ to our data set. Our supervised learning algorithm of choice will take as inputs the real space eigenvectors $\mathbf{v}^{(j)}_i$ of each Hamiltonian $H_i$ in the training set and be trained to learn the winding number $W_i$ of their parent Hamiltonian $H_i$. Therefore, our dataset will consist of eigenvector-winding number pairs $(\mathbf{v}_i^{(j)}, W_i)$.
\item[4)]\textbf{Eigenvector ensembling:} in order to predict the phase of a system described by a particular Hamiltonian we need to take into account how each of its eigenvectors were classified. This amounts to performing ensemble learning on the eigenvectors of each Hamiltonian. In this work we estimate the phase probabilities for each Hamiltonian as the fraction of its eigenvectors that were classified in each phase.
\item[5)] \textbf{Bootstrapping:} We refine the phase probabilities for each Hamiltonian using a bootstrapping procedure, i.e., we repeat steps (1)-(4) $n_\text{exp}$ times, at each round sampling randomly a new training set from our grid in \textbf{t}-space. The final estimated probabilities are then arrived at by averaging the probabilities obtained in each experiment.
\end{enumerate} 

\vspace{.3cm}
Before continuing to the analyses of the SSH systems with the eigenvector ensembling algorithm, it will be timely to digress a moment and peek into the algorithm itself. The focus on eigenvectors (and hence the algorithm's name) as the input data to a machine learning algorithm of choice is a hallmark of the procedure as it differentiates it from related applications of machine learning to the study of phase transitions. The intuition that eigenvectors can be used in replacement of raw Hamiltonians can be grasped when we consider the spectral decomposition of a Hamiltonian $H$,    
\begin{equation}
\label{eigen_decomposition}
H = \sum_{i=1}^{N}\lambda^{(i)} \ket{\mathbf{v}^{(i)}}\bra{\mathbf{v}^{(i)}}
\end{equation}
where $\lambda^{(i)}$ is the eigenenergy corresponding to the eigenstate $\ket{\mathbf{v}^{(i)}}$. It is therefore clear that all information available from a Hamiltonian can be recovered from its spectral decomposition. By expressing the eigenvectors in a basis suitable to a particular problem (e.g. the real space basis chosen in this article), it becomes possible to investigate the properties of a set of Hamiltonians using the coordinates of eigenvectors in the chosen basis as features. Thus the eigenvector ensembling procedure described above provides a broad framework for the implementation of model explainability in applications to data-driven physics. 

\section{Numerical experiments}
\label{numerical_experiments}
We performed two numerical experiments with the eigenvector ensembling algorithm. The first experiment deals with the simplest case, the SSH model with nearest-neighbor hopping (here called SSH 1, figure \ref{ssh1}), while the second experiment uses the SSH model with first and second nearest-neighbor hoppings (here called SSH 2, figure \ref{ssh2}).

In each experiment our grid consisted of \nHam\ Hamiltonians uniformly distributed in the closed square $[-2,2]\times[-2,2]$ in the $t_1$-$t_2$ plane in parameter space. The goal in each experiment is to recover the corresponding phase diagram in 2D (two-dimensional) parameter space, figures \ref{ssh1} and \ref{ssh2}, from local lattice data in the much higher-dimensional real space (100D - in both experiments lattices have 50 unit cells, yielding 100$\times$100 Hamiltonian matrices).

This task is particularly hard near phase transition boundaries, where numerical computation of winding numbers become less stable. For this reason, when sampling the training set we only consider those Hamiltonians in the grid whose numerically computed winding numbers lie in a minimum range of $\epsilon = \epsilonValue$ from the correct winding number values. Therefore, a good performance metric is the accuracy measured at those Hamiltonians near phase transitions that are never used for training, and thus we assign them to the test set. The remaining Hamiltonians in the grid are split into training and validation sets as detailed in the subsections below.

As performance metrics, we report here both accuracy of predicted classes for eigenvectors as well as accuracy of predicted classes for Hamiltonians obtained from eigenvector ensembling. These accuracy scores are to be gauged against the baseline of a system that simply guesses the most frequent class for all Hamiltonians. Checking against this baseline is important because it indicates whether the decision trees are in fact learning the underlying patterns that relate real space coordinates to winding numbers, and therefore whether the associated information entropy signature is meaningful or not.

When generating the Hamiltonians we applied periodic boundary conditions to eliminate border effects. This should make recovering a topological signal from local eigenvector coordinates even harder, since in this case the translational symmetry of the systems should allow for no obvious way to distinguish between unit cells. The choice of periodic boundary conditions also implies that the information recovered from real space data comes from the bulk of the topological systems considered and therefore provides strong evidence for the existence of topological signatures in the bulk of such systems. 

\begin{figure}
\centering
\subfigure[]{\label{figexp1_exp:a}\includegraphics[width=.32\textwidth]{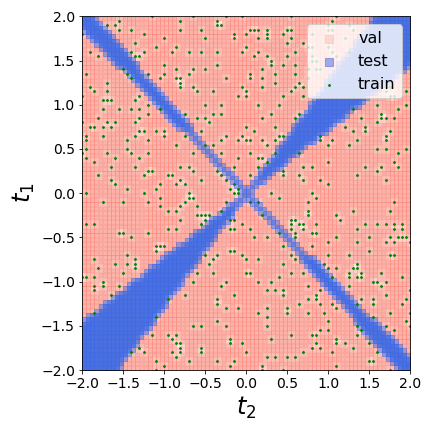}}
\subfigure[]{\label{figexp1_exp:b}\includegraphics[width=.32\textwidth]{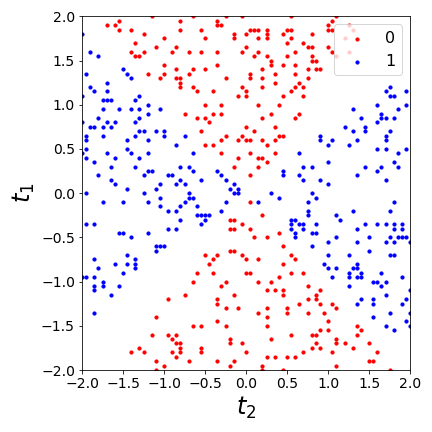}}
\subfigure[]{\label{figexp1_exp:c}\includegraphics[width=.32\textwidth]{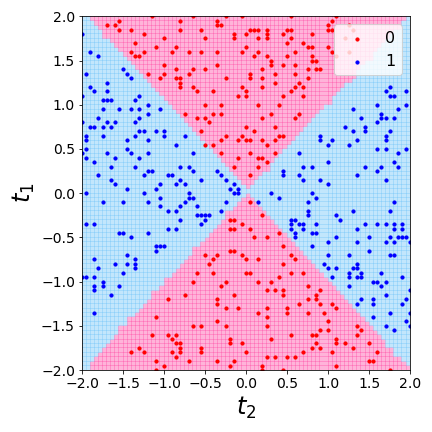}}
\caption{Visualization of a single iteration of experiment 1 (section \ref{experiment_1}) as seen from 2D parameter space. (a) Train/validation/test split. (b) Distribution of winding numbers in the training set. (c) Phase diagram learned from real space lattice data by combining a decision tree with eigenvector ensembling.}
\label{figexp1_exp}
\end{figure}

\begin{figure}
\centering
\subfigure[]{\label{figexp2_exp:a}\includegraphics[width=.32\textwidth]{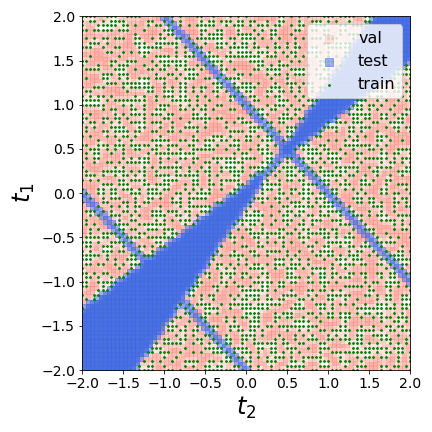}}
\subfigure[]{\label{figexp2_exp:b}\includegraphics[width=.32\textwidth]{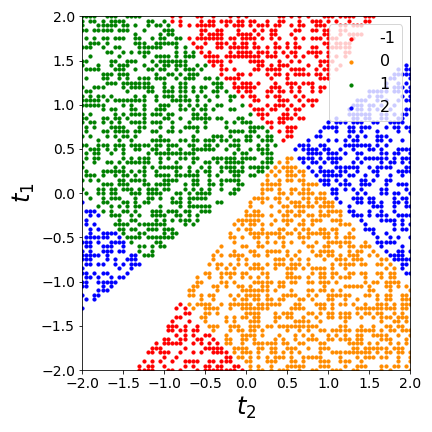}}
\subfigure[]{\label{figexp2_exp:c}\includegraphics[width=.32\textwidth]{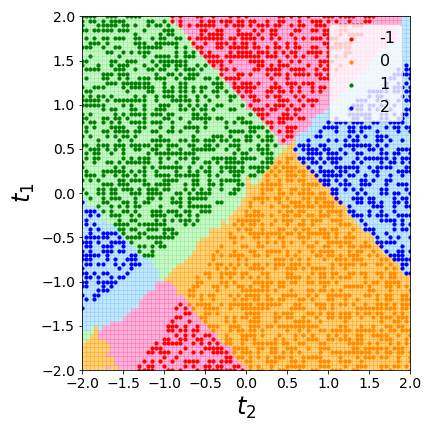}}
\caption{Visualization of a single iteration of experiment 2 (section \ref{experiment_2}) as seen from 2D parameter space. (a) Train/validation/test split. (b) Distribution of winding numbers in the training set. (c) Phase diagram learned from real space lattice data by combining a random forest with eigenvector ensembling.}
\label{figexp2_exp}
\end{figure}

\begin{figure}
\centering
\subfigure[]{\label{ssh1_heatmap_0}\includegraphics[width=.32\textwidth]{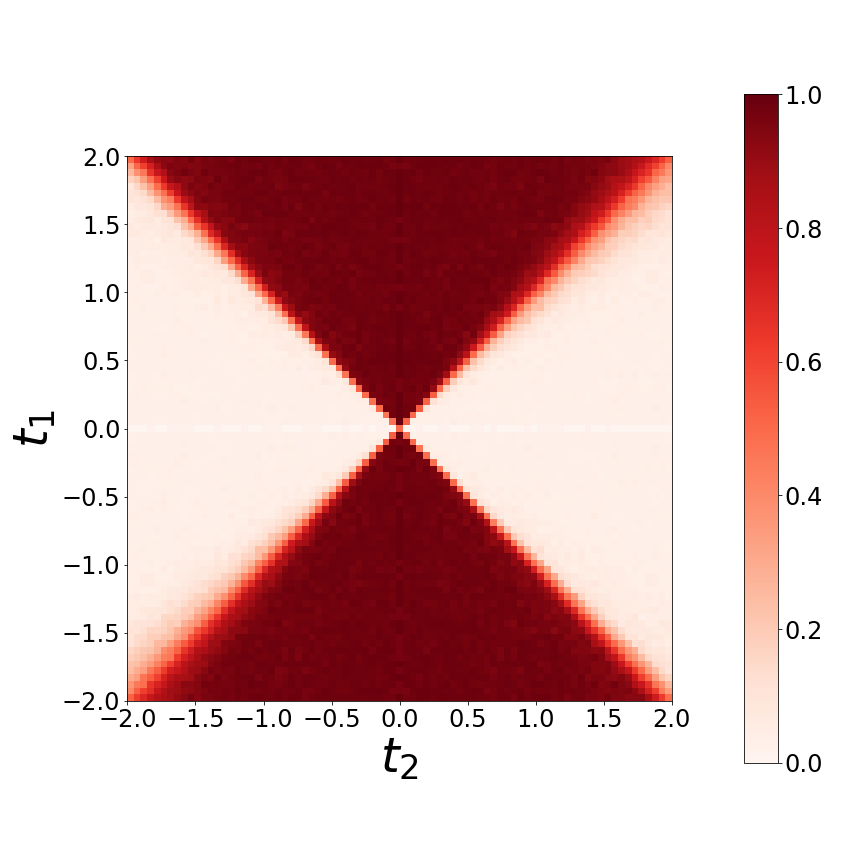}}
\subfigure[]{\label{ssh1_heatmap_1}\includegraphics[width=.32\textwidth]{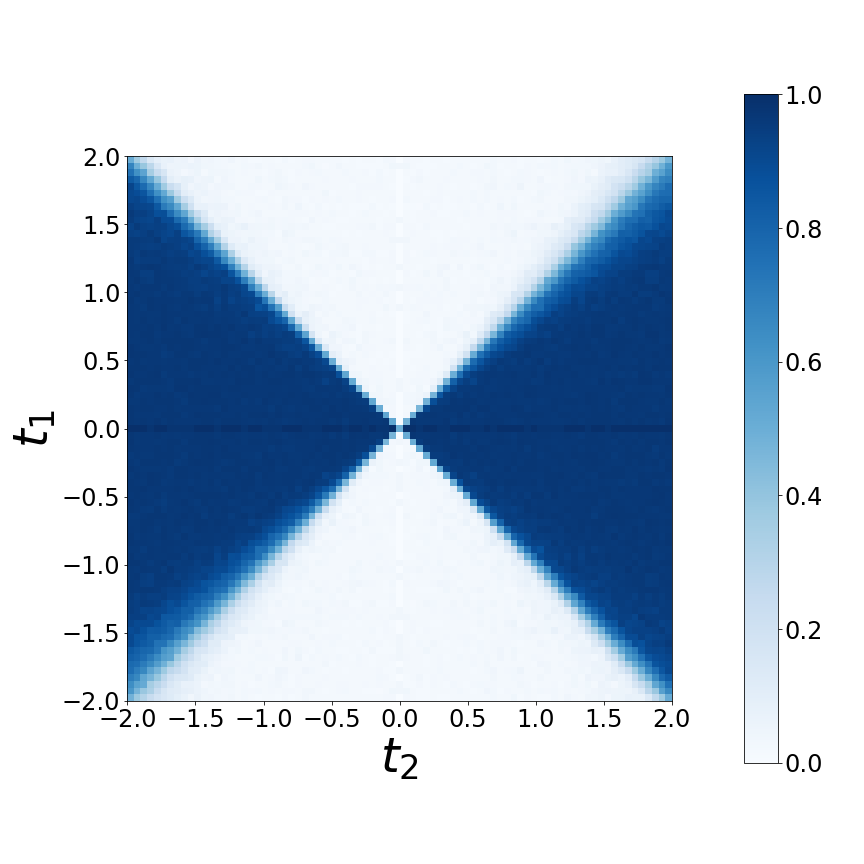}}
\subfigure[]{\label{ssh1_heatmap}\includegraphics[width=.32\textwidth]{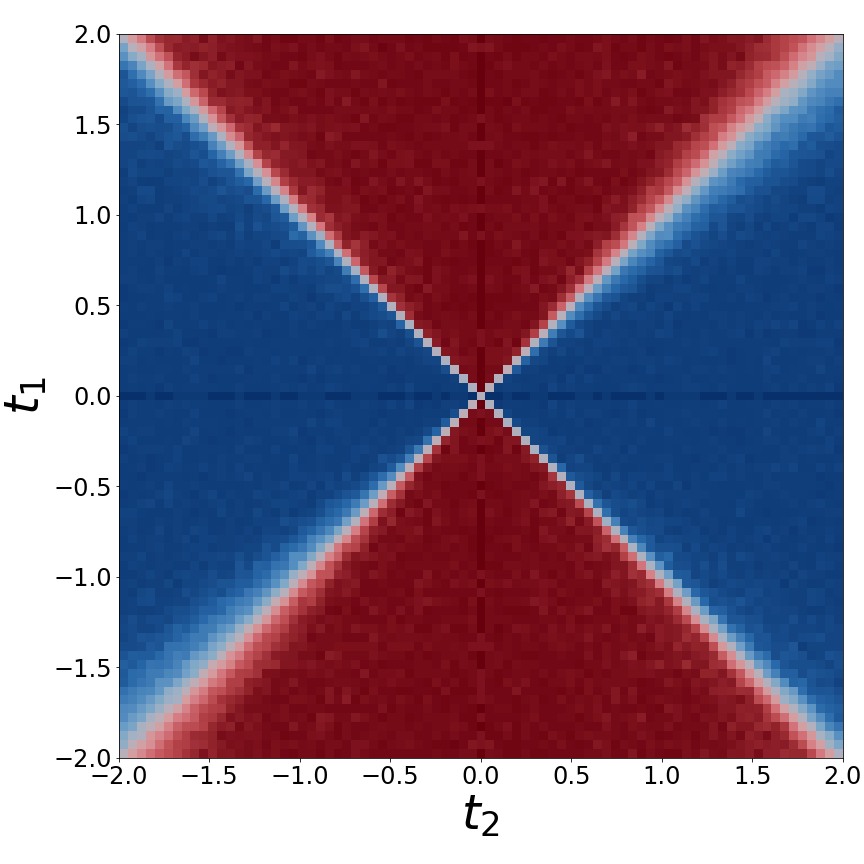}}
\caption{Probability heatmaps learned by a combination of decision trees with eigenvector ensembling from bulk real space eigenvector data in experiment 1 (section \ref{experiment_1}). Heatmaps were averaged across all 100 iterations of the experiment. (a) Probability heatmap showing the probability that a Hamiltonian in the grid has winding number equal to 0.  (b) Probability heatmap showing the probability that a Hamiltonian in the grid has winding number equal to 1. (c) The phase diagram resulting from heatmaps (a) and (b).}
\label{ssh1_heatmaps}
\end{figure}

\begin{figure}
\centering
\subfigure[]{\label{ssh2_heatmap_-1}\includegraphics[width=.32\textwidth]{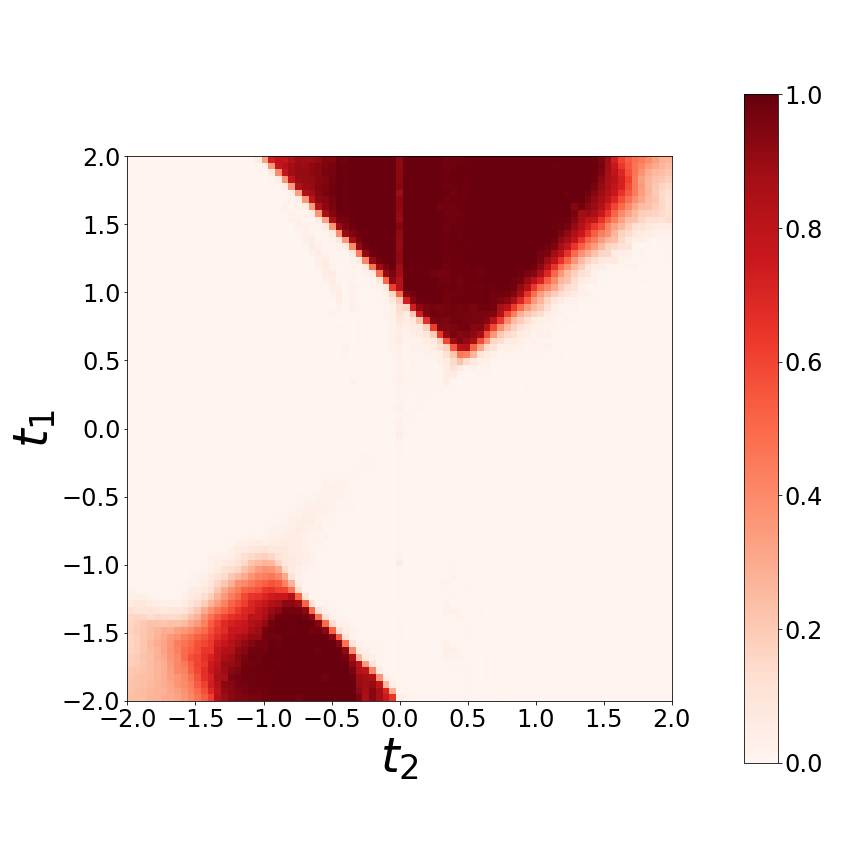}}
\subfigure[]{\label{ssh2_heatmap_0}\includegraphics[width=.32\textwidth]{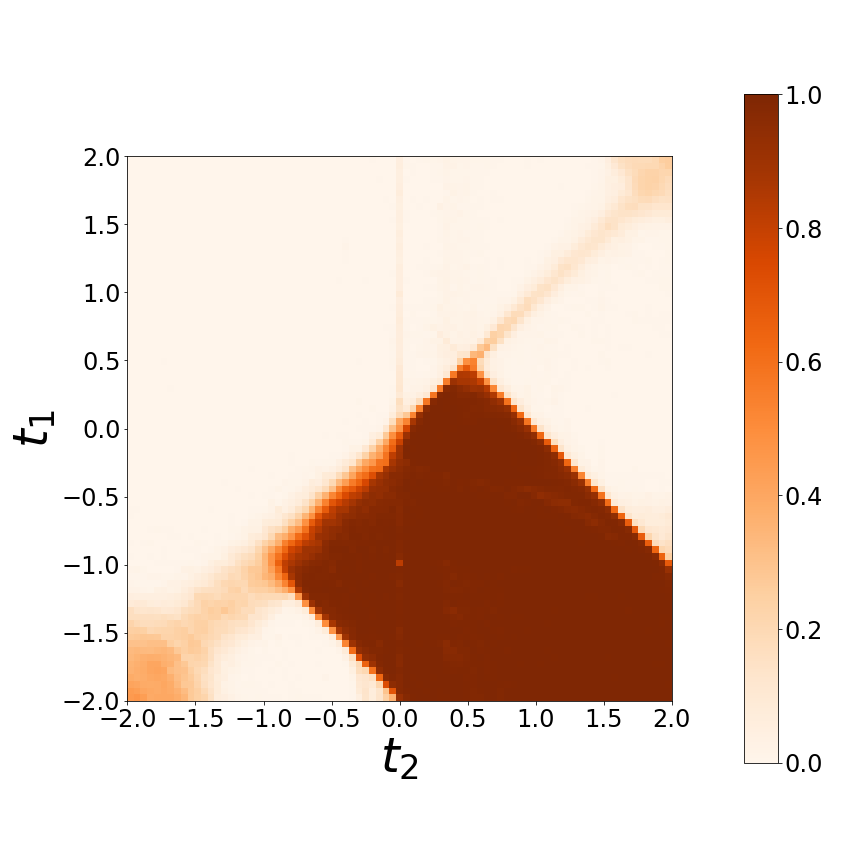}}
\subfigure[]{\label{ssh2_heatmap_1}\includegraphics[width=.32\textwidth]{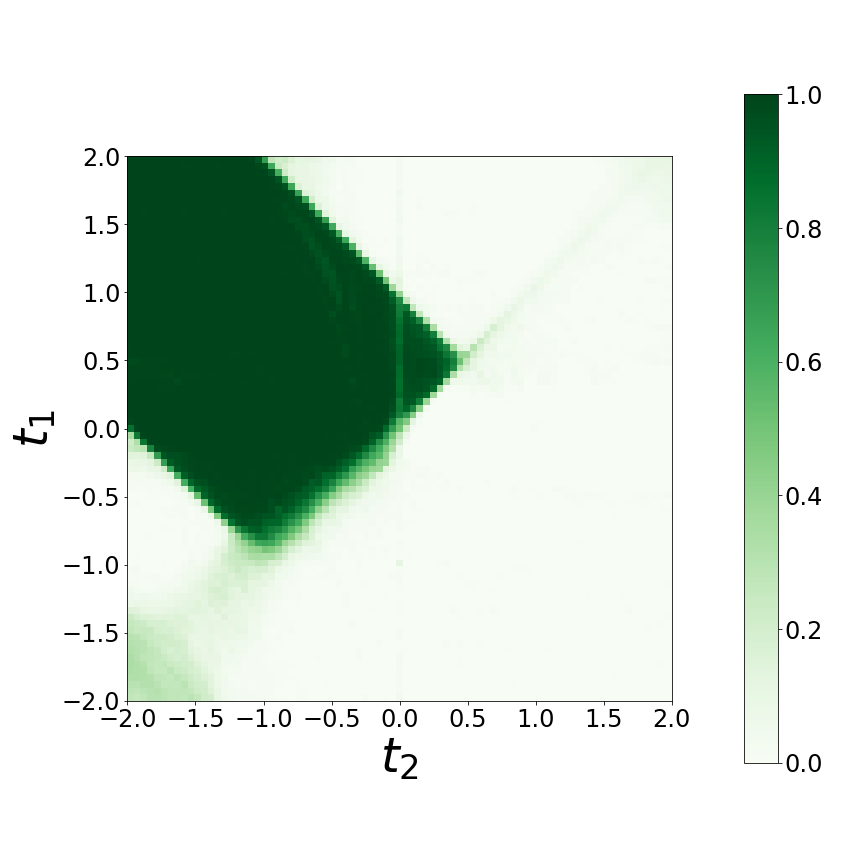}}
\subfigure[]{\label{ssh2_heatmap_2}\includegraphics[width=.32\textwidth]{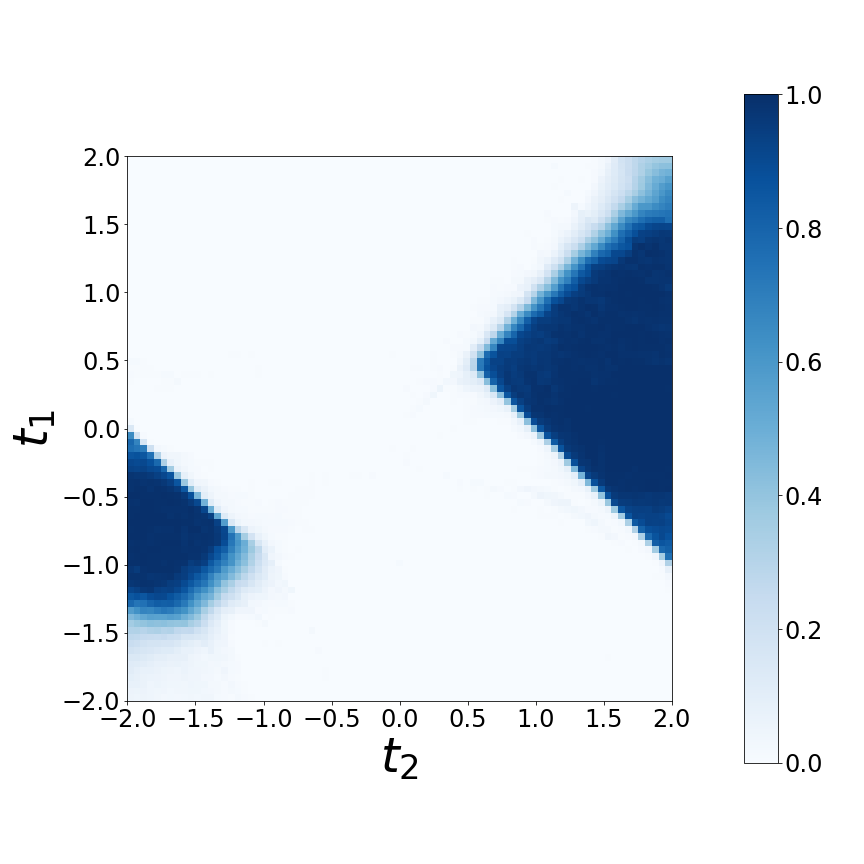}}
\subfigure[]{\label{ssh2_heatmap}\includegraphics[width=.32\textwidth]{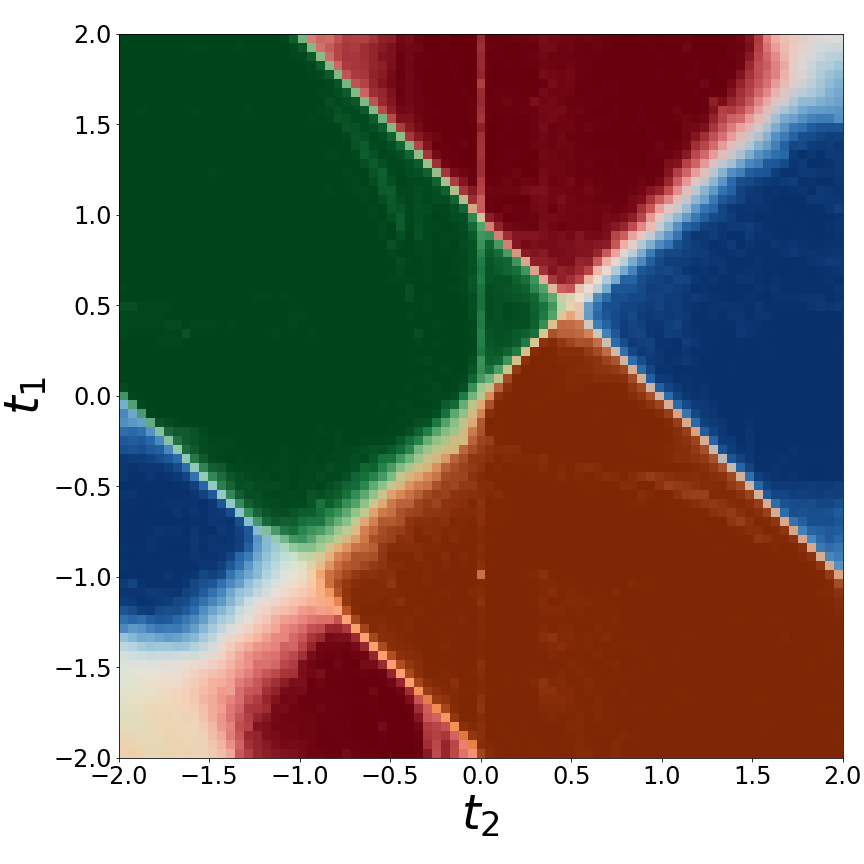}}
\caption{Probability heatmaps learned by a combination of random forests with eigenvector ensembling from bulk real space eigenvector data in experiment 2 (section \ref{experiment_2}). Heatmaps were averaged across all 100 iterations of the experiment. (a) Probability heatmap showing the probability that a Hamiltonian in the grid has winding number equal to -1.  (b) Probability heatmap showing the probability that a Hamiltonian in the grid has winding number equal to 0. (c) Probability heatmap showing the probability that a Hamiltonian in the grid has winding number equal to 1. (d) Probability heatmap showing the probability that a Hamiltonian in the grid has winding number equal to 2. (e) The phase diagram resulting from heatmaps (a)-(d).}
\label{ssh2_heatmaps}
\end{figure}

Figures \ref{figexp1_exp} and \ref{figexp2_exp} respectively illustrate single iterations of experiment 1 (section \ref{experiment_1}) and experiment 2 (section \ref{experiment_2}) as seen from parameter space. The accuracy statistics presented in the following subsections, as well as the probability heatmaps and recovered phase diagrams shown in figures \ref{ssh1_heatmaps} and \ref{ssh2_heatmaps} were obtained after bootstrapping each experiment $n_{exp}$ = \nExp\ times. Thus, each probability heatmap shown in figures \ref{ssh1_heatmap_0}, \ref{ssh1_heatmap_1} and \ref{ssh2_heatmap_-1} to \ref{ssh2_heatmap_2} represents the averaged fraction of eigenvectors of each Hamiltonian in the grid that were classified with a given winding number across 100 experiments. The recovered phase diagrams \ref{ssh1_heatmap} and \ref{ssh2_heatmap} are constructed by superposing the corresponding probability heatmaps. As these figure make clear, the recovered phase diagrams faithfully portray the true phase diagrams in figure \ref{fig:phasediagrams}, with clear phase transition lines appearing in the regions of highest uncertainty.

The numerical experiments with the eigenvector ensembling algorithm described in the next subsections were implemented in Python using the scikit-learn module \cite{scikit-learn,sklearn_api}. 

\subsection{Experiment 1: Learning a first-neighbor hopping SSH model with decision trees}
\label{experiment_1}
Our test set in this experiment contained \sshOneTestNHam\ Hamiltonians (approx. \sshOneTestFracHam\% of all data). Of the remaining \sshOneTrainPlusValNHam\ Hamiltonians, \sshOneTrainNHam\ were randomly assigned to the training set (approx. \sshOneTrainFracHam\%) and \sshOneValNHam\ (approx. \sshOneValFracHam\%) were used to compute validation scores at each iteration. These proportions between training and validation sets are such that approximately 10\%  of Hamiltonians from outside of the test set were used for training at each iteration. The composition of the train + validation set for this experiment was \sshOneTrainPlusValWindZeroFracHam\% of Hamiltonians with winding number $W$ = 0 and \sshOneTrainPlusValWindOneFracHam\% with winding number $W$ = 1. The composition of the test set was \sshOneTestWindZeroFracHam\% of Hamiltonians with winding number $W$ = 0 and \sshOneTestWindOneFracHam\% with winding number $W=1$. Our learning algorithm of choice for this experiment was a simple decision tree model \cite{breiman2017classification}.

The bootstrap allows us to collect several statistics to evaluate performance. In particular, we report mean accuracies on training eigenvectors (\xOneEigTrain), validation eigenvectors (\xOneEigVal) and test eigenvectors (\xOneEigTest). Eigenvector ensembling substantially improved mean accuracies for Hamiltonians. These were \xOneHamTrain\ for training Hamiltonians, \xOneHamVal\ for validation Hamiltonians and \xOneHamTest\ for test Hamiltonians. When compared with the baseline test accuracy of \sshOneHamTestBaseline\ of a system that predicts the whole test set as having winding number $W=1$, the accuracy achieved on test Hamiltonians indicates that the decision trees indeed learned the patterns that relate real space coordinates to winding numbers.

The probability heatmaps and phase diagram learned by the combination of decision trees with eigenvector ensembling used in experiment 1 are shown in figure \ref{ssh1_heatmaps}. 

\subsection{Experiment 2: Learning a first- and second-neighbor hoppings SSH model with random forests}
\label{experiment_2}

This task is considerably more difficult than the previous one due to the higher number of classes and the fact that some of the labels encompass disconnected regions. For this reason, instead of using a single decision tree, we upgraded our model to a random forest \cite{Breiman2001} with \nEst\ decision trees. Our data set consisted of \sshTwoTestNHam\ (\sshTwoTestFracHam\%) test Hamiltonians. The remaining \sshTwoTrainPlusValNHam\ Hamiltonians are randomly split in half between training and validation sets at each iteration, giving \sshTwoTrainNHam\ (\sshTwoTrainFracHam\%) training Hamiltonians and \sshTwoValNHam\ (\sshTwoValFracHam\%) validation Hamiltonians. The distribution of winding numbers for the Hamiltonians in the train + validation set for this experiment was $W$ = -1 (\sshTwoTrainPlusValWindMinusOneFracHam\%), $W$ = 0 (\sshTwoTrainPlusValWindZeroFracHam\%), $W$ = 1 (\sshTwoTrainPlusValWindOneFracHam\%) and $W$ = 2 (\sshTwoTrainPlusValWindTwoFracHam\%). The distribution of winding numbers for the Hamiltonians in the test set was $W$ = -1 (\sshTwoTestWindMinusOneFracHam\%), $W$ = 0 (\sshTwoTestWindZeroFracHam\%), $W$ = 1 (\sshTwoTestWindOneFracHam\%) and $W$ = 2 (\sshTwoTestWindTwoFracHam\%).

Mean accuracies across \nExp\ repetitions of experiment 2 were \xTwoEigTrain\ for training eigenvectors, \xTwoEigVal\ for validation eigenvectors and \xTwoEigTest\ for test eigenvectors. Mean accuracies resulting from eigenvector ensembling were \xTwoHamTrain\ for training Hamiltonians, \xTwoHamVal\ for validation Hamiltonians and \xTwoHamTest\ for test Hamiltonians. The large accuracy gain achieved by eigenvector ensembling in the test set (going from \xTwoEigTest\ eigenvector accuracy to \xTwoHamTest\ Hamiltonian accuracy) attests to its power. The effectiveness of eigenvector ensembling is also evident from the much worse performance (\sshTwoHamTestBaseline) achieved by a baseline system that simply guesses $W=2$ for all test Hamiltonians in this experiment.

The probability heatmaps and phase diagram learned by the combination of random forests with eigenvector ensembling used in experiment 2 are shown in figure \ref{ssh2_heatmaps}.

\section{Information entropy signatures}
\label{information_entropy_signatures}

We now analyze how the algorithm was able to recover a global property of the Hamiltonians (their topological phase) from bulk local features (real space eigenvector coordinates on each lattice site). Alongside the fact that decision trees and random forests are very easy to train and visualize, the other reason that led us to test the eigenvector ensembling algorithm with them was that they allow us to check which features (and thus which lattice sites) were most informative in training.

The (normalized) relevance of a feature is given by how much it reduces a loss function (in this paper, the Shannon information entropy of ensembles of eigenvectors). By averaging normalized relevances as measured by reduction in the information entropy of ensembles of real space eigenvectors across $n_{exp}$ = 100 iterations of both experiment 1 (section \ref{experiment_1}) and experiment 2 (section \ref{experiment_2}) we recovered Shannon entropy signals that reveal which lattice sites were consistently more relevant in learning topological phases from data in real space for each experiment. These signals are the information entropy signatures of each topological phase transition. 

We should briefly comment on the possibility of using the Gini impurity of ensembles of eigenvectors \cite{friedman2001elements} instead of their Shannon entropy as cost function. This would similarly lead us to Gini impurity signatures of topological phase transitions. Given that in the examples analysed in this paper the Gini impurity signatures and the Shannon entropy signatures were very similar, the larger familiarity of a general physics audience with the latter influenced us to choose it over the former. Nevertheless, the question of which split criterion  should be used when training decision trees is as of yet largely undecided \cite{raileanu2004theoretical} and may be of relevance in the analysis of other physical systems than the ones studied here.       

The bar plots in figure \ref{feature_importances} show how informative each lattice site was in learning topological phases for each experiment. They represent the information entropy signatures along the lattices in each SSH system. For experiment 1 (section \ref{experiment_1}), only six lattice sites \sitesExpOne\ corresponding to the two sharp peaks seen in figure \ref{feature_importances_ssh1} contributed approximately 70\% of total reduction in Shannon entropy. Similarly, approximately 30\% of total reduction in the Shannon information entropy of eigenvector data from experiment 2 (section \ref{experiment_2}) was achieved by eighteen lattice sites \sitesExpTwo\ distributed along the three peaks in figure \ref{feature_importances_ssh2}. As we shall see in section \ref{topological_feature_engineering}, the information entropy signatures can be used to compress the topological information in SSH lattices.

The information entropy signatures presented here have some interesting subtleties. Although they give us a visualization of how important each lattice site was in determining the topological phases of Hamiltonians, they actually express a global property of the whole lattice. In section \ref{emergent_information_entropy_wave_functions}, where we develop a quantum formalism for the information entropy signatures obtained in this section, these seemingly antagonistic conceptions shall be harmonized. What is important to emphasize at this point is that an information entropy signature should not be naively taken at face value: a lattice site that appears unimportant in an information entropy signature plot may not be unimportant or void of topological information by itself.

To give a concrete example, reduction in Shannon entropy tends to be distributed among highly correlated variables. This implies that if only a single lattice site in a highly correlated subset is used by a learning algorithm, it will likely inherit most of the reduction in Shannon entropy from the other correlated lattice sites that were not taken into account by the algorithm. The corollary of this fact is that lattice sites that carry redundant information that is also available from other lattice sites tend to have decreased importance in the information entropy signature.  In this regard the information entropy signatures presented here express a summary of relations between lattice sites and are therefore intrinsically global.

Each of the information entropy signatures shown in figure \ref{feature_importances} captures a general pattern that persists regardless of the length of the lattice (i.e., the number of unit cells) used to compute them. In fact, by rerunning each experiment with longer lattices we have verified that the signals in figures \ref{feature_importances_ssh1} and \ref{feature_importances_ssh2} appear to converge to well defined continuous density functions in the macroscopic limit. They are not, therefore, artifacts of particular choices of hyperparameters used to run the eigenvector ensembling algorithm. The information entropy signatures for longer lattices are presented in the section \textbf{Numerical explorations on longer lattices} in the Supplementary Material.

\begin{table}
\begin{tabular}{||c@{\hskip 0.3in} c@{\hskip 0.3in} c@{\hskip 0.3in} c@{\hskip 0.3in} c@{\hskip 0.3in} c||} 
\hline
\SSHSys               &\Feat             &\ValEig            &\TestEig           &\ValHam              &\TestHam         \tableRowHeaderEnd 
1                     &\xOne             &\xOneEigVal        &\xOneEigTest       &\xOneHamVal          &\xOneHamTest       \tableRowEnd
1                     &\xSOne            &\xSOneEigVal       &\xSOneEigTest      &\xSOneHamVal         &\xSOneHamVal       \tableRowEnd
1                     &\xcOne            &\xcOneEigVal       &\xcOneEigTest      &\xcOneHamVal         &\xcOneHamTest      \tableRowEnd
1                     &\xcEOne           &\xcEOneEigVal      &\xcEOneEigTest     &\xcEOneHamVal        &\xcEOneHamTest     \tableRowEnd
1                     &\xcSEOne          &\xcSEOneEigVal     &\xcSEOneEigTest    &\xcSEOneHamVal       &\xcSEOneHamTest    \tableRowEnd
1                     &\xsOne            &\xsOneEigVal       &\xsOneEigTest      &\xsOneHamVal         &\xsOneHamTest      \tableRowEnd
1                     &\xsOOne           &\xsOOneEigVal      &\xsOOneEigTest     &\xsOOneHamVal        &\xsOOneHamTest     \tableRowEnd
1                     &\xsSOOne          &\xsSOOneEigVal     &\xsSOOneEigTest    &\xsSOOneHamVal       &\xsSOOneHamTest    \tableRowEnd
2                     &\xTwo             &\xTwoEigVal        &\xTwoEigTest       &\xTwoHamVal          &\xTwoHamTest       \tableRowEnd
2                     &\xSTwo            &\xSTwoEigVal       &\xSTwoEigTest      &\xSTwoHamVal         &\xSTwoHamVal       \tableRowEnd
2                     &\xcTwo            &\xcTwoEigVal       &\xcTwoEigTest      &\xcTwoHamVal         &\xcTwoHamTest      \tableRowEnd
2                     &\xcETwo           &\xcETwoEigVal      &\xcETwoEigTest     &\xcETwoHamVal        &\xcETwoHamTest     \tableRowEnd
2                     &\xcSETwo          &\xcSETwoEigVal     &\xcSETwoEigTest    &\xcSETwoHamVal       &\xcSETwoHamTest    \tableRowEnd
2                     &\xsTwo            &\xsTwoEigVal       &\xsTwoEigTest      &\xsTwoHamVal         &\xsTwoHamTest      \tableRowEnd
2                     &\xsOTwo           &\xsOTwoEigVal      &\xsOTwoEigTest     &\xsOTwoHamVal        &\xsOTwoHamTest     \tableRowEnd
2                     &\xsSOTwo          &\xsSOTwoEigVal     &\xsSOTwoEigTest    &\xsSOTwoHamVal       &\xsSOTwoHamTest    \tableRowEnd
\hline
\end{tabular}
\caption{\label{accuracy_scores}Accuracy scores of numerical experiments. The features used to train the decision trees (SSH 1) or random forests (SSH 2) in the numerical experiments with SSH systems were as follows. Real space features: all real space lattice sites (\xalpha) (sections \ref{experiment_1} and \ref{experiment_2}); real space lattice sites from subset $\mathcal{S}_{\alpha}$ (\xSalpha) (equation \eqref{lattice_subsets}). DCT topological features: all DCT topological features (\xcalpha) (equation \eqref{DCT}); DCT topological features from subset $\mathcal{E}_\alpha$ (\xcEalpha) (equation \eqref{DCT_subsets}); DCT topological features from subset $\mathcal{E}_\alpha$ computed using only real space lattice sites $\mathcal{S}_{\alpha}$ (\xcSEalpha) (equation \eqref{DCT_very_compressed}). DST topological features: all DST topological features (\xsalpha) (equation \eqref{DST}); DST topological features from subset $\mathcal{O}_\alpha$ (\xsOalpha) (equation \eqref{DST_subsets}); DST topological features from subset $\mathcal{O}_\alpha$ computed using only real space lattice sites $\mathcal{S}_{\alpha}$ (\xsSOalpha) (equation \eqref{DST_very_compressed}). }
\end{table}

\section{Topological feature engineering and lattice compression}
\label{topological_feature_engineering}

In this section, we explore the information entropy signatures obtained in section \ref{information_entropy_signatures} from a signal processing perspective. As we shall see, this will enable us to perform two important unsupervised learning tasks on topological systems: dimensionality reduction and data compression. 

\begin{figure}
\centering
\subfigure[]{\label{feature_importances_ssh1}\includegraphics[width=0.49\textwidth]{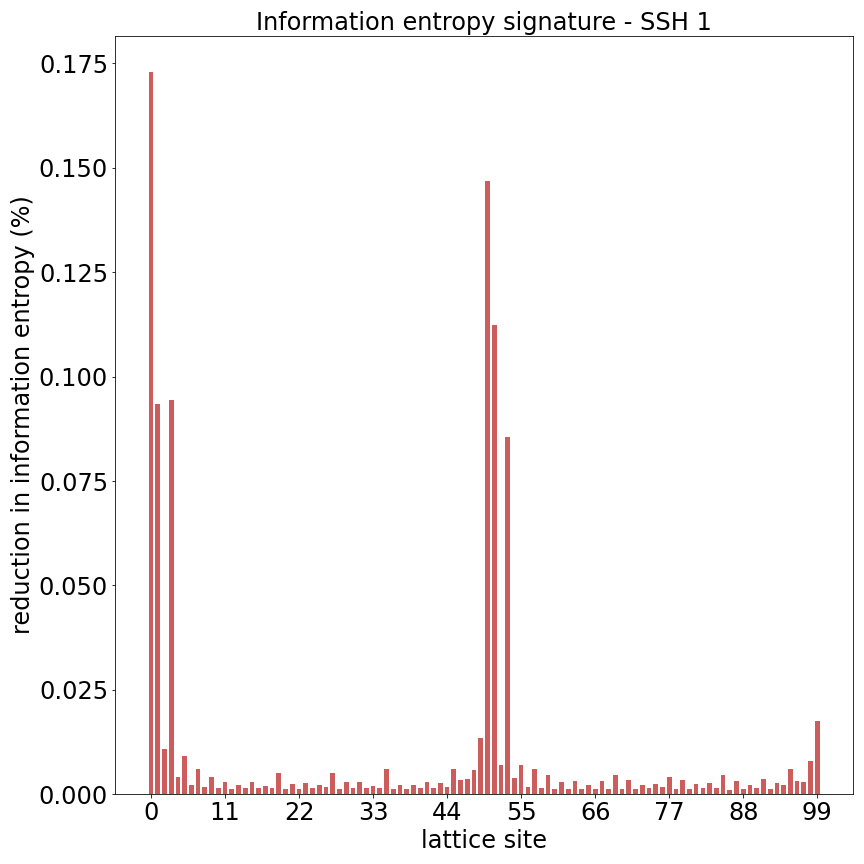}}\quad
\subfigure[]{\label{feature_importances_ssh2}\includegraphics[width=0.49\textwidth]{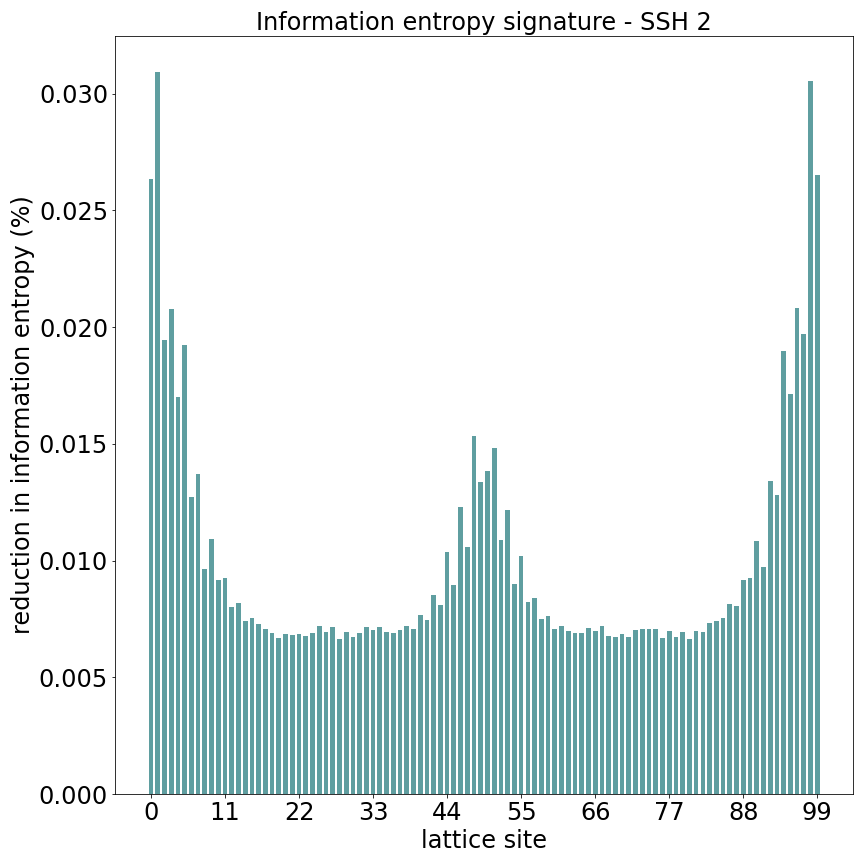}}\quad
\caption{Information entropy signatures of the topological phase transitions from the numerical experiments of section \ref{numerical_experiments}. (a) In experiment 1 (section \ref{experiment_1}), the two sharp peaks in the Shannon entropy signal account for approximately 70\% of reduction in information entropy. (b) In experiment 2 (section \ref{experiment_2}), the three visible peaks account for approximately 30\% of reduction in information entropy.}
\label{feature_importances}
\end{figure}

As a cursory glance at figures \ref{feature_importances_ssh1} and \ref{feature_importances_ssh2} seems to suggest, the existence of peaks and symmetries in information entropy signatures can be exploited as a powerful dimensionality reduction tool: by keeping only the most relevant lattice sites in an information entropy signature, the amount of data needed to characterize the topological phase of a SSH system can be largely decreased with little loss of information. Nevertheless, given that the topological phase of a SSH system is a global property of the whole lattice, it is natural to expect that it should be possible to engineer global features from real space coordinates. Here we show how the symmetries in real space eigenvectors and information entropy signatures can be exploited to engineer new global topological features, leading to an effective compression of topological information.



Given a real space eigenvector $x[m]$, we can compute its coordinates in wavevector space using the Discrete Fourier Transform (DFT),  
\begin{equation}
\label{discrete_fourier_transform}
\hat{x}[n] = \frac{1}{\sqrt{N}}\sum_{m=0}^{N-1}x[m]e^{-i\frac{2\pi}{N}nm}, \quad n=0,\dots,N-1.
\end{equation}
Since the choice of phase of the real space eigenvectors was such that they were all in $\mathbb{R}^{N}$, the eigenvectors in wavevector space computed from equation \eqref{discrete_fourier_transform} will be Hermitian vectors in $\mathbb{C}^N$. The Hermitian symmetry of the wavevector space eigenvectors manifests itself mathematically in closed lattices with periodic boundary conditions as 
\begin{equation}
\label{hermitian_symmetry}
\hat{x}[\overline{k}] = \hat{x}^*[\overline{N-k}], \quad k = 0,\dots, N-1,
\end{equation}
where  we have used the notation $\overline{k} = k \mod N$. 

Equation \eqref{hermitian_symmetry} forks into two natural paths to topological feature engineering. In the first path, we exploit the fact that the real part of $\hat{x}[m]$ is even-symmetric around the reciprocal lattice sites $0$ and $\frac{N}{2}$. This leads us to the Discrete Cosine Transform (DCT) topological features,
\begin{equation}
\label{DCT}
\hat{x}^{c}[n] = x[0] + (-1)^n x[M-1] + \sum_{m=1}^{M-2} 2x[m]\cos\bigg(\frac{\pi}{M-1}nm\bigg), \quad n=0,\dots,M-1, 
\end{equation}
where $M = \frac{N}{2}+1$. The second path capitalizes on the fact that the imaginary part of $\hat{x}[m]$ is odd-symetric around the reciprocal lattice sites $0$ and $\frac{N}{2}$, thus yielding the Discrete Sine Transform (DST) topological features,
\begin{equation}
\label{DST}
\hat{x}^{s}[n] = \sum_{m=0}^{M-1} 2x[m+1]\sin\bigg(\frac{\pi}{M+1}(n+1)(m+1)\bigg), \quad n=0,\dots,M-1,
\end{equation} 
where $M = \frac{N}{2}-1$. 

The topological features in both equations \eqref{DCT} and \eqref{DST} are generated from only half of the real space lattice, i.e. the sites $0 \leq l \leq \frac{N}{2}$. This is due to the fact that each equation assumes that the eigenvectors are even-symmetric (DCT topological features) or odd-symmetric (DST topological features) around the lattice sites $0$ and $\frac{N}{2}$ in real space as well. While these assumptions are strictly true for the eigenvector representations in wavevector space, they are not generally true for the real space representations. Therefore, equations \eqref{DCT} and \eqref{DST} achieve lattice compression by keeping only half of the real space eigenvector coordinates and imposing the corresponding boundary conditions (even symmetry or odd symmetry) on the lattice sites $0$ and $\frac{N}{2}$ to extrapolate the information from one half of the lattice to the other. 

The topological feature engineering techniques described above are commonly employed in several applications of Digital Signal Processing like audio and image processing and, most importantly here, data compression. As equations \eqref{hermitian_symmetry}-\eqref{DST} show, signal transforms such as the DCT and DST profit from the redundance of information arising from the existence of certain symmetries in signals, allowing us to write a signal of length $N$ in terms of at most $M = \frac{N}{2}+1$ features. 
   
We ran several numerical experiments to evaluate if the DCT and DST topological features defined in equations \eqref{DCT} and \eqref{DST}  are able to efficiently encode the topological information existing in SSH lattices. The accuracy scores obtained in each experiment are listed in table \ref{accuracy_scores}, where we also report the accuracy scores of the numerical experiments of section \ref{numerical_experiments}.

The lattice compression strategies tested in this work were: \emph{i)} learning topological phases from only a subset of real space lattice sites; \emph{ii)} learning topological phases from the DCT or DST engineered features of equations \eqref{DCT} and \eqref{DST}; \emph{iii)} learning topological phases from a fraction of the DCT or DST topological features and \emph{iv)} learning topological phases from a fraction of the DCT or DST features, computed from only a fraction of real space lattice sites. 

In strategy \emph{i)} the lattice sites used in the SSH 1 and SSH 2 systems were
\begin{equation}
\label{lattice_subsets}
\mathcal{S}_1 = \sitesSOne \quad \text{and} \quad  \mathcal{S}_2 =\sitesSTwo.   
\end{equation}
Note that for the nearest-neighbor SSH systems $\mathcal{S}_1$ corresponds to the four most informative sites such that $0\leq l \leq 50$ as indicated by the information entropy signature in figure \ref{feature_importances_ssh1}. Similarly, for the first- and second- nearest neighbors SSH systems $\mathcal{S}_2$ corresponds to the twelve most informative sites such that $0\leq l \leq 50$ in the corresponding information entropy signature in figure \ref{feature_importances_ssh2}. In table \ref{accuracy_scores}, the features used in strategy \emph{i)} are referred to as \xSOne and \xSTwo, according to the SSH systems they relate to.  

In strategy \emph{ii)}, the full set of DCT or DST topological features written in equations \eqref{DCT} and \eqref{DST} were used. These features are denoted in table \ref{accuracy_scores} by \xcOne, \xsOne, \xcTwo, \xsTwo, according to which set of topological features and SSH systems they refer to. 

Similarly to strategy \emph{i)}, in strategy \emph{iii)} we selected the most informative DCT topological features of both SSH 1 and SSH 2 systems that were obtained from strategy \emph{ii)},
\begin{equation}
\label{DCT_subsets}
\mathcal{E}_1 = \sitesEOne \quad \text{and} \quad  \mathcal{E}_2 =\sitesETwo \qquad \text{(DCT)}  
\end{equation}
and the most informative DST topological features of both systems that were obtained from strategy \emph{ii)} as well,
\begin{equation}
\label{DST_subsets}
\mathcal{O}_1 =  \sitesOOne \quad \text{and} \quad  \mathcal{O}_2 = \sitesOTwo \qquad \text{(DST)}.  
\end{equation}
The features in strategy \emph{iii)} are denoted in table \ref{accuracy_scores} by \xcEOne, \xsOOne, \xcETwo, \xsOTwo,   according to the topological features and wavevector space subset used with each SSH system. 

The most aggressive lattice compression strategy tested in this work was strategy \emph{iv)}. It consists of using the DCT (DST) topological features of equation \eqref{DCT_subsets} (equation \eqref{DST_subsets}), but having computed these topological features using only the real space lattice sites given in equation \eqref{lattice_subsets}. Thus in strategy \emph{iv)} a lossy compression is performed both on real space features and the engineered DCT (DST) topological features. Mathematically, we can express the topological features used in strategy \emph{iv)} as follows:
\begin{equation}
\label{DCT_very_compressed}
\hat{x}_{\mathcal{S}, \mathcal{E}}^c[n] = \indicator{\mathcal{S}}[0]x[0] + (-1)^n\indicator{\mathcal{S}}[M-1] x[M-1] + \sum_{m \in \mathcal{S}^*} 2x[m]\cos\bigg(\frac{\pi}{M-1}nm\bigg), \quad n \in \mathcal{E} \qquad \text{(DCT)} 
\end{equation}
\begin{equation}
\label{DST_very_compressed}
\hat{x}_{\mathcal{S}, \mathcal{O}}^s[n] = \sum_{m \in \mathcal{S}}2x[m+1]\sin\bigg(\frac{\pi}{M+1}(n+1)(m+1)\bigg), \quad n \in \mathcal{O} \qquad \text{(DST)}
\end{equation}
where in equation \eqref{DCT_very_compressed} the notation $\mathbbm{1}_{\mathcal{S}}[l]$ stands for the indicator function of the lattice subset $\mathcal{S}$ and $\mathcal{S}^*$ is the complement of $\{0,M-1\}$ with respect to $\mathcal{S}$, 
\begin{equation}
\label{indicator_function}
\indicator{\mathcal{S}}[l] = \begin{cases}
      1 & \text{if } l \in \mathcal{S} \\
      0 & \text{otherwise}
    \end{cases}, \qquad \mathcal{S}^* = \mathcal{S}{\backslash} \{0,M-1\}.
\end{equation}
The features engineered in strategy \emph{iv)} are denoted in table \ref{accuracy_scores} by \xcSEOne, \xsSOOne, \xcSETwo, \xsSOTwo, again referencing the type of topological features used, which components in real and wavevector space engineered them and the appropriate SSH systems. 

The results shown in table \ref{accuracy_scores} bring some startling surprises. For example, the topological phase transition boundaries of SSH 1 systems can be learned using only the four real space lattice sites $\mathcal{S}_1$ with virtually no loss in accuracy, as seen from eigenvector and Hamiltonian accuracy scores for the features \xSOne. The same appears to be true to SSH 2 systems, where the twelve real space lattice sites $\mathcal{S}_2$ corresponding to the features \xSTwo\ in the table produce accuracy scores at near the same level as using the whole lattice.    

Even more striking is the performance achieved by the compressed DCT topological features \xcSEalpha\ defined in equation \eqref{DCT_very_compressed}. For SSH 1 systems, they perform on par with using the full set of real space features \xOne, while for SSH 2 systems a small loss in accuracy is incurred relative to the full set of real space features \xTwo.

Another interesting insight comes from comparing the accuracy scores obtained with the DCT topological features versus the  DST topological features. The latter have poorer performance than the former, as is indicated by the sharp drops in accuracy scores obtained with the DST topological features in both SSH 1 and SSH 2 systems. This may be related to the fact that the odd-symmetric boundary conditions imposed on DST topological features imply discarding the lattice sites 0 and $50$, which correspond to sharp peaks in the information entropy signatures of figures \ref{feature_importances_ssh1} and \ref{feature_importances_ssh2}.  

The accuracy scores obtained with the real space features \xSalpha\ and the DCT topological features \xcSEalpha, both of which use information from a small fraction of real space lattice sites, demonstrate that learning topological phases from local real space data in the bulk is still possible even for small subsets of lattice sites. In this sense, key topological information can be said to be localized on few sites in the lattice. We refer the reader to the section \textbf{Learning topological phases from real space data} in the Supplementary Material for a discussion of how this is possible.       

\section{Emergent information entropy wave functions}
\label{emergent_information_entropy_wave_functions}

The information entropy signatures that we have been investigating pose an immediate theoretical question: how can a signal that is locally defined arise from a global property of the whole SSH lattice?    
In this section we venture into a theoretical exploration of the information entropy signatures in the hope of elucidating this issue. Our goal is to arrive at a theoretical framework that will allow us to interpret the information entropy signatures in terms of quantum mechanics.  

We can think of the Shannon information entropy signatures in figures \ref{feature_importances_ssh1} and \ref{feature_importances_ssh2} as discrete information entropy mass functions that, in the continuum (i.e. macroscopic) limit of an infinite chain, lead to local entropy density functions along the lattices, which themselves become 1D manifolds. By mapping the lattice to a  partition of the 1D manifold, the cumulative distribution of topological information in the continuum limit will be given by

\begin{equation}
\label{entropy_density}
F_S(x) = \int_{\ell}^{(x)}\rho_S(x')dx'
\end{equation}
where $\rho_S(x)$ is the local information entropy density function in the continuum limit and $x$ is defined by the coordinate system specified on the 1D manifold $\ell$. The index $S$ is meant to emphasize that in this paper we have used Shannon's definition of entropy to arrive at the information entropy signatures as opposed to e.g. Gini impurity. 

Our use of periodic boundary conditions implies that the coordinate $x$ should be defined on the circle $ S^1 = [0, 1]/R$, where $R$ is the equivalence relation in $[0,1]$ defining the circle $S^1$ , 
\begin{equation}
x \text{ } R \text{ } y \iff x=y \text{ or } (x,y)\in \{(0,1), (1,0)\}.
\end{equation}
For open boundary conditions, the spatial coordinate $x$ is defined on the closed interval $[0,1]$ or, in the case of infinite systems, $\mathbb{R}$. However, for the sake of generality, we shall continue to use the caligraphic $\ell$ to denote an arbitrary 1D manifold in this section. 

Given the quantum nature of the phase transitions being discussed, the information entropy density function $\rho_S(x)$ can be naturally interpreted as the squared magnitude of a spatial information entropy wave function,

\begin{equation}
\label{shannon_wave_function}
\rho_S(x) = |\psi_S(x)|^2,
\end{equation}
the local density of topological information available from a single point in the 1D manifold then being expressed in bra-ket notation by Born's rule,  

\begin{equation}
\label{born_rule_spatial}
\rho_S(x) = |\braket{x|\psi_S}|^2.
\end{equation}

The counterpart of the spatial information entropy wave function $\psi_S(x)$ in wavevector space is its Fourier transform 
\begin{equation}
\label{fourier_shannon_wave_function}
\hat{\psi}_S(k) = \int_{\ell}\psi_S(x)e^{-2\pi ikx}dx
\end{equation}
from which the information entropy density function in wavevector space can be computed,
\begin{equation}
\label{shannon_wave_function_wavevector_space}
\hat{\rho}_S(k) = |\hat{\psi}_S(k)|^2.
\end{equation}

The interpretation of information entropy signatures in terms of information entropy wave functions opens several avenues of investigation of possible connections between exotic states of matter and quantum information theory.  Here we explore its most forthright corollary, which is the establishment of uncertainty relations for topological phase transitions. 

Let us denote the mean, the variance and the entropy associated with the probability distribution $\rho_S$ by $\mu_{\rho_S}$, $\sigma_{\rho_S}^2 $ and $H_{\rho_S}$ respectively. Explicitly, we have 
\begin{subequations}
\label{statistics}
\begin{equation}
\label{mean}
\mu_{\rho_S} = \int_{\ell}x\rho_S(x)dx,
\end{equation}
\begin{equation}
\label{variance}
\sigma_{\rho_S}^2 = \int_{\ell}(x-\mu_{\rho_S})^2\rho_S(x)dx,\quad \text{and}
\end{equation}
\begin{equation}
\label{entropy}
H_{\rho_S} = \int_{\ell}\rho_S(x)\ln\big(\rho_S(x)\big) dx,
\end{equation}
\end{subequations}
with analogous equations for the wavevector space counterparts $\mu_{\hat{\rho}_S}$, $\sigma_{\hat{\rho}_S}^2 $ and $H_{\hat{\rho}_S}$ in terms of $\hat{\rho}(k)$.

In possession of this quantum formalism, we may write topological versions of two canonical uncertainty relations that bind together the real space and wavevector space information entropy density functions \eqref{shannon_wave_function} and \eqref{shannon_wave_function_wavevector_space}: 

\begin{enumerate}
\item[i)] \emph{the Heisenberg uncertainty principle}
\begin{equation}
\label{heisenberg}
\sigma_{\rho_S} \cdot \sigma_{\hat{\rho}_S} \geq \frac{1}{4\pi}, \quad \text{and }
\end{equation}
\item[ii)] \emph{the Hirschman entropic uncertainty}
\begin{equation}
\label{hirschman}
H_{\rho_S} + H_{\hat{\rho}_S} \geq \ln\Big(\frac{e}{2}\Big). 
\end{equation}
\end{enumerate}

The information entropy density function $\rho_S$ devised in this section furnishes a physics-grounded interpretation of the information entropy signatures obtained in section \ref{information_entropy_signatures} from sheer data analysis of finite SSH systems in real space. In particular, the uncertainty relations \eqref{heisenberg} and \eqref{hirschman} express concisely the trade-off between the localizability of information in topological phase transitions in real space and wavevector space. 

Perhaps the fundamental consequence of interpreting the information entropy density function $\rho_S(x)$ as the probability distribution resulting from an information entropy wave function $\psi_S(x)$ defined on a 1D manifold is that it reconciles the apparently conflicting notions of a local topological signal arising from a global property of the SSH systems. Indeed, while $\psi_S(x)$ is defined locally at every point of the 1D manifold, it is a single, global wave function encoding the spatial distribution of topological information of an ensemble of SSH Hamiltonians close to phase transition boundaries in parameter space. Therefore, the information entropy wave function $\psi_S(x)$ and its corresponding information entropy density function $\rho_S(x)$ can be pictured as emergent properties of an ensemble of quantum many-body systems near phase transitions.    

\section{Discussion}
\label{discussion}

Given the increasing complexity of systems studied in condensed matter physics and the rising demand for materials with exotic and robust properties to power future technological progress, it is only expected that data-driven approaches to physics will grow in demand. Our work represents a step in this direction, as we have devised (section \ref{the_eigenvector_ensembling_algorithm}) and implemented (section \ref{numerical_experiments}) a data-driven approach to the discovery of previously unknown properties of topological materials from real space data.

By starting from eigenvector data generated from the simulation of SSH systems in real space, proposing an approach based on eigenvector ensembling and decision trees and using model explainability to uncover the information entropy signatures presented in this article, and then exploring the numerical and theoretical possibilities offered by the information entropy signatures, our work exemplifies a full cycle of data-driven physics and illustrates how the interactions between machine learning and physics can be enriching to both disciplines.      
 
The development of data-driven methods based on real space lattice data will be particularly relevant to the study of disordered systems in condensed matter. Such systems usually break translational symmetry and therefore are not amenable to canonical wavevector space methods. Thus, the discovery and engineering of topological features from real space data as demonstrated in this work carries great promise to the theoretical investigation of these systems.

Furthermore, as is generally the case in engineering, the evolution of quantum technologies such as quantum computing and quantum communication  will likely depend on a delicate balance between simplicity and robustness of components such as topological qubits. On the simplicity side, engineers try to build their systems with as little redundancy as possible to reduce design complexity, whilst for robustness redundancy is a necessary commodity to ensure error-correction within the system. We expect that information theoretic approaches to quantum materials such as the one advanced by this paper shall eventually become a staple of quantum engineering.  

As we have seen, the use of real space data enabled us to investigate how topological information is spatially distributed in SSH systems. This was demonstrated by the information entropy signatures of section \ref{information_entropy_signatures}, which were recovered from the Shannon entropy of ensembles of eigenvectors in each numerical experiment executed in section \ref{numerical_experiments} and led us to the new topological features of section \ref{topological_feature_engineering} and the emergent information entropy wave functions of section \ref{emergent_information_entropy_wave_functions}. The existence of such signals that can be recovered from data from many distinct physical systems but are hard to conceptualize from sheer theoretical reasoning provides a clear example of how machine learning and model explainability can be important tools in the investigation of quantum materials.

The accuracy scores obtained in the numerical experiments performed in this paper were comparable to those reported in \cite{PhysRevLett.120.066401}, where dense and convolutional neural networks were trained on wavevector space data to predict the winding numbers of SSH Hamiltonians via supervised learning. This high accuracy level serves as a strong evidence that the entropy signatures presented here indeed express where topological information is most readily available in the SSH lattices investigated.

This paper should also be contrasted with \cite{zhang2020interpreting}, where the subject of investigation is the interpretability of neural network models trained to recognize topological phase transitions in some condensed matter systems. In \cite{zhang2020interpreting}, interesting visualizations are shown demonstrating that the patterns captured by a single-layer feedforward neural network indeed map directly to known physical quantities that are relevant to the problems at hand. We agree that such tasks should be called \emph{model interpretability}, as in that case the authors introspect into their models to make sure that they are learning patterns of physical pertinence to the systems being investigated. In our paper we preferred the term \emph{model explainability}, as we used similar model introspection tools to propose previously unknown concepts and properties of the physical systems being investigated. While the nuances in the semantics of these two terms are the subject of often heated philosophical debates in the artificial intelligence community, this choice of nomenclature suits the practical application of these model introspection techniques to physics well.

Recent works have demonstrated the existence of local topological markers in real space that carry important information on the topological state of a system \cite{PhysRevB.84.241106,caio2019topological}. Given the new DCT and DST topological features introduced in section \ref{topological_feature_engineering} which were shown to carry relevant topological information and the theoretical interpretation of the topological signals in terms of information entropy wave functions given in section \ref{emergent_information_entropy_wave_functions}, the results presented here suggest a new road for the theoretical exploration of local topological markers in terms of information theory as well. Whether there is any relationship between the local topological markers of \cite{PhysRevB.84.241106,caio2019topological} and the information entropy wave functions discussed here is left for speculation.
 
The eigenvector ensembling algorithm employed in this work is likely to have further applications in data-driven physics. This is because most of physics is based on eigenvector decomposition, and statistical physics itself can be seen as an application of similar ensembling principles. 

As a concrete example, the study of several many-body systems of current interest in condensed matter physics is hindered by their large dimensionality. This problem, known as \emph{the curse of dimensionality} in the scientific computing community, arises from the necessity of collecting or processing exponentially larger amounts of data as the feature space dimensionality of a problem grows. An approach based on eigenvector ensembling can be of use in such situations both as a dimensionality reduction tool and as a sampling strategy. The first case was illustrated in this work, where it was shown that relevant topological information of SSH systems can be retrieved from few sites in a lattice, which can be exploited as a dimensionality reduction strategy. The latter case, which was not explored here, also poses interesting possibilities, such as sampling eigenstates according to a desired distribution in Monte Carlo simulations of condensed matter systems. Indeed, sampling eigenvectors from a carefully designed probability distribution can ultimately lead to a great reduction in dimensionality  while still capturing all the relevant physics of a system. We therefore expect that a much broader class of data-driven physics problems could benefit from the techniques described in this paper.

Another interesting prospect is the combination of eigenvector ensembling with unsupervised learning algorithms. In the paper, our preference for decision trees and random forests was based on their powerful and accessible model explainability aptitudes. This choice was made in conformity with our main purpose, which was to exploit model explainability tools to investigate how topological information is distributed along a spatial lattice in SSH systems. Nevertheless, the eigenvector ensembling procedure we described here is flexible and can easily be repurposed for other supervised or unsupervised learning tasks.

One final comment should be made about the flourishing relationship between physics and machine learning. In this work we have demonstrated how a machine learning approach can provide new insights into complex physical phenomena of current interest. The other direction of this relationship (physics enhancing understanding in machine learning) is equally important. As the need for ever more powerful machine learning algorithms continues to grow, the development of mathematical frameworks for understanding general data spaces (i.e., a physics of data) will be of crucial relevance. This pursuit is seen in many theoretical works investigating the intriguing connections between geometry, topology and data \cite{carlsson2009topology}$^-$\cite{belkin2003problems}. The detailed study of data generated by physical models with non-trivial geometrical and topological properties such as the SSH model may provide invaluable insights into the structure and shape of real world high-dimensional data, since these models usually underscore well known mathematical frameworks behind the data generating process, a feature that is often absent from machine learning applications. Thus, far from being restricted to applications in physics, the study of the topological and geometrical properties of data sets generated by physical models will also be of great value to the machine learning and artificial intelligence communities.  

\begin{thebibliography}{63}%
\makeatletter
\providecommand \@ifxundefined [1]{%
 \@ifx{#1\undefined}
}%
\providecommand \@ifnum [1]{%
 \ifnum #1\expandafter \@firstoftwo
 \else \expandafter \@secondoftwo
 \fi
}%
\providecommand \@ifx [1]{%
 \ifx #1\expandafter \@firstoftwo
 \else \expandafter \@secondoftwo
 \fi
}%
\providecommand \natexlab [1]{#1}%
\providecommand \enquote  [1]{``#1''}%
\providecommand \bibnamefont  [1]{#1}%
\providecommand \bibfnamefont [1]{#1}%
\providecommand \citenamefont [1]{#1}%
\providecommand \href@noop [0]{\@secondoftwo}%
\providecommand \href [0]{\begingroup \@sanitize@url \@href}%
\providecommand \@href[1]{\@@startlink{#1}\@@href}%
\providecommand \@@href[1]{\endgroup#1\@@endlink}%
\providecommand \@sanitize@url [0]{\catcode `\\12\catcode `\$12\catcode
  `\&12\catcode `\#12\catcode `\^12\catcode `\_12\catcode `\%12\relax}%
\providecommand \@@startlink[1]{}%
\providecommand \@@endlink[0]{}%
\providecommand \url  [0]{\begingroup\@sanitize@url \@url }%
\providecommand \@url [1]{\endgroup\@href {#1}{\urlprefix }}%
\providecommand \urlprefix  [0]{URL }%
\providecommand \Eprint [0]{\href }%
\providecommand \doibase [0]{http://dx.doi.org/}%
\providecommand \selectlanguage [0]{\@gobble}%
\providecommand \bibinfo  [0]{\@secondoftwo}%
\providecommand \bibfield  [0]{\@secondoftwo}%
\providecommand \translation [1]{[#1]}%
\providecommand \BibitemOpen [0]{}%
\providecommand \bibitemStop [0]{}%
\providecommand \bibitemNoStop [0]{.\EOS\space}%
\providecommand \EOS [0]{\spacefactor3000\relax}%
\providecommand \BibitemShut  [1]{\csname bibitem#1\endcsname}%
\let\auto@bib@innerbib\@empty
\bibitem [{\citenamefont {Hasan}\ and\ \citenamefont
  {Kane}(2010)}]{RevModPhys.82.3045}%
  \BibitemOpen
  \bibfield  {author} {\bibinfo {author} {\bibfnamefont {M.~Z.}\ \bibnamefont
  {Hasan}}\ and\ \bibinfo {author} {\bibfnamefont {C.~L.}\ \bibnamefont
  {Kane}},\ }\href@noop {} {\bibfield  {journal} {\bibinfo  {journal} {Rev.
  Mod. Phys.}\ }\textbf {\bibinfo {volume} {82}},\ \bibinfo {pages} {3045}
  (\bibinfo {year} {2010})}\BibitemShut {NoStop}%
\bibitem [{\citenamefont {Continentino}(2017)}]{CONTINENTINO2017A1}%
  \BibitemOpen
  \bibfield  {author} {\bibinfo {author} {\bibfnamefont {M.~A.}\ \bibnamefont
  {Continentino}},\ }\href@noop {} {\bibfield  {journal} {\bibinfo  {journal}
  {Physica B: Condensed Matter}\ }\textbf {\bibinfo {volume} {505}},\ \bibinfo
  {pages} {A1 } (\bibinfo {year} {2017})}\BibitemShut {NoStop}%
\bibitem [{\citenamefont {Puel}\ \emph {et~al.}(2017)\citenamefont {Puel},
  \citenamefont {Sacramento},\ and\ \citenamefont
  {Continentino}}]{PhysRevB.95.094509}%
  \BibitemOpen
  \bibfield  {author} {\bibinfo {author} {\bibfnamefont {T.~O.}\ \bibnamefont
  {Puel}}, \bibinfo {author} {\bibfnamefont {P.~D.}\ \bibnamefont
  {Sacramento}}, \ and\ \bibinfo {author} {\bibfnamefont {M.~A.}\ \bibnamefont
  {Continentino}},\ }\href@noop {} {\bibfield  {journal} {\bibinfo  {journal}
  {Phys. Rev. B}\ }\textbf {\bibinfo {volume} {95}},\ \bibinfo {pages} {094509}
  (\bibinfo {year} {2017})}\BibitemShut {NoStop}%
\bibitem [{\citenamefont {Griffith}\ and\ \citenamefont
  {Continentino}(2018)}]{PhysRevE.97.012107}%
  \BibitemOpen
  \bibfield  {author} {\bibinfo {author} {\bibfnamefont {M.~A.}\ \bibnamefont
  {Griffith}}\ and\ \bibinfo {author} {\bibfnamefont {M.~A.}\ \bibnamefont
  {Continentino}},\ }\href@noop {} {\bibfield  {journal} {\bibinfo  {journal}
  {Phys. Rev. E}\ }\textbf {\bibinfo {volume} {97}},\ \bibinfo {pages} {012107}
  (\bibinfo {year} {2018})}\BibitemShut {NoStop}%
\bibitem [{\citenamefont {Ryu}\ \emph {et~al.}(2010)\citenamefont {Ryu},
  \citenamefont {Schnyder}, \citenamefont {Furusaki},\ and\ \citenamefont
  {Ludwig}}]{ryu2010topological}%
  \BibitemOpen
  \bibfield  {author} {\bibinfo {author} {\bibfnamefont {S.}~\bibnamefont
  {Ryu}}, \bibinfo {author} {\bibfnamefont {A.~P.}\ \bibnamefont {Schnyder}},
  \bibinfo {author} {\bibfnamefont {A.}~\bibnamefont {Furusaki}}, \ and\
  \bibinfo {author} {\bibfnamefont {A.~W.}\ \bibnamefont {Ludwig}},\
  }\href@noop {} {\bibfield  {journal} {\bibinfo  {journal} {New Journal of
  Physics}\ }\textbf {\bibinfo {volume} {12}},\ \bibinfo {pages} {065010}
  (\bibinfo {year} {2010})}\BibitemShut {NoStop}%
\bibitem [{\citenamefont {Atala}\ \emph {et~al.}(2013)\citenamefont {Atala},
  \citenamefont {Aidelsburger}, \citenamefont {Barreiro}, \citenamefont
  {Abanin}, \citenamefont {Kitagawa}, \citenamefont {Demler},\ and\
  \citenamefont {Bloch}}]{atala2013direct}%
  \BibitemOpen
  \bibfield  {author} {\bibinfo {author} {\bibfnamefont {M.}~\bibnamefont
  {Atala}}, \bibinfo {author} {\bibfnamefont {M.}~\bibnamefont {Aidelsburger}},
  \bibinfo {author} {\bibfnamefont {J.~T.}\ \bibnamefont {Barreiro}}, \bibinfo
  {author} {\bibfnamefont {D.}~\bibnamefont {Abanin}}, \bibinfo {author}
  {\bibfnamefont {T.}~\bibnamefont {Kitagawa}}, \bibinfo {author}
  {\bibfnamefont {E.}~\bibnamefont {Demler}}, \ and\ \bibinfo {author}
  {\bibfnamefont {I.}~\bibnamefont {Bloch}},\ }\href@noop {} {\bibfield
  {journal} {\bibinfo  {journal} {Nature Physics}\ }\textbf {\bibinfo {volume}
  {9}},\ \bibinfo {pages} {795} (\bibinfo {year} {2013})}\BibitemShut {NoStop}%
\bibitem [{\citenamefont {Stuhl}\ \emph {et~al.}(2015)\citenamefont {Stuhl},
  \citenamefont {Lu}, \citenamefont {Aycock}, \citenamefont {Genkina},\ and\
  \citenamefont {Spielman}}]{Stuhl1514}%
  \BibitemOpen
  \bibfield  {author} {\bibinfo {author} {\bibfnamefont {B.~K.}\ \bibnamefont
  {Stuhl}}, \bibinfo {author} {\bibfnamefont {H.-I.}\ \bibnamefont {Lu}},
  \bibinfo {author} {\bibfnamefont {L.~M.}\ \bibnamefont {Aycock}}, \bibinfo
  {author} {\bibfnamefont {D.}~\bibnamefont {Genkina}}, \ and\ \bibinfo
  {author} {\bibfnamefont {I.~B.}\ \bibnamefont {Spielman}},\ }\href@noop {}
  {\bibfield  {journal} {\bibinfo  {journal} {Science}\ }\textbf {\bibinfo
  {volume} {349}},\ \bibinfo {pages} {1514} (\bibinfo {year}
  {2015})}\BibitemShut {NoStop}%
\bibitem [{\citenamefont {Leder}\ \emph {et~al.}(2016)\citenamefont {Leder},
  \citenamefont {Grossert}, \citenamefont {Sitta}, \citenamefont {Genske},
  \citenamefont {Rosch},\ and\ \citenamefont {Weitz}}]{leder2016real}%
  \BibitemOpen
  \bibfield  {author} {\bibinfo {author} {\bibfnamefont {M.}~\bibnamefont
  {Leder}}, \bibinfo {author} {\bibfnamefont {C.}~\bibnamefont {Grossert}},
  \bibinfo {author} {\bibfnamefont {L.}~\bibnamefont {Sitta}}, \bibinfo
  {author} {\bibfnamefont {M.}~\bibnamefont {Genske}}, \bibinfo {author}
  {\bibfnamefont {A.}~\bibnamefont {Rosch}}, \ and\ \bibinfo {author}
  {\bibfnamefont {M.}~\bibnamefont {Weitz}},\ }\href@noop {} {\bibfield
  {journal} {\bibinfo  {journal} {Nature communications}\ }\textbf {\bibinfo
  {volume} {7}},\ \bibinfo {pages} {13112} (\bibinfo {year}
  {2016})}\BibitemShut {NoStop}%
\bibitem [{\citenamefont {Goldman}\ \emph {et~al.}(2016)\citenamefont
  {Goldman}, \citenamefont {Budich},\ and\ \citenamefont
  {Zoller}}]{goldman2016topological}%
  \BibitemOpen
  \bibfield  {author} {\bibinfo {author} {\bibfnamefont {N.}~\bibnamefont
  {Goldman}}, \bibinfo {author} {\bibfnamefont {J.}~\bibnamefont {Budich}}, \
  and\ \bibinfo {author} {\bibfnamefont {P.}~\bibnamefont {Zoller}},\
  }\href@noop {} {\bibfield  {journal} {\bibinfo  {journal} {Nature Physics}\
  }\textbf {\bibinfo {volume} {12}},\ \bibinfo {pages} {639} (\bibinfo {year}
  {2016})}\BibitemShut {NoStop}%
\bibitem [{\citenamefont {Meier}\ \emph {et~al.}(2016)\citenamefont {Meier},
  \citenamefont {An},\ and\ \citenamefont {Gadway}}]{meier2016observation}%
  \BibitemOpen
  \bibfield  {author} {\bibinfo {author} {\bibfnamefont {E.~J.}\ \bibnamefont
  {Meier}}, \bibinfo {author} {\bibfnamefont {F.~A.}\ \bibnamefont {An}}, \
  and\ \bibinfo {author} {\bibfnamefont {B.}~\bibnamefont {Gadway}},\
  }\href@noop {} {\bibfield  {journal} {\bibinfo  {journal} {Nature
  communications}\ }\textbf {\bibinfo {volume} {7}},\ \bibinfo {pages} {13986}
  (\bibinfo {year} {2016})}\BibitemShut {NoStop}%
\bibitem [{\citenamefont {Hafezi}\ \emph {et~al.}(2013)\citenamefont {Hafezi},
  \citenamefont {Mittal}, \citenamefont {Fan}, \citenamefont {Migdall},\ and\
  \citenamefont {Taylor}}]{hafezi2013imaging}%
  \BibitemOpen
  \bibfield  {author} {\bibinfo {author} {\bibfnamefont {M.}~\bibnamefont
  {Hafezi}}, \bibinfo {author} {\bibfnamefont {S.}~\bibnamefont {Mittal}},
  \bibinfo {author} {\bibfnamefont {J.}~\bibnamefont {Fan}}, \bibinfo {author}
  {\bibfnamefont {A.}~\bibnamefont {Migdall}}, \ and\ \bibinfo {author}
  {\bibfnamefont {J.}~\bibnamefont {Taylor}},\ }\href@noop {} {\bibfield
  {journal} {\bibinfo  {journal} {Nature Photonics}\ }\textbf {\bibinfo
  {volume} {7}},\ \bibinfo {pages} {1001} (\bibinfo {year} {2013})}\BibitemShut
  {NoStop}%
\bibitem [{\citenamefont {Lu}\ \emph {et~al.}(2016)\citenamefont {Lu},
  \citenamefont {Joannopoulos},\ and\ \citenamefont
  {Solja{\v{c}}i{\'c}}}]{lu2016topological}%
  \BibitemOpen
  \bibfield  {author} {\bibinfo {author} {\bibfnamefont {L.}~\bibnamefont
  {Lu}}, \bibinfo {author} {\bibfnamefont {J.~D.}\ \bibnamefont
  {Joannopoulos}}, \ and\ \bibinfo {author} {\bibfnamefont {M.}~\bibnamefont
  {Solja{\v{c}}i{\'c}}},\ }\href@noop {} {\bibfield  {journal} {\bibinfo
  {journal} {Nature Physics}\ }\textbf {\bibinfo {volume} {12}},\ \bibinfo
  {pages} {626} (\bibinfo {year} {2016})}\BibitemShut {NoStop}%
\bibitem [{\citenamefont {Peano}\ \emph {et~al.}(2015)\citenamefont {Peano},
  \citenamefont {Brendel}, \citenamefont {Schmidt},\ and\ \citenamefont
  {Marquardt}}]{PhysRevX.5.031011}%
  \BibitemOpen
  \bibfield  {author} {\bibinfo {author} {\bibfnamefont {V.}~\bibnamefont
  {Peano}}, \bibinfo {author} {\bibfnamefont {C.}~\bibnamefont {Brendel}},
  \bibinfo {author} {\bibfnamefont {M.}~\bibnamefont {Schmidt}}, \ and\
  \bibinfo {author} {\bibfnamefont {F.}~\bibnamefont {Marquardt}},\ }\href@noop
  {} {\bibfield  {journal} {\bibinfo  {journal} {Phys. Rev. X}\ }\textbf
  {\bibinfo {volume} {5}},\ \bibinfo {pages} {031011} (\bibinfo {year}
  {2015})}\BibitemShut {NoStop}%
\bibitem [{\citenamefont {Kitagawa}\ \emph {et~al.}(2012)\citenamefont
  {Kitagawa}, \citenamefont {Broome}, \citenamefont {Fedrizzi}, \citenamefont
  {Rudner}, \citenamefont {Berg}, \citenamefont {Kassal}, \citenamefont
  {Aspuru-Guzik}, \citenamefont {Demler},\ and\ \citenamefont
  {White}}]{kitagawa2012observation}%
  \BibitemOpen
  \bibfield  {author} {\bibinfo {author} {\bibfnamefont {T.}~\bibnamefont
  {Kitagawa}}, \bibinfo {author} {\bibfnamefont {M.~A.}\ \bibnamefont
  {Broome}}, \bibinfo {author} {\bibfnamefont {A.}~\bibnamefont {Fedrizzi}},
  \bibinfo {author} {\bibfnamefont {M.~S.}\ \bibnamefont {Rudner}}, \bibinfo
  {author} {\bibfnamefont {E.}~\bibnamefont {Berg}}, \bibinfo {author}
  {\bibfnamefont {I.}~\bibnamefont {Kassal}}, \bibinfo {author} {\bibfnamefont
  {A.}~\bibnamefont {Aspuru-Guzik}}, \bibinfo {author} {\bibfnamefont
  {E.}~\bibnamefont {Demler}}, \ and\ \bibinfo {author} {\bibfnamefont {A.~G.}\
  \bibnamefont {White}},\ }\href@noop {} {\bibfield  {journal} {\bibinfo
  {journal} {Nature communications}\ }\textbf {\bibinfo {volume} {3}},\
  \bibinfo {pages} {882} (\bibinfo {year} {2012})}\BibitemShut {NoStop}%
\bibitem [{\citenamefont {Cardano}\ \emph {et~al.}(2016)\citenamefont
  {Cardano}, \citenamefont {Maffei}, \citenamefont {Massa}, \citenamefont
  {Piccirillo}, \citenamefont {De~Lisio}, \citenamefont {De~Filippis},
  \citenamefont {Cataudella}, \citenamefont {Santamato},\ and\ \citenamefont
  {Marrucci}}]{cardano2016statistical}%
  \BibitemOpen
  \bibfield  {author} {\bibinfo {author} {\bibfnamefont {F.}~\bibnamefont
  {Cardano}}, \bibinfo {author} {\bibfnamefont {M.}~\bibnamefont {Maffei}},
  \bibinfo {author} {\bibfnamefont {F.}~\bibnamefont {Massa}}, \bibinfo
  {author} {\bibfnamefont {B.}~\bibnamefont {Piccirillo}}, \bibinfo {author}
  {\bibfnamefont {C.}~\bibnamefont {De~Lisio}}, \bibinfo {author}
  {\bibfnamefont {G.}~\bibnamefont {De~Filippis}}, \bibinfo {author}
  {\bibfnamefont {V.}~\bibnamefont {Cataudella}}, \bibinfo {author}
  {\bibfnamefont {E.}~\bibnamefont {Santamato}}, \ and\ \bibinfo {author}
  {\bibfnamefont {L.}~\bibnamefont {Marrucci}},\ }\href@noop {} {\bibfield
  {journal} {\bibinfo  {journal} {Nature communications}\ }\textbf {\bibinfo
  {volume} {7}},\ \bibinfo {pages} {11439} (\bibinfo {year}
  {2016})}\BibitemShut {NoStop}%
\bibitem [{\citenamefont {Flurin}\ \emph {et~al.}(2017)\citenamefont {Flurin},
  \citenamefont {Ramasesh}, \citenamefont {Hacohen-Gourgy}, \citenamefont
  {Martin}, \citenamefont {Yao},\ and\ \citenamefont
  {Siddiqi}}]{PhysRevX.7.031023}%
  \BibitemOpen
  \bibfield  {author} {\bibinfo {author} {\bibfnamefont {E.}~\bibnamefont
  {Flurin}}, \bibinfo {author} {\bibfnamefont {V.~V.}\ \bibnamefont
  {Ramasesh}}, \bibinfo {author} {\bibfnamefont {S.}~\bibnamefont
  {Hacohen-Gourgy}}, \bibinfo {author} {\bibfnamefont {L.~S.}\ \bibnamefont
  {Martin}}, \bibinfo {author} {\bibfnamefont {N.~Y.}\ \bibnamefont {Yao}}, \
  and\ \bibinfo {author} {\bibfnamefont {I.}~\bibnamefont {Siddiqi}},\
  }\href@noop {} {\bibfield  {journal} {\bibinfo  {journal} {Phys. Rev. X}\
  }\textbf {\bibinfo {volume} {7}},\ \bibinfo {pages} {031023} (\bibinfo {year}
  {2017})}\BibitemShut {NoStop}%
\bibitem [{\citenamefont {Soluyanov}\ \emph {et~al.}(2015)\citenamefont
  {Soluyanov}, \citenamefont {Gresch}, \citenamefont {Wang}, \citenamefont
  {Wu}, \citenamefont {Troyer}, \citenamefont {Dai},\ and\ \citenamefont
  {Bernevig}}]{soluyanov2015type}%
  \BibitemOpen
  \bibfield  {author} {\bibinfo {author} {\bibfnamefont {A.~A.}\ \bibnamefont
  {Soluyanov}}, \bibinfo {author} {\bibfnamefont {D.}~\bibnamefont {Gresch}},
  \bibinfo {author} {\bibfnamefont {Z.}~\bibnamefont {Wang}}, \bibinfo {author}
  {\bibfnamefont {Q.}~\bibnamefont {Wu}}, \bibinfo {author} {\bibfnamefont
  {M.}~\bibnamefont {Troyer}}, \bibinfo {author} {\bibfnamefont
  {X.}~\bibnamefont {Dai}}, \ and\ \bibinfo {author} {\bibfnamefont {B.~A.}\
  \bibnamefont {Bernevig}},\ }\href@noop {} {\bibfield  {journal} {\bibinfo
  {journal} {Nature}\ }\textbf {\bibinfo {volume} {527}},\ \bibinfo {pages}
  {495} (\bibinfo {year} {2015})}\BibitemShut {NoStop}%
\bibitem [{\citenamefont {Lv}\ \emph {et~al.}(2015)\citenamefont {Lv},
  \citenamefont {Weng}, \citenamefont {Fu}, \citenamefont {Wang}, \citenamefont
  {Miao}, \citenamefont {Ma}, \citenamefont {Richard}, \citenamefont {Huang},
  \citenamefont {Zhao}, \citenamefont {Chen}, \citenamefont {Fang},
  \citenamefont {Dai}, \citenamefont {Qian},\ and\ \citenamefont
  {Ding}}]{PhysRevX.5.031013}%
  \BibitemOpen
  \bibfield  {author} {\bibinfo {author} {\bibfnamefont {B.~Q.}\ \bibnamefont
  {Lv}}, \bibinfo {author} {\bibfnamefont {H.~M.}\ \bibnamefont {Weng}},
  \bibinfo {author} {\bibfnamefont {B.~B.}\ \bibnamefont {Fu}}, \bibinfo
  {author} {\bibfnamefont {X.~P.}\ \bibnamefont {Wang}}, \bibinfo {author}
  {\bibfnamefont {H.}~\bibnamefont {Miao}}, \bibinfo {author} {\bibfnamefont
  {J.}~\bibnamefont {Ma}}, \bibinfo {author} {\bibfnamefont {P.}~\bibnamefont
  {Richard}}, \bibinfo {author} {\bibfnamefont {X.~C.}\ \bibnamefont {Huang}},
  \bibinfo {author} {\bibfnamefont {L.~X.}\ \bibnamefont {Zhao}}, \bibinfo
  {author} {\bibfnamefont {G.~F.}\ \bibnamefont {Chen}}, \bibinfo {author}
  {\bibfnamefont {Z.}~\bibnamefont {Fang}}, \bibinfo {author} {\bibfnamefont
  {X.}~\bibnamefont {Dai}}, \bibinfo {author} {\bibfnamefont {T.}~\bibnamefont
  {Qian}}, \ and\ \bibinfo {author} {\bibfnamefont {H.}~\bibnamefont {Ding}},\
  }\href@noop {} {\bibfield  {journal} {\bibinfo  {journal} {Phys. Rev. X}\
  }\textbf {\bibinfo {volume} {5}},\ \bibinfo {pages} {031013} (\bibinfo {year}
  {2015})}\BibitemShut {NoStop}%
\bibitem [{\citenamefont {Su}\ \emph {et~al.}(1979)\citenamefont {Su},
  \citenamefont {Schrieffer},\ and\ \citenamefont
  {Heeger}}]{PhysRevLett.42.1698}%
  \BibitemOpen
  \bibfield  {author} {\bibinfo {author} {\bibfnamefont {W.~P.}\ \bibnamefont
  {Su}}, \bibinfo {author} {\bibfnamefont {J.~R.}\ \bibnamefont {Schrieffer}},
  \ and\ \bibinfo {author} {\bibfnamefont {A.~J.}\ \bibnamefont {Heeger}},\
  }\href@noop {} {\bibfield  {journal} {\bibinfo  {journal} {Phys. Rev. Lett.}\
  }\textbf {\bibinfo {volume} {42}},\ \bibinfo {pages} {1698} (\bibinfo {year}
  {1979})}\BibitemShut {NoStop}%
\bibitem [{\citenamefont {Maffei}\ \emph {et~al.}(2018)\citenamefont {Maffei},
  \citenamefont {Dauphin}, \citenamefont {Cardano}, \citenamefont
  {Lewenstein},\ and\ \citenamefont {Massignan}}]{maffei2018topological}%
  \BibitemOpen
  \bibfield  {author} {\bibinfo {author} {\bibfnamefont {M.}~\bibnamefont
  {Maffei}}, \bibinfo {author} {\bibfnamefont {A.}~\bibnamefont {Dauphin}},
  \bibinfo {author} {\bibfnamefont {F.}~\bibnamefont {Cardano}}, \bibinfo
  {author} {\bibfnamefont {M.}~\bibnamefont {Lewenstein}}, \ and\ \bibinfo
  {author} {\bibfnamefont {P.}~\bibnamefont {Massignan}},\ }\href@noop {}
  {\bibfield  {journal} {\bibinfo  {journal} {New Journal of Physics}\ }\textbf
  {\bibinfo {volume} {20}},\ \bibinfo {pages} {013023} (\bibinfo {year}
  {2018})}\BibitemShut {NoStop}%
\bibitem [{\citenamefont {Heeger}(2001)}]{RevModPhys.73.681}%
  \BibitemOpen
  \bibfield  {author} {\bibinfo {author} {\bibfnamefont {A.~J.}\ \bibnamefont
  {Heeger}},\ }\href@noop {} {\bibfield  {journal} {\bibinfo  {journal} {Rev.
  Mod. Phys.}\ }\textbf {\bibinfo {volume} {73}},\ \bibinfo {pages} {681}
  (\bibinfo {year} {2001})}\BibitemShut {NoStop}%
\bibitem [{\citenamefont {Kane}\ and\ \citenamefont
  {Lubensky}(2014)}]{kane2014topological}%
  \BibitemOpen
  \bibfield  {author} {\bibinfo {author} {\bibfnamefont {C.}~\bibnamefont
  {Kane}}\ and\ \bibinfo {author} {\bibfnamefont {T.}~\bibnamefont
  {Lubensky}},\ }\href@noop {} {\bibfield  {journal} {\bibinfo  {journal}
  {Nature Physics}\ }\textbf {\bibinfo {volume} {10}},\ \bibinfo {pages} {39}
  (\bibinfo {year} {2014})}\BibitemShut {NoStop}%
\bibitem [{\citenamefont {Chen}\ \emph {et~al.}(2014)\citenamefont {Chen},
  \citenamefont {Upadhyaya},\ and\ \citenamefont {Vitelli}}]{Chen13004}%
  \BibitemOpen
  \bibfield  {author} {\bibinfo {author} {\bibfnamefont {B.~G.-g.}\
  \bibnamefont {Chen}}, \bibinfo {author} {\bibfnamefont {N.}~\bibnamefont
  {Upadhyaya}}, \ and\ \bibinfo {author} {\bibfnamefont {V.}~\bibnamefont
  {Vitelli}},\ }\href@noop {} {\bibfield  {journal} {\bibinfo  {journal}
  {Proceedings of the National Academy of Sciences}\ }\textbf {\bibinfo
  {volume} {111}},\ \bibinfo {pages} {13004} (\bibinfo {year}
  {2014})}\BibitemShut {NoStop}%
\bibitem [{\citenamefont {Carrasquilla}\ and\ \citenamefont
  {Melko}(2017)}]{carrasquilla2017machine}%
  \BibitemOpen
  \bibfield  {author} {\bibinfo {author} {\bibfnamefont {J.}~\bibnamefont
  {Carrasquilla}}\ and\ \bibinfo {author} {\bibfnamefont {R.~G.}\ \bibnamefont
  {Melko}},\ }\href@noop {} {\bibfield  {journal} {\bibinfo  {journal} {Nature
  Physics}\ }\textbf {\bibinfo {volume} {13}},\ \bibinfo {pages} {431}
  (\bibinfo {year} {2017})}\BibitemShut {NoStop}%
\bibitem [{\citenamefont {Ch'ng}\ \emph {et~al.}(2017)\citenamefont {Ch'ng},
  \citenamefont {Carrasquilla}, \citenamefont {Melko},\ and\ \citenamefont
  {Khatami}}]{PhysRevX.7.031038}%
  \BibitemOpen
  \bibfield  {author} {\bibinfo {author} {\bibfnamefont {K.}~\bibnamefont
  {Ch'ng}}, \bibinfo {author} {\bibfnamefont {J.}~\bibnamefont {Carrasquilla}},
  \bibinfo {author} {\bibfnamefont {R.~G.}\ \bibnamefont {Melko}}, \ and\
  \bibinfo {author} {\bibfnamefont {E.}~\bibnamefont {Khatami}},\ }\href@noop
  {} {\bibfield  {journal} {\bibinfo  {journal} {Phys. Rev. X}\ }\textbf
  {\bibinfo {volume} {7}},\ \bibinfo {pages} {031038} (\bibinfo {year}
  {2017})}\BibitemShut {NoStop}%
\bibitem [{\citenamefont {Wang}(2016)}]{PhysRevB.94.195105}%
  \BibitemOpen
  \bibfield  {author} {\bibinfo {author} {\bibfnamefont {L.}~\bibnamefont
  {Wang}},\ }\href@noop {} {\bibfield  {journal} {\bibinfo  {journal} {Phys.
  Rev. B}\ }\textbf {\bibinfo {volume} {94}},\ \bibinfo {pages} {195105}
  (\bibinfo {year} {2016})}\BibitemShut {NoStop}%
\bibitem [{\citenamefont {Broecker}\ \emph {et~al.}(2017)\citenamefont
  {Broecker}, \citenamefont {Carrasquilla}, \citenamefont {Melko},\ and\
  \citenamefont {Trebst}}]{broecker2017machine}%
  \BibitemOpen
  \bibfield  {author} {\bibinfo {author} {\bibfnamefont {P.}~\bibnamefont
  {Broecker}}, \bibinfo {author} {\bibfnamefont {J.}~\bibnamefont
  {Carrasquilla}}, \bibinfo {author} {\bibfnamefont {R.~G.}\ \bibnamefont
  {Melko}}, \ and\ \bibinfo {author} {\bibfnamefont {S.}~\bibnamefont
  {Trebst}},\ }\href@noop {} {\bibfield  {journal} {\bibinfo  {journal}
  {Scientific reports}\ }\textbf {\bibinfo {volume} {7}},\ \bibinfo {pages}
  {8823} (\bibinfo {year} {2017})}\BibitemShut {NoStop}%
\bibitem [{\citenamefont {Van~Nieuwenburg}\ \emph {et~al.}(2017)\citenamefont
  {Van~Nieuwenburg}, \citenamefont {Liu},\ and\ \citenamefont
  {Huber}}]{van2017learning}%
  \BibitemOpen
  \bibfield  {author} {\bibinfo {author} {\bibfnamefont {E.~P.}\ \bibnamefont
  {Van~Nieuwenburg}}, \bibinfo {author} {\bibfnamefont {Y.-H.}\ \bibnamefont
  {Liu}}, \ and\ \bibinfo {author} {\bibfnamefont {S.~D.}\ \bibnamefont
  {Huber}},\ }\href@noop {} {\bibfield  {journal} {\bibinfo  {journal} {Nature
  Physics}\ }\textbf {\bibinfo {volume} {13}},\ \bibinfo {pages} {435}
  (\bibinfo {year} {2017})}\BibitemShut {NoStop}%
\bibitem [{\citenamefont {Zhang}\ \emph {et~al.}(2018)\citenamefont {Zhang},
  \citenamefont {Shen},\ and\ \citenamefont {Zhai}}]{PhysRevLett.120.066401}%
  \BibitemOpen
  \bibfield  {author} {\bibinfo {author} {\bibfnamefont {P.}~\bibnamefont
  {Zhang}}, \bibinfo {author} {\bibfnamefont {H.}~\bibnamefont {Shen}}, \ and\
  \bibinfo {author} {\bibfnamefont {H.}~\bibnamefont {Zhai}},\ }\href@noop {}
  {\bibfield  {journal} {\bibinfo  {journal} {Phys. Rev. Lett.}\ }\textbf
  {\bibinfo {volume} {120}},\ \bibinfo {pages} {066401} (\bibinfo {year}
  {2018})}\BibitemShut {NoStop}%
\bibitem [{\citenamefont {Sun}\ \emph {et~al.}(2018)\citenamefont {Sun},
  \citenamefont {Yi}, \citenamefont {Zhang}, \citenamefont {Shen},\ and\
  \citenamefont {Zhai}}]{PhysRevB.98.085402}%
  \BibitemOpen
  \bibfield  {author} {\bibinfo {author} {\bibfnamefont {N.}~\bibnamefont
  {Sun}}, \bibinfo {author} {\bibfnamefont {J.}~\bibnamefont {Yi}}, \bibinfo
  {author} {\bibfnamefont {P.}~\bibnamefont {Zhang}}, \bibinfo {author}
  {\bibfnamefont {H.}~\bibnamefont {Shen}}, \ and\ \bibinfo {author}
  {\bibfnamefont {H.}~\bibnamefont {Zhai}},\ }\href@noop {} {\bibfield
  {journal} {\bibinfo  {journal} {Phys. Rev. B}\ }\textbf {\bibinfo {volume}
  {98}},\ \bibinfo {pages} {085402} (\bibinfo {year} {2018})}\BibitemShut
  {NoStop}%
\bibitem [{\citenamefont {Suchsland}\ and\ \citenamefont
  {Wessel}(2018)}]{PhysRevB.97.174435}%
  \BibitemOpen
  \bibfield  {author} {\bibinfo {author} {\bibfnamefont {P.}~\bibnamefont
  {Suchsland}}\ and\ \bibinfo {author} {\bibfnamefont {S.}~\bibnamefont
  {Wessel}},\ }\href@noop {} {\bibfield  {journal} {\bibinfo  {journal} {Phys.
  Rev. B}\ }\textbf {\bibinfo {volume} {97}},\ \bibinfo {pages} {174435}
  (\bibinfo {year} {2018})}\BibitemShut {NoStop}%
\bibitem [{\citenamefont {Zhang}\ and\ \citenamefont
  {Kim}(2017)}]{PhysRevLett.118.216401}%
  \BibitemOpen
  \bibfield  {author} {\bibinfo {author} {\bibfnamefont {Y.}~\bibnamefont
  {Zhang}}\ and\ \bibinfo {author} {\bibfnamefont {E.-A.}\ \bibnamefont
  {Kim}},\ }\href@noop {} {\bibfield  {journal} {\bibinfo  {journal} {Phys.
  Rev. Lett.}\ }\textbf {\bibinfo {volume} {118}},\ \bibinfo {pages} {216401}
  (\bibinfo {year} {2017})}\BibitemShut {NoStop}%
\bibitem [{\citenamefont {Venderley}\ \emph {et~al.}(2018)\citenamefont
  {Venderley}, \citenamefont {Khemani},\ and\ \citenamefont
  {Kim}}]{PhysRevLett.120.257204}%
  \BibitemOpen
  \bibfield  {author} {\bibinfo {author} {\bibfnamefont {J.}~\bibnamefont
  {Venderley}}, \bibinfo {author} {\bibfnamefont {V.}~\bibnamefont {Khemani}},
  \ and\ \bibinfo {author} {\bibfnamefont {E.-A.}\ \bibnamefont {Kim}},\
  }\href@noop {} {\bibfield  {journal} {\bibinfo  {journal} {Phys. Rev. Lett.}\
  }\textbf {\bibinfo {volume} {120}},\ \bibinfo {pages} {257204} (\bibinfo
  {year} {2018})}\BibitemShut {NoStop}%
\bibitem [{\citenamefont {Ohtsuki}\ and\ \citenamefont
  {Ohtsuki}(2017)}]{ohtsuki2017deep}%
  \BibitemOpen
  \bibfield  {author} {\bibinfo {author} {\bibfnamefont {T.}~\bibnamefont
  {Ohtsuki}}\ and\ \bibinfo {author} {\bibfnamefont {T.}~\bibnamefont
  {Ohtsuki}},\ }\href@noop {} {\bibfield  {journal} {\bibinfo  {journal}
  {Journal of the Physical Society of Japan}\ }\textbf {\bibinfo {volume}
  {86}},\ \bibinfo {pages} {044708} (\bibinfo {year} {2017})}\BibitemShut
  {NoStop}%
\bibitem [{\citenamefont {Yoshioka}\ \emph {et~al.}(2018)\citenamefont
  {Yoshioka}, \citenamefont {Akagi},\ and\ \citenamefont
  {Katsura}}]{PhysRevB.97.205110}%
  \BibitemOpen
  \bibfield  {author} {\bibinfo {author} {\bibfnamefont {N.}~\bibnamefont
  {Yoshioka}}, \bibinfo {author} {\bibfnamefont {Y.}~\bibnamefont {Akagi}}, \
  and\ \bibinfo {author} {\bibfnamefont {H.}~\bibnamefont {Katsura}},\
  }\href@noop {} {\bibfield  {journal} {\bibinfo  {journal} {Phys. Rev. B}\
  }\textbf {\bibinfo {volume} {97}},\ \bibinfo {pages} {205110} (\bibinfo
  {year} {2018})}\BibitemShut {NoStop}%
\bibitem [{\citenamefont {Deng}\ \emph {et~al.}(2017)\citenamefont {Deng},
  \citenamefont {Li},\ and\ \citenamefont {Das~Sarma}}]{PhysRevB.96.195145}%
  \BibitemOpen
  \bibfield  {author} {\bibinfo {author} {\bibfnamefont {D.-L.}\ \bibnamefont
  {Deng}}, \bibinfo {author} {\bibfnamefont {X.}~\bibnamefont {Li}}, \ and\
  \bibinfo {author} {\bibfnamefont {S.}~\bibnamefont {Das~Sarma}},\ }\href@noop
  {} {\bibfield  {journal} {\bibinfo  {journal} {Phys. Rev. B}\ }\textbf
  {\bibinfo {volume} {96}},\ \bibinfo {pages} {195145} (\bibinfo {year}
  {2017})}\BibitemShut {NoStop}%
\bibitem [{\citenamefont {Huembeli}\ \emph {et~al.}(2018)\citenamefont
  {Huembeli}, \citenamefont {Dauphin},\ and\ \citenamefont
  {Wittek}}]{PhysRevB.97.134109}%
  \BibitemOpen
  \bibfield  {author} {\bibinfo {author} {\bibfnamefont {P.}~\bibnamefont
  {Huembeli}}, \bibinfo {author} {\bibfnamefont {A.}~\bibnamefont {Dauphin}}, \
  and\ \bibinfo {author} {\bibfnamefont {P.}~\bibnamefont {Wittek}},\
  }\href@noop {} {\bibfield  {journal} {\bibinfo  {journal} {Phys. Rev. B}\
  }\textbf {\bibinfo {volume} {97}},\ \bibinfo {pages} {134109} (\bibinfo
  {year} {2018})}\BibitemShut {NoStop}%
\bibitem [{\citenamefont {Carvalho}\ \emph {et~al.}(2018)\citenamefont
  {Carvalho}, \citenamefont {Garc\'{\i}a-Mart\'{\i}nez}, \citenamefont {Lado},\
  and\ \citenamefont {Fern\'andez-Rossier}}]{PhysRevB.97.115453}%
  \BibitemOpen
  \bibfield  {author} {\bibinfo {author} {\bibfnamefont {D.}~\bibnamefont
  {Carvalho}}, \bibinfo {author} {\bibfnamefont {N.~A.}\ \bibnamefont
  {Garc\'{\i}a-Mart\'{\i}nez}}, \bibinfo {author} {\bibfnamefont {J.~L.}\
  \bibnamefont {Lado}}, \ and\ \bibinfo {author} {\bibfnamefont
  {J.}~\bibnamefont {Fern\'andez-Rossier}},\ }\href@noop {} {\bibfield
  {journal} {\bibinfo  {journal} {Phys. Rev. B}\ }\textbf {\bibinfo {volume}
  {97}},\ \bibinfo {pages} {115453} (\bibinfo {year} {2018})}\BibitemShut
  {NoStop}%
\bibitem [{\citenamefont {Zhang}\ \emph {et~al.}(2017)\citenamefont {Zhang},
  \citenamefont {Melko},\ and\ \citenamefont {Kim}}]{PhysRevB.96.245119}%
  \BibitemOpen
  \bibfield  {author} {\bibinfo {author} {\bibfnamefont {Y.}~\bibnamefont
  {Zhang}}, \bibinfo {author} {\bibfnamefont {R.~G.}\ \bibnamefont {Melko}}, \
  and\ \bibinfo {author} {\bibfnamefont {E.-A.}\ \bibnamefont {Kim}},\
  }\href@noop {} {\bibfield  {journal} {\bibinfo  {journal} {Phys. Rev. B}\
  }\textbf {\bibinfo {volume} {96}},\ \bibinfo {pages} {245119} (\bibinfo
  {year} {2017})}\BibitemShut {NoStop}%
\bibitem [{\citenamefont {Rodriguez-Nieva}\ and\ \citenamefont
  {Scheurer}(2018)}]{rodriguez2018identifying}%
  \BibitemOpen
  \bibfield  {author} {\bibinfo {author} {\bibfnamefont {J.~F.}\ \bibnamefont
  {Rodriguez-Nieva}}\ and\ \bibinfo {author} {\bibfnamefont {M.~S.}\
  \bibnamefont {Scheurer}},\ }\href@noop {} {\bibfield  {journal} {\bibinfo
  {journal} {arXiv preprint arXiv:1805.05961}\ } (\bibinfo {year}
  {2018})}\BibitemShut {NoStop}%
\bibitem [{\citenamefont {Gilpin}\ \emph {et~al.}(2018)\citenamefont {Gilpin},
  \citenamefont {Bau}, \citenamefont {Yuan}, \citenamefont {Bajwa},
  \citenamefont {Specter},\ and\ \citenamefont {Kagal}}]{gilpin2018explaining}%
  \BibitemOpen
  \bibfield  {author} {\bibinfo {author} {\bibfnamefont {L.~H.}\ \bibnamefont
  {Gilpin}}, \bibinfo {author} {\bibfnamefont {D.}~\bibnamefont {Bau}},
  \bibinfo {author} {\bibfnamefont {B.~Z.}\ \bibnamefont {Yuan}}, \bibinfo
  {author} {\bibfnamefont {A.}~\bibnamefont {Bajwa}}, \bibinfo {author}
  {\bibfnamefont {M.}~\bibnamefont {Specter}}, \ and\ \bibinfo {author}
  {\bibfnamefont {L.}~\bibnamefont {Kagal}},\ }in\ \href@noop {} {\emph
  {\bibinfo {booktitle} {2018 IEEE 5th International Conference on data science
  and advanced analytics (DSAA)}}}\ (\bibinfo {organization} {IEEE},\ \bibinfo
  {year} {2018})\ pp.\ \bibinfo {pages} {80--89}\BibitemShut {NoStop}%
\bibitem [{\citenamefont {Do{\v{s}}ilovi{\'c}}\ \emph
  {et~al.}(2018)\citenamefont {Do{\v{s}}ilovi{\'c}}, \citenamefont
  {Br{\v{c}}i{\'c}},\ and\ \citenamefont
  {Hlupi{\'c}}}]{dovsilovic2018explainable}%
  \BibitemOpen
  \bibfield  {author} {\bibinfo {author} {\bibfnamefont {F.~K.}\ \bibnamefont
  {Do{\v{s}}ilovi{\'c}}}, \bibinfo {author} {\bibfnamefont {M.}~\bibnamefont
  {Br{\v{c}}i{\'c}}}, \ and\ \bibinfo {author} {\bibfnamefont {N.}~\bibnamefont
  {Hlupi{\'c}}},\ }in\ \href@noop {} {\emph {\bibinfo {booktitle} {2018 41st
  International convention on information and communication technology,
  electronics and microelectronics (MIPRO)}}}\ (\bibinfo {organization}
  {IEEE},\ \bibinfo {year} {2018})\ pp.\ \bibinfo {pages}
  {0210--0215}\BibitemShut {NoStop}%
\bibitem [{\citenamefont {Roscher}\ \emph {et~al.}(2020)\citenamefont
  {Roscher}, \citenamefont {Bohn}, \citenamefont {Duarte},\ and\ \citenamefont
  {Garcke}}]{roscher2020explainable}%
  \BibitemOpen
  \bibfield  {author} {\bibinfo {author} {\bibfnamefont {R.}~\bibnamefont
  {Roscher}}, \bibinfo {author} {\bibfnamefont {B.}~\bibnamefont {Bohn}},
  \bibinfo {author} {\bibfnamefont {M.~F.}\ \bibnamefont {Duarte}}, \ and\
  \bibinfo {author} {\bibfnamefont {J.}~\bibnamefont {Garcke}},\ }\href@noop {}
  {\bibfield  {journal} {\bibinfo  {journal} {IEEE Access}\ }\textbf {\bibinfo
  {volume} {8}},\ \bibinfo {pages} {42200} (\bibinfo {year}
  {2020})}\BibitemShut {NoStop}%
\bibitem [{\citenamefont {Friedman}\ \emph {et~al.}(2001)\citenamefont
  {Friedman}, \citenamefont {Hastie},\ and\ \citenamefont
  {Tibshirani}}]{friedman2001elements}%
  \BibitemOpen
  \bibfield  {author} {\bibinfo {author} {\bibfnamefont {J.}~\bibnamefont
  {Friedman}}, \bibinfo {author} {\bibfnamefont {T.}~\bibnamefont {Hastie}}, \
  and\ \bibinfo {author} {\bibfnamefont {R.}~\bibnamefont {Tibshirani}},\
  }\href@noop {} {\emph {\bibinfo {title} {The elements of statistical
  learning}}}\ (\bibinfo  {publisher} {Springer New York, NY, USA},\ \bibinfo
  {year} {2001})\BibitemShut {NoStop}%
\bibitem [{\citenamefont {Pedregosa}\ \emph {et~al.}(2011)\citenamefont
  {Pedregosa}, \citenamefont {Varoquaux}, \citenamefont {Gramfort},
  \citenamefont {Michel}, \citenamefont {Thirion}, \citenamefont {Grisel},
  \citenamefont {Blondel}, \citenamefont {Prettenhofer}, \citenamefont {Weiss},
  \citenamefont {Dubourg}, \citenamefont {Vanderplas}, \citenamefont {Passos},
  \citenamefont {Cournapeau}, \citenamefont {Brucher}, \citenamefont {Perrot},\
  and\ \citenamefont {Duchesnay}}]{scikit-learn}%
  \BibitemOpen
  \bibfield  {author} {\bibinfo {author} {\bibfnamefont {F.}~\bibnamefont
  {Pedregosa}}, \bibinfo {author} {\bibfnamefont {G.}~\bibnamefont
  {Varoquaux}}, \bibinfo {author} {\bibfnamefont {A.}~\bibnamefont {Gramfort}},
  \bibinfo {author} {\bibfnamefont {V.}~\bibnamefont {Michel}}, \bibinfo
  {author} {\bibfnamefont {B.}~\bibnamefont {Thirion}}, \bibinfo {author}
  {\bibfnamefont {O.}~\bibnamefont {Grisel}}, \bibinfo {author} {\bibfnamefont
  {M.}~\bibnamefont {Blondel}}, \bibinfo {author} {\bibfnamefont
  {P.}~\bibnamefont {Prettenhofer}}, \bibinfo {author} {\bibfnamefont
  {R.}~\bibnamefont {Weiss}}, \bibinfo {author} {\bibfnamefont
  {V.}~\bibnamefont {Dubourg}}, \bibinfo {author} {\bibfnamefont
  {J.}~\bibnamefont {Vanderplas}}, \bibinfo {author} {\bibfnamefont
  {A.}~\bibnamefont {Passos}}, \bibinfo {author} {\bibfnamefont
  {D.}~\bibnamefont {Cournapeau}}, \bibinfo {author} {\bibfnamefont
  {M.}~\bibnamefont {Brucher}}, \bibinfo {author} {\bibfnamefont
  {M.}~\bibnamefont {Perrot}}, \ and\ \bibinfo {author} {\bibfnamefont
  {E.}~\bibnamefont {Duchesnay}},\ }\href@noop {} {\bibfield  {journal}
  {\bibinfo  {journal} {Journal of Machine Learning Research}\ }\textbf
  {\bibinfo {volume} {12}},\ \bibinfo {pages} {2825} (\bibinfo {year}
  {2011})}\BibitemShut {NoStop}%
\bibitem [{\citenamefont {Buitinck}\ \emph {et~al.}(2013)\citenamefont
  {Buitinck}, \citenamefont {Louppe}, \citenamefont {Blondel}, \citenamefont
  {Pedregosa}, \citenamefont {Mueller}, \citenamefont {Grisel}, \citenamefont
  {Niculae}, \citenamefont {Prettenhofer}, \citenamefont {Gramfort},
  \citenamefont {Grobler}, \citenamefont {Layton}, \citenamefont {VanderPlas},
  \citenamefont {Joly}, \citenamefont {Holt},\ and\ \citenamefont
  {Varoquaux}}]{sklearn_api}%
  \BibitemOpen
  \bibfield  {author} {\bibinfo {author} {\bibfnamefont {L.}~\bibnamefont
  {Buitinck}}, \bibinfo {author} {\bibfnamefont {G.}~\bibnamefont {Louppe}},
  \bibinfo {author} {\bibfnamefont {M.}~\bibnamefont {Blondel}}, \bibinfo
  {author} {\bibfnamefont {F.}~\bibnamefont {Pedregosa}}, \bibinfo {author}
  {\bibfnamefont {A.}~\bibnamefont {Mueller}}, \bibinfo {author} {\bibfnamefont
  {O.}~\bibnamefont {Grisel}}, \bibinfo {author} {\bibfnamefont
  {V.}~\bibnamefont {Niculae}}, \bibinfo {author} {\bibfnamefont
  {P.}~\bibnamefont {Prettenhofer}}, \bibinfo {author} {\bibfnamefont
  {A.}~\bibnamefont {Gramfort}}, \bibinfo {author} {\bibfnamefont
  {J.}~\bibnamefont {Grobler}}, \bibinfo {author} {\bibfnamefont
  {R.}~\bibnamefont {Layton}}, \bibinfo {author} {\bibfnamefont
  {J.}~\bibnamefont {VanderPlas}}, \bibinfo {author} {\bibfnamefont
  {A.}~\bibnamefont {Joly}}, \bibinfo {author} {\bibfnamefont {B.}~\bibnamefont
  {Holt}}, \ and\ \bibinfo {author} {\bibfnamefont {G.}~\bibnamefont
  {Varoquaux}},\ }in\ \href@noop {} {\emph {\bibinfo {booktitle} {ECML PKDD
  Workshop: Languages for Data Mining and Machine Learning}}}\ (\bibinfo {year}
  {2013})\ pp.\ \bibinfo {pages} {108--122}\BibitemShut {NoStop}%
\bibitem [{\citenamefont {Breiman}\ \emph {et~al.}(1984)\citenamefont
  {Breiman}, \citenamefont {Friedman}, \citenamefont {Stone},\ and\
  \citenamefont {Olshen}}]{breiman2017classification}%
  \BibitemOpen
  \bibfield  {author} {\bibinfo {author} {\bibfnamefont {L.}~\bibnamefont
  {Breiman}}, \bibinfo {author} {\bibfnamefont {J.}~\bibnamefont {Friedman}},
  \bibinfo {author} {\bibfnamefont {C.~J.}\ \bibnamefont {Stone}}, \ and\
  \bibinfo {author} {\bibfnamefont {R.~A.}\ \bibnamefont {Olshen}},\
  }\href@noop {} {\emph {\bibinfo {title} {Classification and regression
  trees}}}\ (\bibinfo  {publisher} {Chapman and Hall/CRC, London, UK},\
  \bibinfo {year} {1984})\BibitemShut {NoStop}%
\bibitem [{\citenamefont {Breiman}(2001)}]{Breiman2001}%
  \BibitemOpen
  \bibfield  {author} {\bibinfo {author} {\bibfnamefont {L.}~\bibnamefont
  {Breiman}},\ }\href@noop {} {\bibfield  {journal} {\bibinfo  {journal}
  {Machine Learning}\ }\textbf {\bibinfo {volume} {45}},\ \bibinfo {pages} {5}
  (\bibinfo {year} {2001})}\BibitemShut {NoStop}%
\bibitem [{\citenamefont {Raileanu}\ and\ \citenamefont
  {Stoffel}(2004)}]{raileanu2004theoretical}%
  \BibitemOpen
  \bibfield  {author} {\bibinfo {author} {\bibfnamefont {L.~E.}\ \bibnamefont
  {Raileanu}}\ and\ \bibinfo {author} {\bibfnamefont {K.}~\bibnamefont
  {Stoffel}},\ }\href@noop {} {\bibfield  {journal} {\bibinfo  {journal}
  {Annals of Mathematics and Artificial Intelligence}\ }\textbf {\bibinfo
  {volume} {41}},\ \bibinfo {pages} {77} (\bibinfo {year} {2004})}\BibitemShut
  {NoStop}%
\bibitem [{\citenamefont {Zhang}\ \emph {et~al.}(2020)\citenamefont {Zhang},
  \citenamefont {Ginsparg},\ and\ \citenamefont {Kim}}]{zhang2020interpreting}%
  \BibitemOpen
  \bibfield  {author} {\bibinfo {author} {\bibfnamefont {Y.}~\bibnamefont
  {Zhang}}, \bibinfo {author} {\bibfnamefont {P.}~\bibnamefont {Ginsparg}}, \
  and\ \bibinfo {author} {\bibfnamefont {E.-A.}\ \bibnamefont {Kim}},\
  }\href@noop {} {\bibfield  {journal} {\bibinfo  {journal} {Physical Review
  Research}\ }\textbf {\bibinfo {volume} {2}},\ \bibinfo {pages} {023283}
  (\bibinfo {year} {2020})}\BibitemShut {NoStop}%
\bibitem [{\citenamefont {Bianco}\ and\ \citenamefont
  {Resta}(2011)}]{PhysRevB.84.241106}%
  \BibitemOpen
  \bibfield  {author} {\bibinfo {author} {\bibfnamefont {R.}~\bibnamefont
  {Bianco}}\ and\ \bibinfo {author} {\bibfnamefont {R.}~\bibnamefont {Resta}},\
  }\href@noop {} {\bibfield  {journal} {\bibinfo  {journal} {Phys. Rev. B}\
  }\textbf {\bibinfo {volume} {84}},\ \bibinfo {pages} {241106} (\bibinfo
  {year} {2011})}\BibitemShut {NoStop}%
\bibitem [{\citenamefont {Caio}\ \emph {et~al.}(2019)\citenamefont {Caio},
  \citenamefont {M{\"o}ller}, \citenamefont {Cooper},\ and\ \citenamefont
  {Bhaseen}}]{caio2019topological}%
  \BibitemOpen
  \bibfield  {author} {\bibinfo {author} {\bibfnamefont {M.~D.}\ \bibnamefont
  {Caio}}, \bibinfo {author} {\bibfnamefont {G.}~\bibnamefont {M{\"o}ller}},
  \bibinfo {author} {\bibfnamefont {N.~R.}\ \bibnamefont {Cooper}}, \ and\
  \bibinfo {author} {\bibfnamefont {M.}~\bibnamefont {Bhaseen}},\ }\href@noop
  {} {\bibfield  {journal} {\bibinfo  {journal} {Nature Physics}\ ,\ \bibinfo
  {pages} {1}} (\bibinfo {year} {2019})}\BibitemShut {NoStop}%
\bibitem [{\citenamefont {Carlsson}(2009)}]{carlsson2009topology}%
  \BibitemOpen
  \bibfield  {author} {\bibinfo {author} {\bibfnamefont {G.}~\bibnamefont
  {Carlsson}},\ }\href@noop {} {\bibfield  {journal} {\bibinfo  {journal}
  {Bulletin of the American Mathematical Society}\ }\textbf {\bibinfo {volume}
  {46}},\ \bibinfo {pages} {255} (\bibinfo {year} {2009})}\BibitemShut
  {NoStop}%
\bibitem [{\citenamefont {Wasserman}(2018)}]{wasserman2018topological}%
  \BibitemOpen
  \bibfield  {author} {\bibinfo {author} {\bibfnamefont {L.}~\bibnamefont
  {Wasserman}},\ }\href@noop {} {\bibfield  {journal} {\bibinfo  {journal}
  {Annual Review of Statistics and Its Application}\ }\textbf {\bibinfo
  {volume} {5}},\ \bibinfo {pages} {501} (\bibinfo {year} {2018})}\BibitemShut
  {NoStop}%
\bibitem [{\citenamefont {Wang}\ \emph {et~al.}(2005)\citenamefont {Wang},
  \citenamefont {Zhang},\ and\ \citenamefont {Zha}}]{wang2005adaptive}%
  \BibitemOpen
  \bibfield  {author} {\bibinfo {author} {\bibfnamefont {J.}~\bibnamefont
  {Wang}}, \bibinfo {author} {\bibfnamefont {Z.}~\bibnamefont {Zhang}}, \ and\
  \bibinfo {author} {\bibfnamefont {H.}~\bibnamefont {Zha}},\ }in\ \href@noop
  {} {\emph {\bibinfo {booktitle} {Advances in neural information processing
  systems}}}\ (\bibinfo {year} {2005})\ pp.\ \bibinfo {pages}
  {1473--1480}\BibitemShut {NoStop}%
\bibitem [{\citenamefont {Lin}\ and\ \citenamefont
  {Zha}(2008)}]{lin2008riemannian}%
  \BibitemOpen
  \bibfield  {author} {\bibinfo {author} {\bibfnamefont {T.}~\bibnamefont
  {Lin}}\ and\ \bibinfo {author} {\bibfnamefont {H.}~\bibnamefont {Zha}},\
  }\href@noop {} {\bibfield  {journal} {\bibinfo  {journal} {IEEE Transactions
  on Pattern Analysis and Machine Intelligence}\ }\textbf {\bibinfo {volume}
  {30}},\ \bibinfo {pages} {796} (\bibinfo {year} {2008})}\BibitemShut
  {NoStop}%
\bibitem [{\citenamefont {Belkin}(2003)}]{belkin2003problems}%
  \BibitemOpen
  \bibfield  {author} {\bibinfo {author} {\bibfnamefont {M.}~\bibnamefont
  {Belkin}},\ }\href@noop {} {\emph {\bibinfo {title} {Problems of learning on
  manifolds}}}\ (\bibinfo  {publisher} {The University of Chicago, Chicago, IL,
  USA},\ \bibinfo {year} {2003})\BibitemShut {NoStop}%
\bibitem [{\citenamefont {Asb{\'o}th}\ \emph {et~al.}(2016)\citenamefont
  {Asb{\'o}th}, \citenamefont {Oroszl{\'a}ny},\ and\ \citenamefont
  {P{\'a}lyi}}]{asboth2016short}%
  \BibitemOpen
  \bibfield  {author} {\bibinfo {author} {\bibfnamefont {J.~K.}\ \bibnamefont
  {Asb{\'o}th}}, \bibinfo {author} {\bibfnamefont {L.}~\bibnamefont
  {Oroszl{\'a}ny}}, \ and\ \bibinfo {author} {\bibfnamefont {A.}~\bibnamefont
  {P{\'a}lyi}},\ }\href@noop {} {\bibfield  {journal} {\bibinfo  {journal}
  {Lecture notes in physics}\ }\textbf {\bibinfo {volume} {919}} (\bibinfo
  {year} {2016})}\BibitemShut {NoStop}%
\bibitem [{\citenamefont {Goodfellow}\ \emph {et~al.}(2016)\citenamefont
  {Goodfellow}, \citenamefont {Bengio},\ and\ \citenamefont
  {Courville}}]{goodfellow2016deep}%
  \BibitemOpen
  \bibfield  {author} {\bibinfo {author} {\bibfnamefont {I.}~\bibnamefont
  {Goodfellow}}, \bibinfo {author} {\bibfnamefont {Y.}~\bibnamefont {Bengio}},
  \ and\ \bibinfo {author} {\bibfnamefont {A.}~\bibnamefont {Courville}},\
  }\href@noop {} {\emph {\bibinfo {title} {Deep learning}}}\ (\bibinfo
  {publisher} {MIT press, Cambridge, MA, USA},\ \bibinfo {year}
  {2016})\BibitemShut {NoStop}%
\bibitem [{\citenamefont {Bishop}(2006)}]{bishop2006pattern}%
  \BibitemOpen
  \bibfield  {author} {\bibinfo {author} {\bibfnamefont {C.~M.}\ \bibnamefont
  {Bishop}},\ }\href@noop {} {\emph {\bibinfo {title} {Pattern recognition and
  machine learning}}}\ (\bibinfo  {publisher} {Springer New York, NY, USA},\
  \bibinfo {year} {2006})\BibitemShut {NoStop}%
\bibitem [{\citenamefont {Cayton}(2005)}]{cayton2005algorithms}%
  \BibitemOpen
  \bibfield  {author} {\bibinfo {author} {\bibfnamefont {L.}~\bibnamefont
  {Cayton}},\ }\href@noop {} {\bibfield  {journal} {\bibinfo  {journal} {Univ.
  of California at San Diego Tech. Rep}\ }\textbf {\bibinfo {volume} {12}},\
  \bibinfo {pages} {1} (\bibinfo {year} {2005})}\BibitemShut {NoStop}%
\bibitem [{\citenamefont {Narayanan}\ and\ \citenamefont
  {Mitter}(2010)}]{narayanan2010sample}%
  \BibitemOpen
  \bibfield  {author} {\bibinfo {author} {\bibfnamefont {H.}~\bibnamefont
  {Narayanan}}\ and\ \bibinfo {author} {\bibfnamefont {S.}~\bibnamefont
  {Mitter}},\ }in\ \href@noop {} {\emph {\bibinfo {booktitle} {Advances in
  Neural Information Processing Systems}}}\ (\bibinfo {year} {2010})\ pp.\
  \bibinfo {pages} {1786--1794}\BibitemShut {NoStop}%
\bibitem [{\citenamefont {Rifai}\ \emph {et~al.}(2011)\citenamefont {Rifai},
  \citenamefont {Dauphin}, \citenamefont {Vincent}, \citenamefont {Bengio},\
  and\ \citenamefont {Muller}}]{rifai2011manifold}%
  \BibitemOpen
  \bibfield  {author} {\bibinfo {author} {\bibfnamefont {S.}~\bibnamefont
  {Rifai}}, \bibinfo {author} {\bibfnamefont {Y.~N.}\ \bibnamefont {Dauphin}},
  \bibinfo {author} {\bibfnamefont {P.}~\bibnamefont {Vincent}}, \bibinfo
  {author} {\bibfnamefont {Y.}~\bibnamefont {Bengio}}, \ and\ \bibinfo {author}
  {\bibfnamefont {X.}~\bibnamefont {Muller}},\ }in\ \href@noop {} {\emph
  {\bibinfo {booktitle} {Advances in Neural Information Processing Systems}}}\
  (\bibinfo {year} {2011})\ pp.\ \bibinfo {pages} {2294--2302}\BibitemShut
  {NoStop}%
\end{thebibliography}


%


\section*{Acknowledgements}

We thank S. E. Rowley, J. F. de Oliveira, T. Micklitz and M. A. Continentino for insightful discussions and S. E. Rowley for carefully reading the manuscript and suggesting improvements. N. L. Holanda acknowledges financial support from CENPES/Petrobr\'as/CBPF. M. A. R. Griffith acknowledges financial support from Capes. N. L. Holanda is grateful to the Theory of Condensed Matter and Quantum Materials groups at the Cavendish Laboratory and the Quantum Information Group at CBPF.

\section*{Author contributions}

Both authors of this work contributed equally to its realization at all stages.

\section*{Competing financial interests}

The authors declare no competing financial or non-financial interests.

\section*{Additional information}

Correspondence and requests for materials should be addressed to N. L. Holanda. 



\begin{figure}
  \centering
  \includegraphics[width=.37\textwidth]{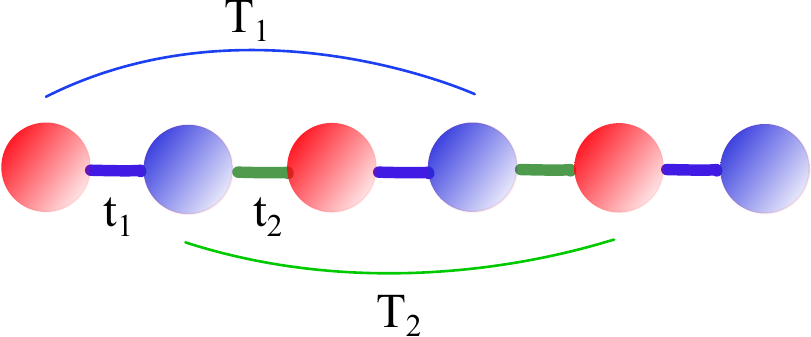}
  \caption{Schematic set-up of a 1D SSH system. Here $t_1$ and $t_2$ are nearest-neighbor hoppings while $T_1$ and $T_2$ are second nearest-neighbor hoppings.}
\label{fig:model}
\end{figure}

\begin{figure}
  \centering
  \includegraphics[width=.37\textwidth]{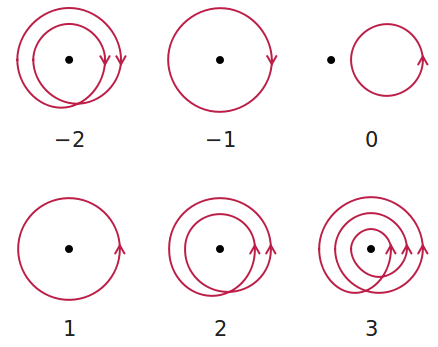}
  \caption{Winding number. The winding number of a closed, oriented curve with respect to a reference point is a topological invariant that counts how many times the curve winds around the point. Picture credits: Jim Belk, public domain.}
\label{fig:winding}
\end{figure}

\section*{Supplementary Material}
\subsection*{The SSH model}
\label{sshapp}

The SSH model \cite{asboth2016short} describes the movement of free electrons along a dimerized chain whose basic units consist of two distinct atoms. This movement, usually called ``hopping'' in the literature, can be made either between atoms in a unit cell or between unit cells, and the allowed hopping rules for a given system completely determine its Hamiltonian. This is because the kinetic energies of the electrons are parameterized by a vector of real numbers $\mathbf{t}$ that also encodes hopping terms, thus allowing for a compact mathematical description of a Hamiltonian in terms of creation/annihilation operators as
\begin{equation}
\label{SSH}
\mathbf{H}(\mathbf{t})=\mathbf{c}^{\dagger}H(\mathbf{t})\mathbf{c}
\end{equation}
where the column vector
\begin{equation*}
\mathbf{c} =\Big(c^{A}_1,c^{B}_1,\cdots,c^{A}_\frac{N}{2},c^{B}_\frac{N}{2}\Big)^T
\end{equation*}
contains annihilation operators $c^{A(B)}_p$ that erase electrons at atom A (B) and lattice site $p$ and similarly the row vector
\begin{equation*}
\mathbf{c}^\dagger =\Big(c^{A\dagger}_1,c^{B\dagger}_1,\cdots,c^{A\dagger}_\frac{N}{2},c^{B\dagger}_\frac{N}{2}\Big)
\end{equation*}
contains creation operators $c^{A(B)\dagger}_p$ that produce electrons at atom A (B) and lattice site $p$. Please note that $N$ is twice the number of unit cells in the chain and therefore an even integer.

The convenience of equation \eqref{SSH} is that all information about a system such as its eigenstates and eigenenergies can be recovered from the $N\times N$ matrix $H(\mathbf{t})$. We can thus think of the vectors $\mathbf{t}$ in parameter space as very compact representations of SSH models: each point in $\mathbf{t}$-space can be mapped to a $N\times N$ matrix $H(\mathbf{t})$ whose eigenvectors and eigenvalues can then be computed, as is usually done in quantum mechanics. As an example, a general matrix $H(\mathbf{t})$ describing a SSH system with hoppings between nearest and second nearest neighbors is given by
\begin{equation}\label{hmatrix}
H(t_1,t_2,T_1,T_2)=\left(
  \begin{array}{cccccc}
    0 & t_1 & 0 & T_1 & 0 & \cdots \\
    t_1 & 0 & t_2& 0 & T_2 & \cdots\\
    0 & t_2 & 0 & t_1 & 0 & \cdots\\
    T_1 & 0 & t_1 & 0 & t_2 & \cdots\\
    0 & T_2 & 0 & t_2 &  0& \cdots\\
    \vdots & \vdots    & \vdots & \vdots &  \vdots & \ddots\\
  \end{array}
  \right)_{N \times N}.
\end{equation}
Knowing the vector $\mathbf{t}=(t_1,t_2,T_1,T_2)$ corresponding to a particular system described by the Hamiltonian in equation \eqref{hmatrix} should suffice to compute any of its physical properties, including its topological phase. Figure \ref{fig:model} depicts a SSH system described by equation \eqref{hmatrix}.

The reason why topological materials garnered so much interest in recent years is that their physical properties are topologically robust.
This means that these properties are stable under continuous (i.e., adiabatic) mathematical operations performed on the system's underlying wave functions. This topological robustness is expressed theoretically in terms of a topological invariant that characterizes different phases of a system. In the particular case of the SSH model, the topological invariant used to classify the topological phases is the winding number.

The winding number is a topological property of any closed, oriented curve that measures how many times the curve winds around a point that does not belong to itself. It can be any integer and is usually chosen to be positive when the curve winds in counterclockwise motion with respect to the reference point (equivalently, when a closed, oriented curve winds in clockwise motion around a reference point its winding number is negative). Figure \ref{fig:winding} shows several closed, oriented curves and their winding numbers computed with respect to a given point.

An interesting property of topological invariants like the winding number is that they are a global feature of geometric objects: for each of the curves in figure \ref{fig:winding} for example the winding number is a property of the whole curve that cannot be defined locally for each of its points. This fundamental characteristic of topological invariants makes the study of topological phases of matter from local lattice data a challenging task.   

For SSH systems with translational symmetry like the finite systems with periodic boundary conditions investigated in the article, the winding number is usually computed in wavevector space via
\begin{equation}\label{winding_number}
W=\frac{1}{4\pi i}\int_0^{2\pi} dk Tr( \sigma_3 H(k)^{-1} \partial_k H(k)),
\end{equation}
where $H(k)$ is the kernel in wavevector space of a Hamiltonian $\mathbf{H}$$($$\mathbf{t}$$)$ and $\sigma_3$ is the chiral operator. Equation \eqref{winding_number} can be evaluated for several Hamiltonians by varying the parameter $\mathbf{t}$, resulting in phase diagrams in parameter space like the ones shown in figure 1 in the article. 

\begin{figure}
\centering
\subfigure[]{\label{eng_feat_ssh1:a}\includegraphics[width=.32\textwidth]{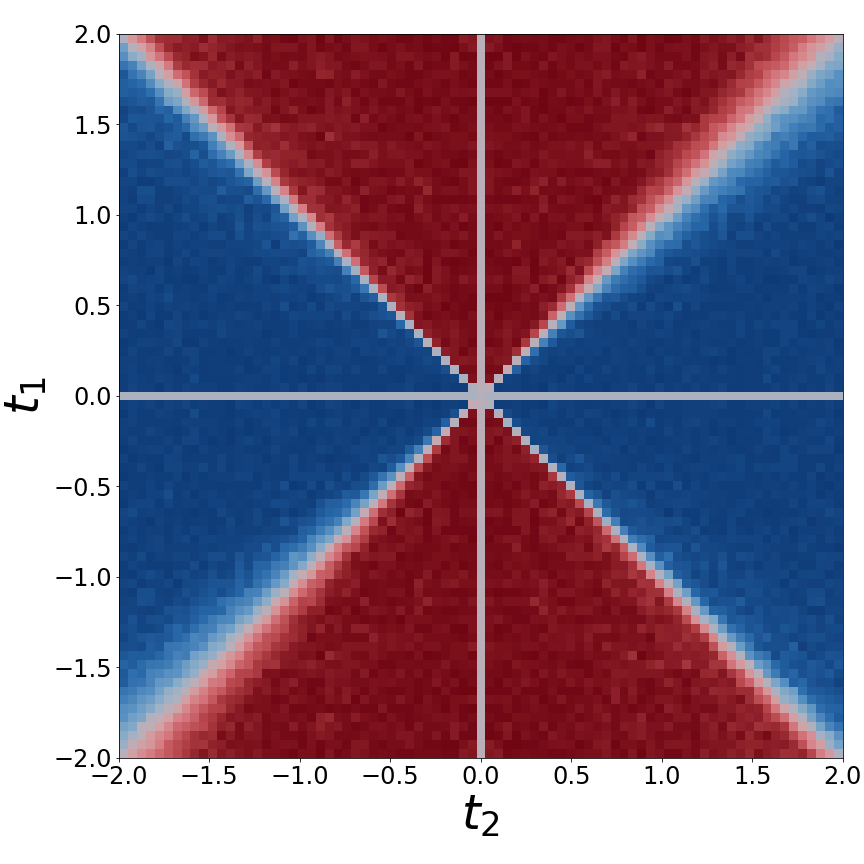}}
\subfigure[]{\label{eng_feat_ssh1:b}\includegraphics[width=.32\textwidth]{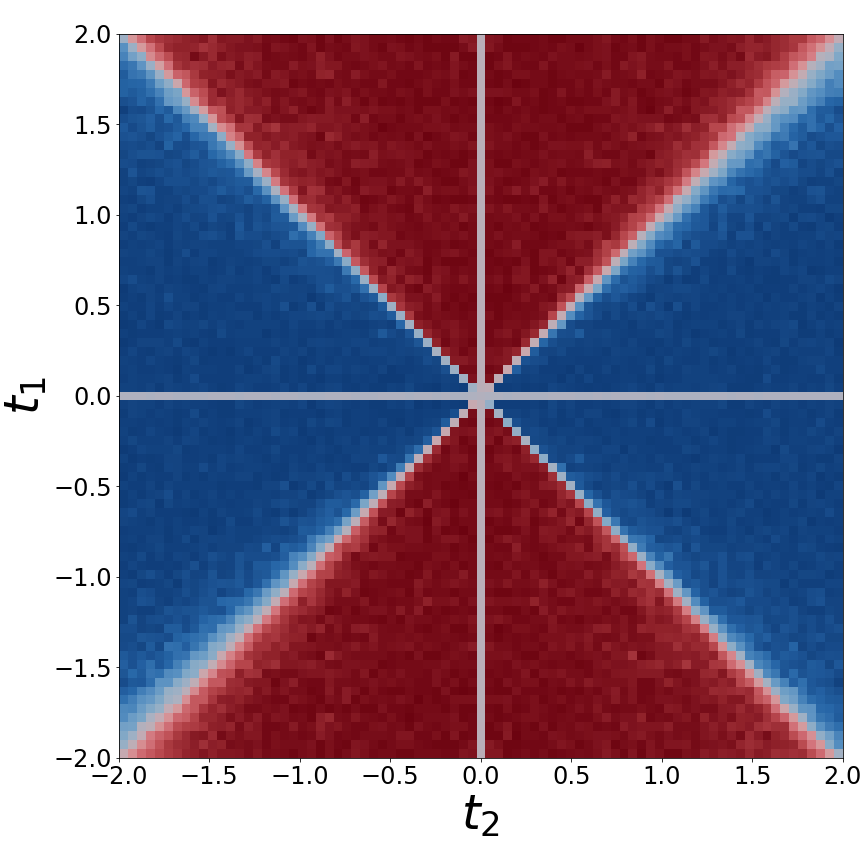}}
\subfigure[]{\label{eng_feat_ssh1:c}\includegraphics[width=.32\textwidth]{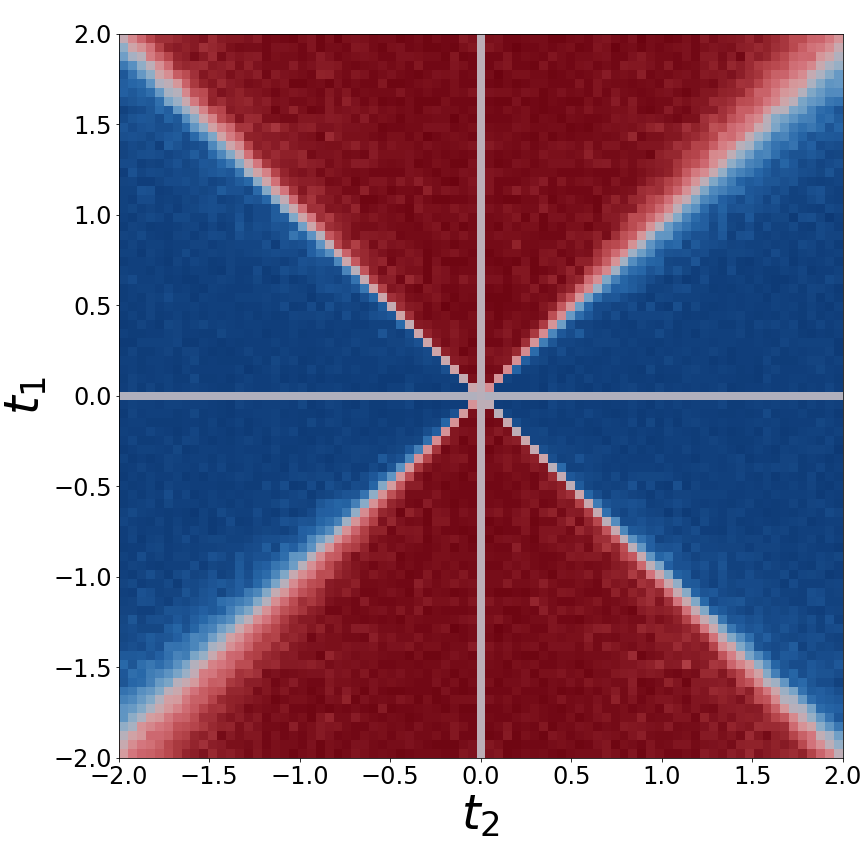}}
\caption{Phase diagrams learned from compressed representations of the SSH 1 system. (a) Phase diagram learned using the real space features \xSOne.  (b) Phase diagram learned from the DCT topological features \xcSEOne. (b) Phase diagram learned from the DST topological features \xsSOOne.}
\label{eng_feat_ssh1}
\end{figure}

\begin{figure}
\centering
\subfigure[]{\label{eng_feat_ssh2:a}\includegraphics[width=.32\textwidth]{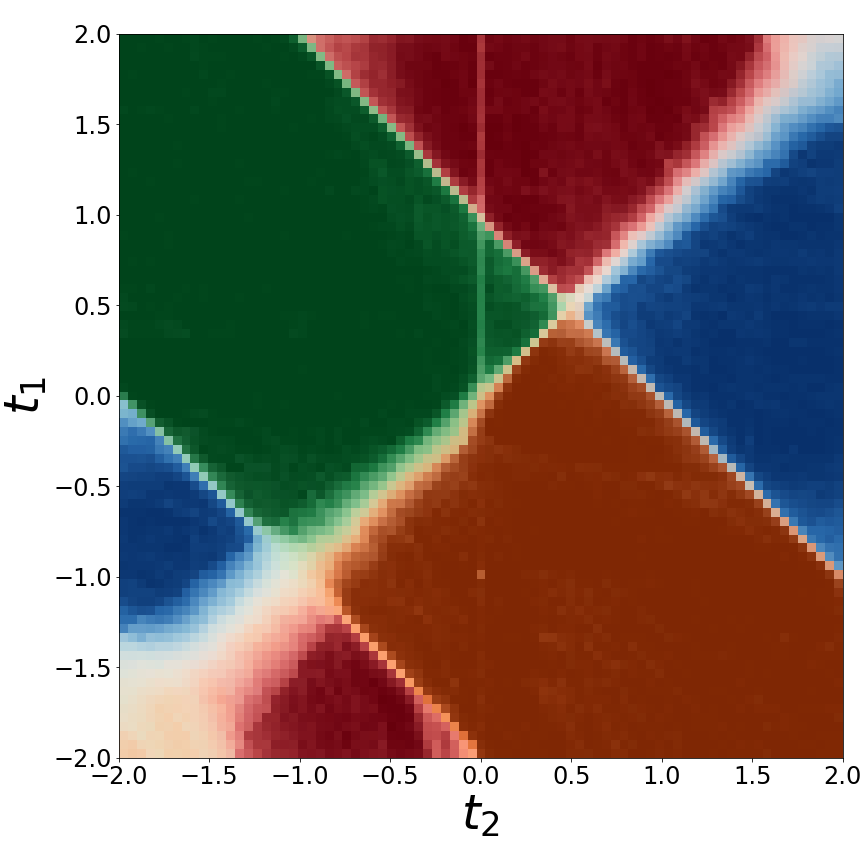}}
\subfigure[]{\label{eng_feat_ssh2:b}\includegraphics[width=.32\textwidth]{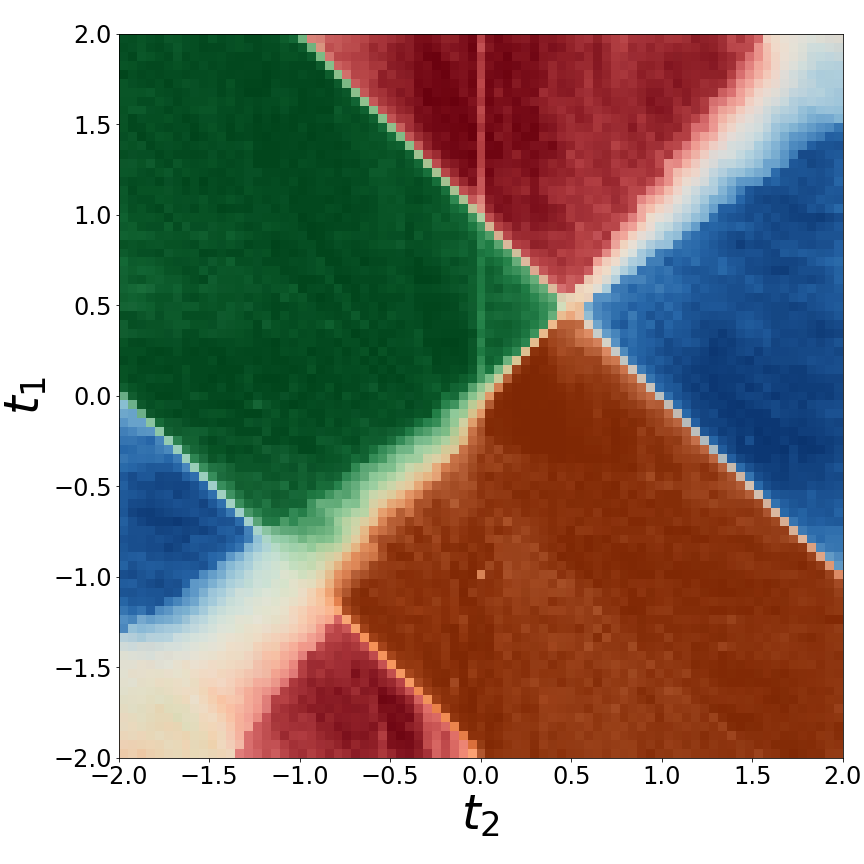}}
\subfigure[]{\label{eng_feat_ssh2:c}\includegraphics[width=.32\textwidth]{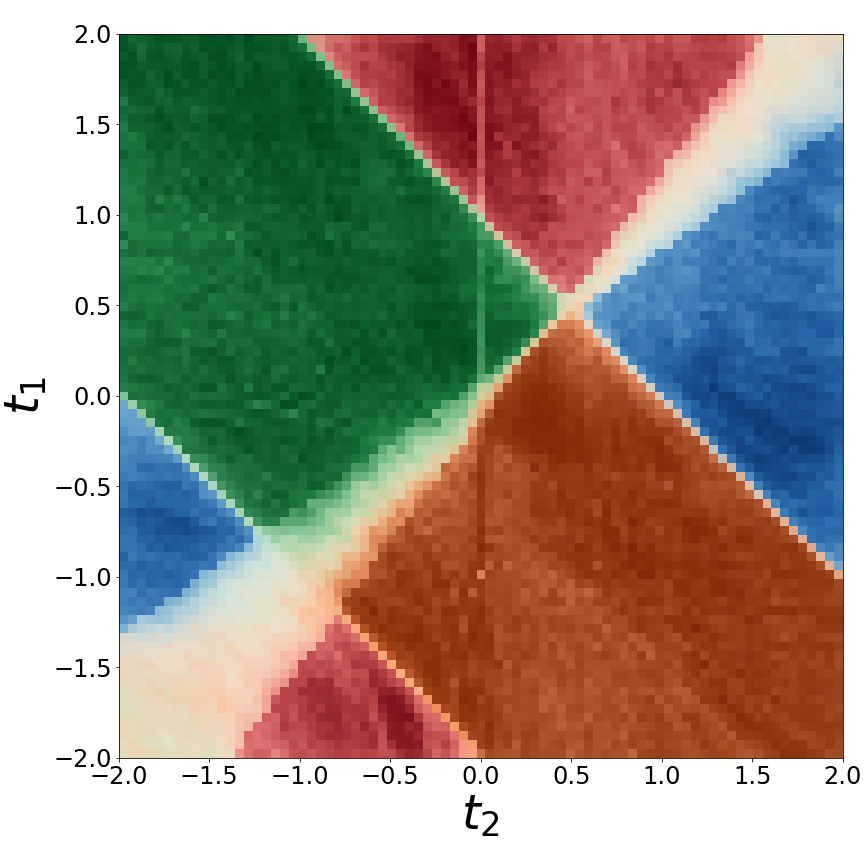}}
\caption{Phase diagrams learned from compressed representations of the SSH 2 system. (a) Phase diagram learned using the real space features \xSTwo.  (b) Phase diagram learned from the DCT topological features \xcSETwo. (b) Phase diagram learned from the DST topological features \xsSOTwo.}
\label{eng_feat_ssh2}
\end{figure}

\subsection*{Learning topological phases from real space data}


We now discuss the principles that lead us to the intuition that learning topological phases from local real space data should be possible.

First, while the topological invariants that characterize distinct topological phases are usually computed in wavevector space, the topological properties of a Hamiltonian are the same regardless of the basis in Hilbert space used to represent it. Thus, information on the topological phase of a Hamiltonian should still be available when it is represented in real lattice space. 

Second, even though the topological properties of a Hamiltonian are global, meaning that in general they cannot be said to be localized at a particular lattice site, in parameter space topology is indeed a local property: knowing the vector $\mathbf{t}$ associated with a Hamiltonian completely determines its topological phase.

In a data-driven approach, locality is often exploited by means of a local constancy hypothesis \cite{goodfellow2016deep}. Mathematically, this policy prescribes the value of a function $W(\mathbf{t}')$ at points where it is unknown in a vicinity of a data point $\mathbf{t}$ as approximately equal to its known value $W(\mathbf{t})$,
\begin{equation}
\label{localconstancy}
W(\mathbf{t} + \boldsymbol{\delta}) \approx W(\mathbf{t}).
\end{equation}

That such a policy will be successful in classifying topological phases in parameter space can be visualized in figure 1 in the article 
where we draw phase diagrams of SSH models with first-neighbor (figure 1(a)) and first- and second-neighbor (figure 1(b)) hoppings. In figure 1(a) for example, it is clear that knowing a particular Hamiltonian $H(t_1,t_2)$ with winding number $W = 0$ (that is, in one of the red regions) means that there is a small neighborhood around $(t_1,t_2)$ in which all Hamiltonians  belong to the same topological phase. Were we able to collect data on the topological phases of several Hamiltonians in parameter space, the problem of learning phase boundaries in a supervised setting would reduce to a standard problem of curve estimation which could be tackled with conventional machine learning algorithms.

It does not immediately follow, however, that the same strategy will be successful in real space. Indeed, at first sight it may appear that a local constancy policy should be able to easily exploit locality in real space through the diagonalization maps $v^{(j,l)}: \mathbb{R}^h \rightarrow \mathbb{R}^{N}$,

\begin{equation}
\label{parameter_to_real}
\mathbf{t} = (t_1,..,t_h) \rightarrow \Big(v^{(j,1)}(t_1,...,t_h),...,v^{(j,N)}(t_1,...,t_h)\Big) = \mathbf{v}^{(j)}(\mathbf{t}) 
\end{equation}
where $j$ $=$ $1$, ..., $N$ and $\mathbf{v}^{(j)}$$($$\mathbf{t}$$)$ is an eigenvector of the Hamiltonian $H$$($$\mathbf{t}$$)$. The trouble with this reasoning is that it disregards the high dimensionality of real space, i.e., the fact that $h$ $\ll$ $N$.

This fact is well illustrated by the numerical experiments discussed in the article. Although it may seem from figures 2(b) and 3(b) (from the article) in 2D parameter space that we have used a large number of data points for this learning task, it is important to note that in the numerical experiments the decision trees have taken as inputs 100D eigenvectors in real lattice space. In such high-dimensional spaces, the data should be much sparser.

The difficulty arising from machine learning problems in high-dimensional spaces is commonly referred to as \emph{the curse of dimensionality} \cite{bishop2006pattern}. It essentially expresses the fact that the amount of data needed to ensure a machine learning algorithm will generalize well out of its training set grows exponentially with the dimensionality of feature space.

These apparently conflicting facets of our learning problem are harmonized by the manifold hypothesis \cite{cayton2005algorithms,narayanan2010sample}: even though the eigenvectors exist in a high-dimensional space (100D in the numerical experiments), they are actually much lower-dimensional surfaces (2D in the numerical experiments) embedded in this space. Furthermore, the different classes in our problem correspond to different submanifolds as can easily be seen in parameter space (this is often referred to as the manifold hypothesis for classification \cite{rifai2011manifold}). As we have demonstrated in the article, only a small fraction of the 2D surfaces (i.e., eigenvector lattice coordinates $v^{(j,l)}(t_1,t_2)$) were needed to retrieve the 2D parameter space phase diagrams from the 100D real space data, which reveals that there is a lot of redundancy in the information content of real space eigenvectors. This fact can be exploited to generate compressed representations of SSH systems that still carry most of the relevant topological information, as can be seen in figures \ref{eng_feat_ssh1} and \ref{eng_feat_ssh2}.  

\subsection*{Numerical explorations on longer lattices}

A natural question that comes to mind regarding the information entropy signatures presented here is whether these signals are artifacts of the numerical procedure used to generate them.

The eigenvector ensembling algorithm used in this work contains steps for generating data (step 1), sampling training data (step 2) and training a supervised learning algorithm on the sampled training data (step 3). Each of these steps can generate misleading artifacts that do not represent real properties of the physical systems we are investigating.

Artifacts resulting from randomization in the eigenvector ensembling algorithm can be traced to sampling (step 2) or any random components in the supervised learning algorithm used (step 3). As an example, random forests allocate subsets of features stochastically to each of its decision trees, thus generating a randomization effect. The bootstrapping (step 5) is designed to remove artifacts originating from randomization in the eigenvector ensembling algorithm.

There are still artifacts that might arise from the hyperparameters used to generate the data. These hyperparameters include lattice size and grid specifications (i.e., choices regarding the discretization of parameter space). Such artifacts can only be controlled for by bootstrapping over different hyperparameter settings, which may be computationally prohibitive. Intuitively, the hyperparameter we identify as likely having the most noticeable effect in the information entropy signatures is lattice size.

\begin{figure}
\centering
\subfigure[]{\label{feature_importances_ssh1_140}\includegraphics[width=.32\textwidth]{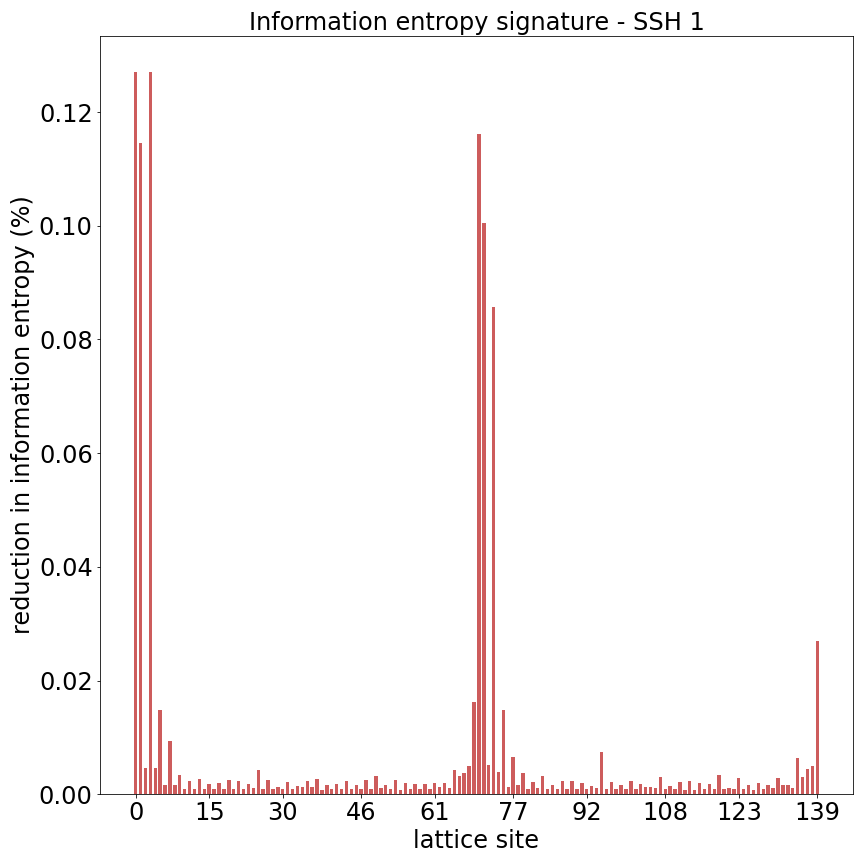}}
\subfigure[]{\label{feature_importances_ssh1_180}\includegraphics[width=.32\textwidth]{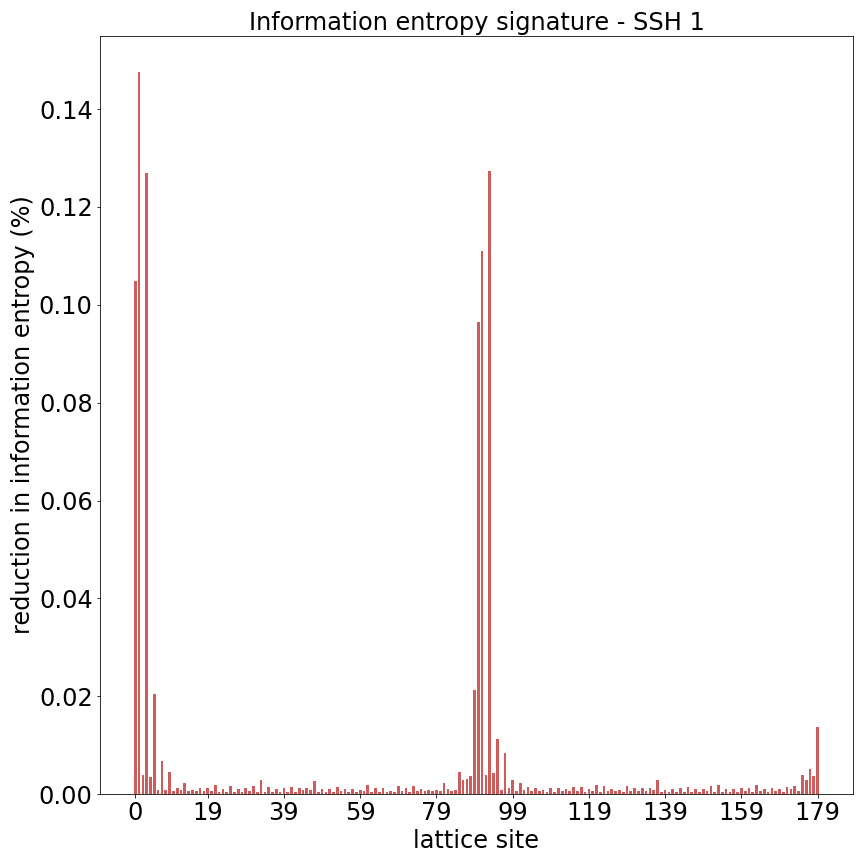}}
\subfigure[]{\label{feature_importances_ssh1_220}\includegraphics[width=.32\textwidth]{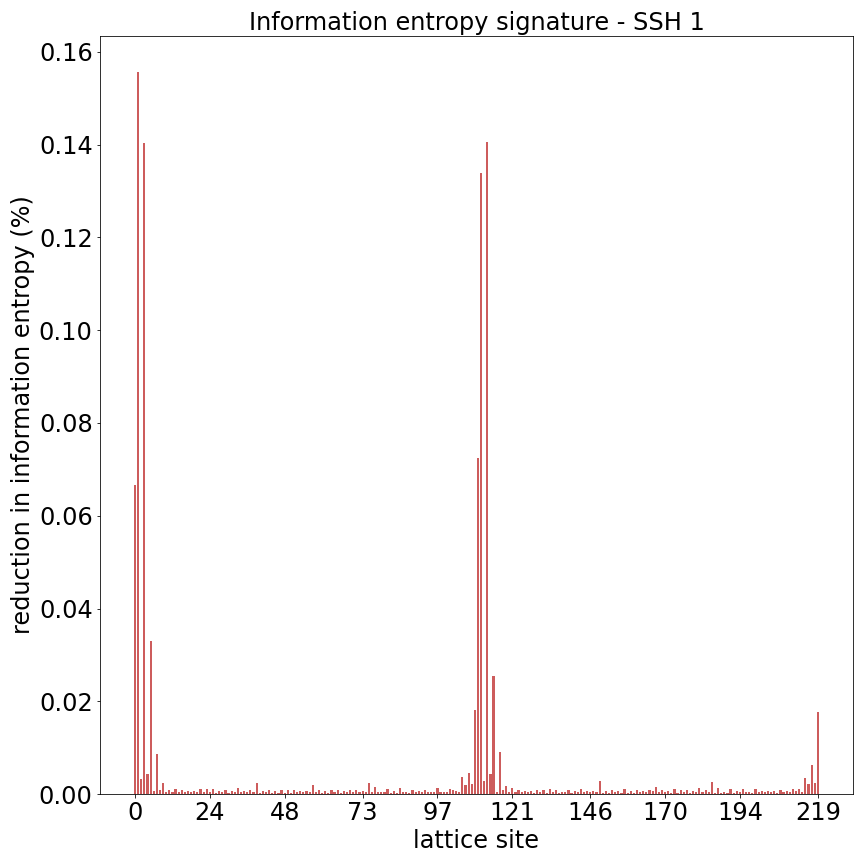}}
\caption{Information entropy signatures obtained for experiment 1 with higher lattice sizes. (a) Information entropy signature for 70 unit cells. (b) Information entropy signature for 90 unit cells. (c) Information entropy signature for 110 unit cells.}
\label{feature_importances_ssh1_longer_lattices}
\end{figure}

\begin{figure}
\centering
\subfigure[]{\label{feature_importances_ssh2_140}\includegraphics[width=.32\textwidth]{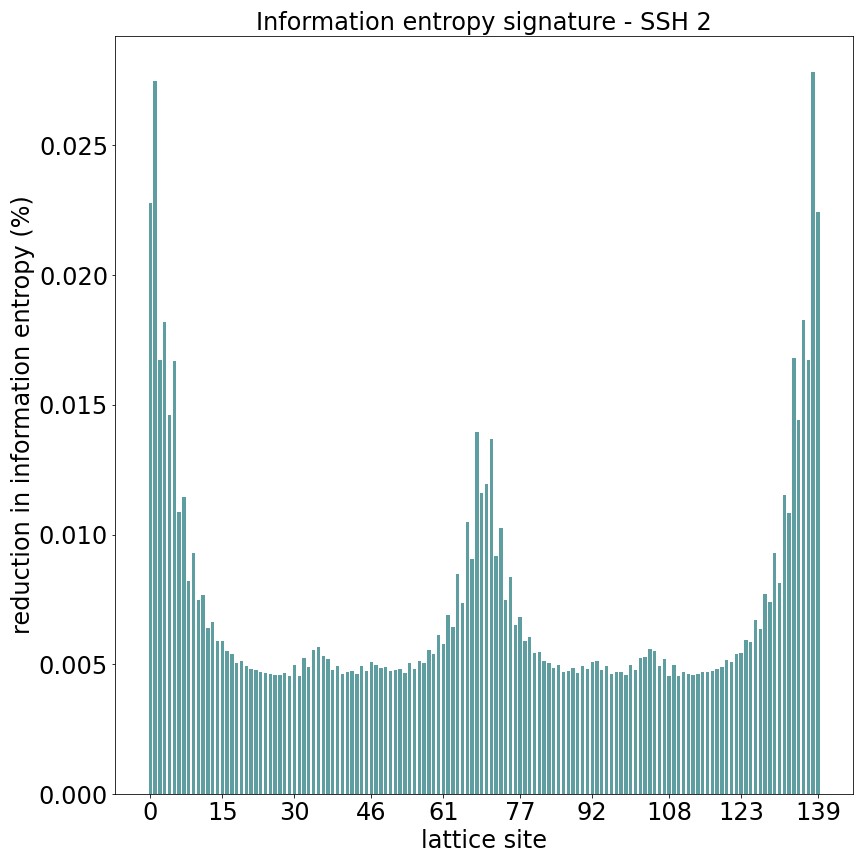}}
\subfigure[]{\label{feature_importances_ssh2_180}\includegraphics[width=.32\textwidth]{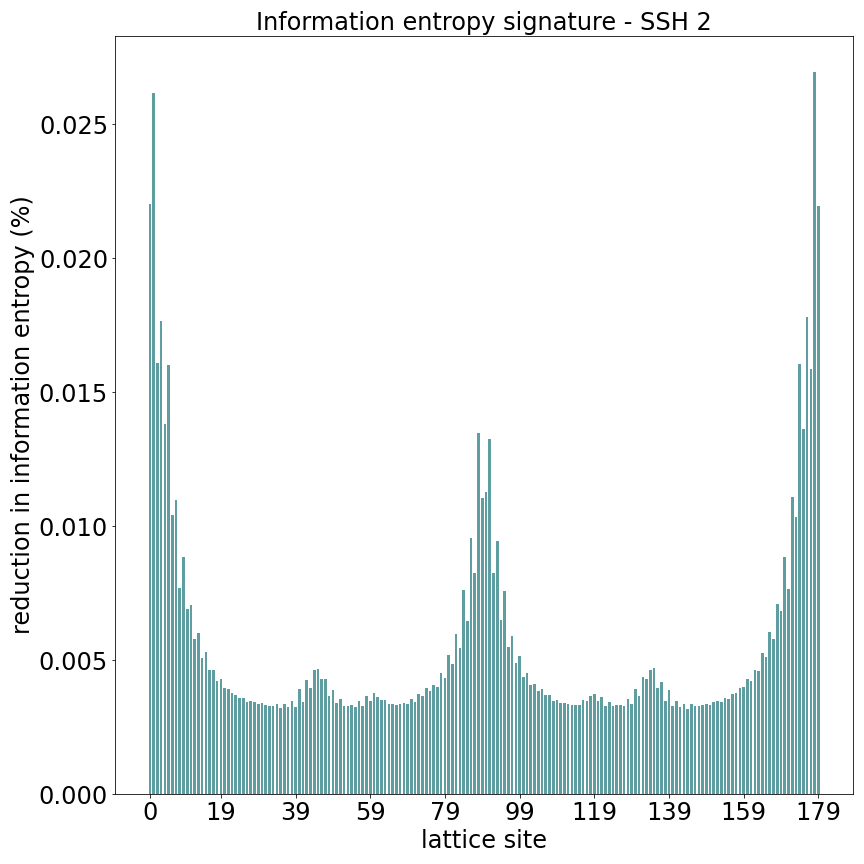}}
\subfigure[]{\label{feature_importances_ssh2_220}\includegraphics[width=.32\textwidth]{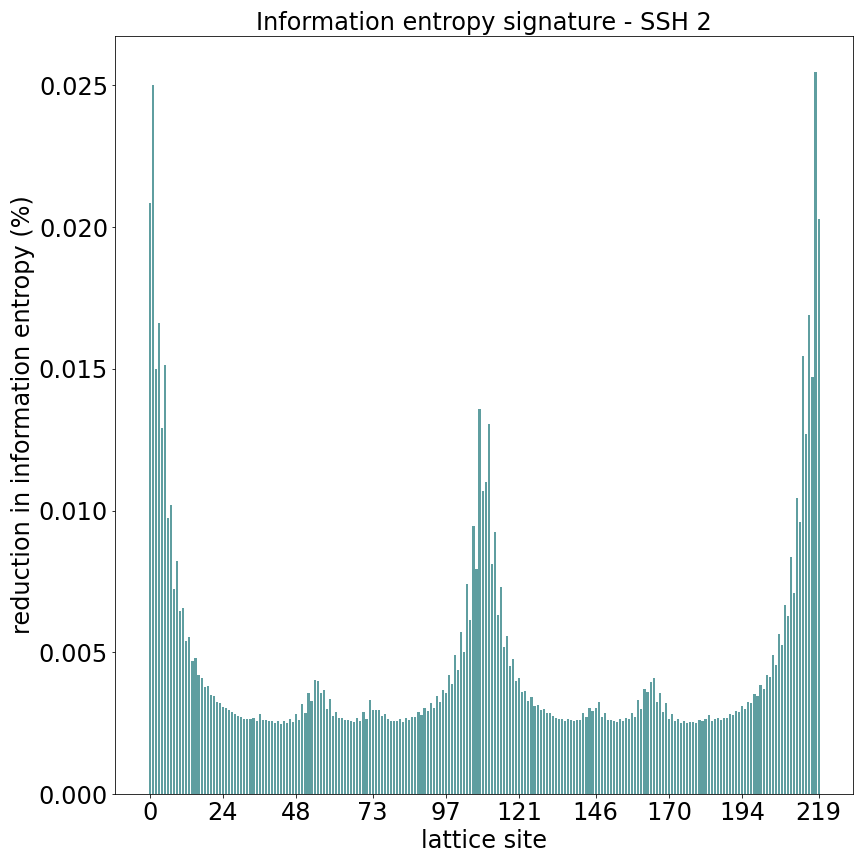}}
\caption{Information entropy signatures obtained for experiment 2 with higher lattice sizes. (a) Information entropy signature for 70 unit cells. (b) Information entropy signature for 90 unit cells. (c) Information entropy signature for 110 unit cells.}
\label{feature_importances_ssh2_longer_lattices}
\end{figure}

All results presented in the article were obtained for lattices with 50 unit cells. Each cell contains two different atoms, thus leading to 100$\times$100 Hamiltonian matrices and their corresponding eigenvectors in $\mathbb{R}^{100}$. Here we present the information entropy signatures obtained for lattices with 70, 90 and 110 unit cells for both experiments (figures \ref{feature_importances_ssh1_longer_lattices} and \ref{feature_importances_ssh2_longer_lattices}). These signals were generated in exactly the same way as the information entropy signatures obtained for 50 unit cells (i.e., after bootstrapping $n_{exp}$ = 100 times and averaging lattice site relevances across all iterations).

It is clear from figures \ref{feature_importances_ssh1_longer_lattices} and \ref{feature_importances_ssh2_longer_lattices} that the patterns seen with 50 unit cells (figures 6(a) and 6(b) in the article) are stable across higher lattice sizes, with finer details emerging in the signals as the lattice size is increased. This naturally leads us to speculation about the continuum limit of information entropy signatures, a matter formally explored in the article.  


\pagebreak 

\begin{figure}
\centering
\subfigure[]{\label{cumulative_entropy_ssh1_140}\includegraphics[width=.32\textwidth]{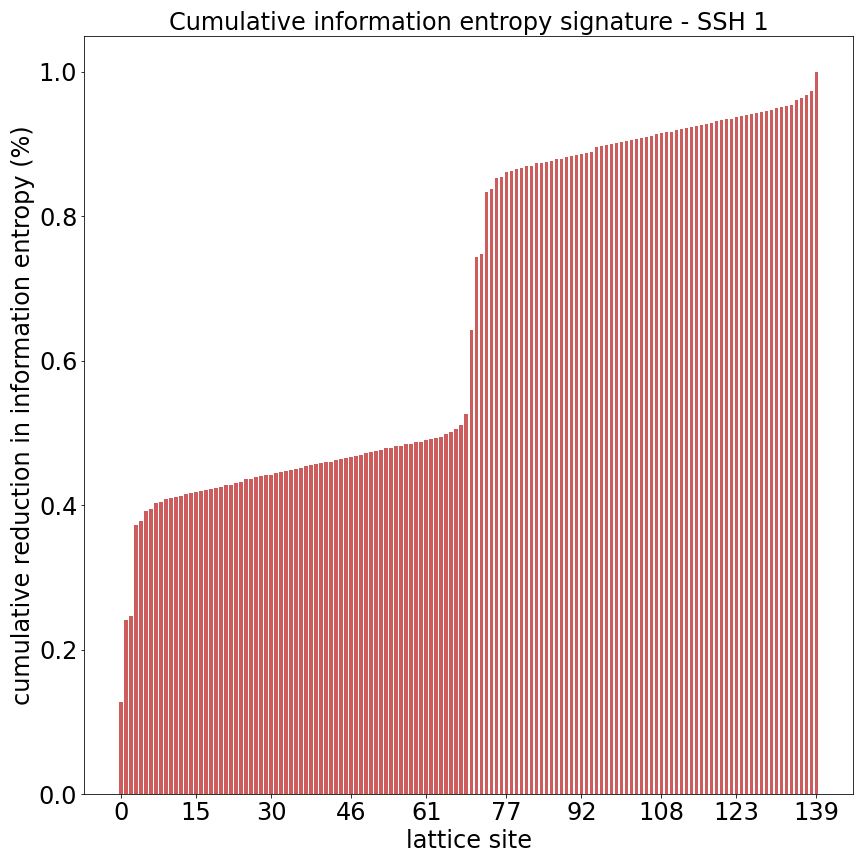}}
\subfigure[]{\label{cumulative_entropy_ssh1_180}\includegraphics[width=.32\textwidth]{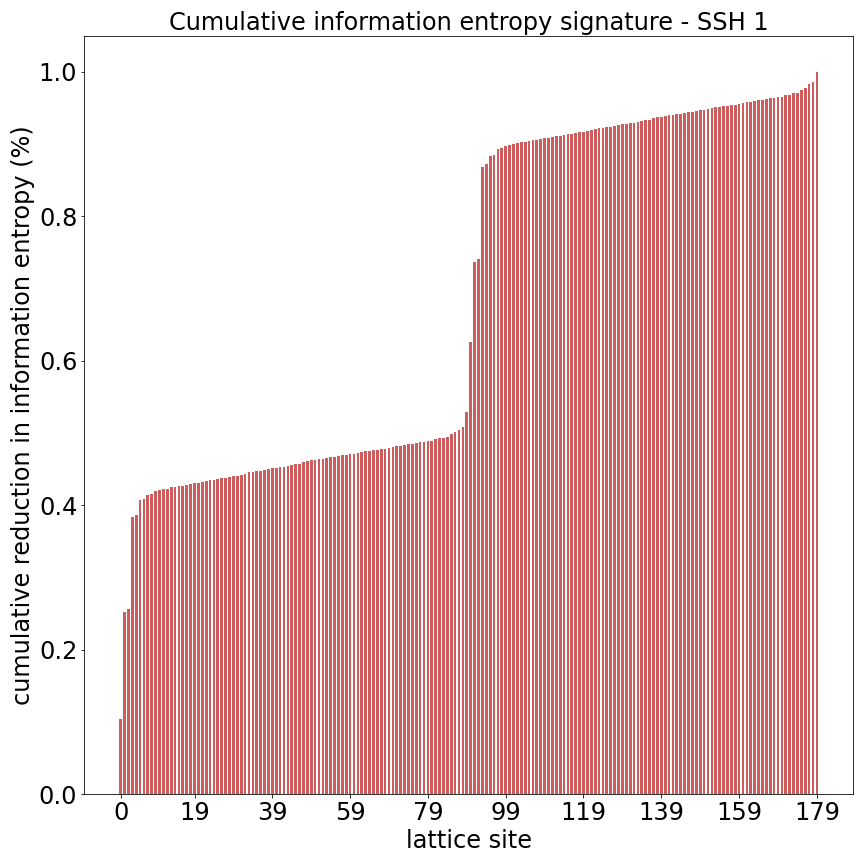}}
\subfigure[]{\label{cumulative_entropy_ssh1_220}\includegraphics[width=.32\textwidth]{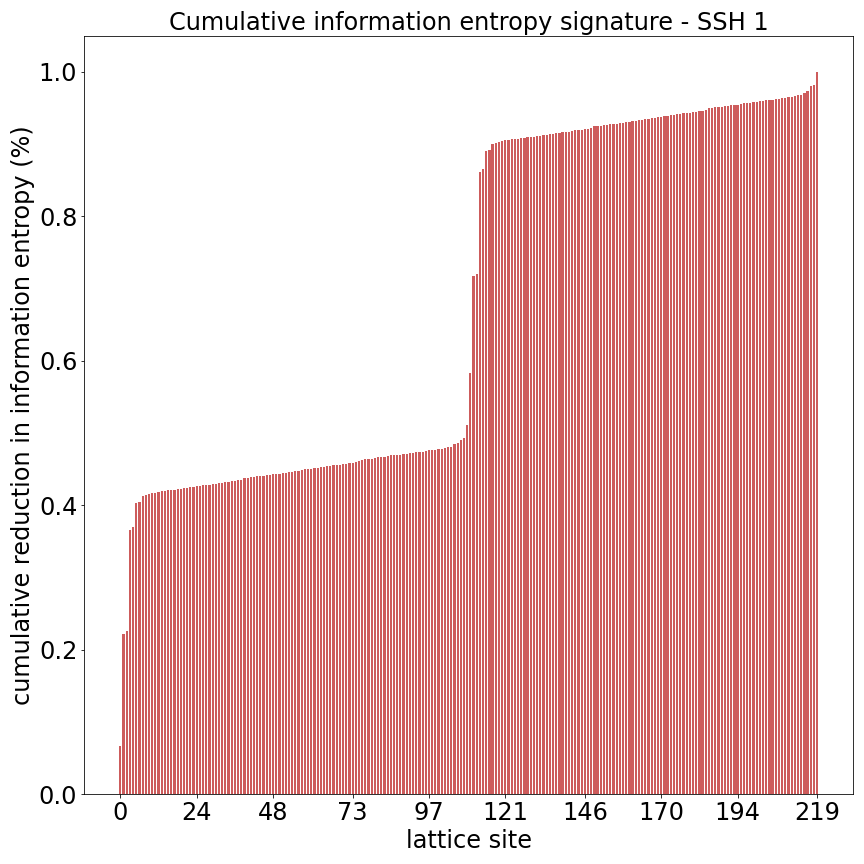}}
\caption{Cumulative entropy distributions for higher lattice sizes in experiment 1. (a) Cumulative entropy distribution for 70 unit cells. (b) Cumulative entropy distribution for 90 unit cells. (c) Cumulative entropy distribution for 110 unit cells.}
\label{cdf1}
\end{figure}

\begin{figure}
\centering
\subfigure[]{\label{cumulative_entropy_ssh2_140}\includegraphics[width=.32\textwidth]{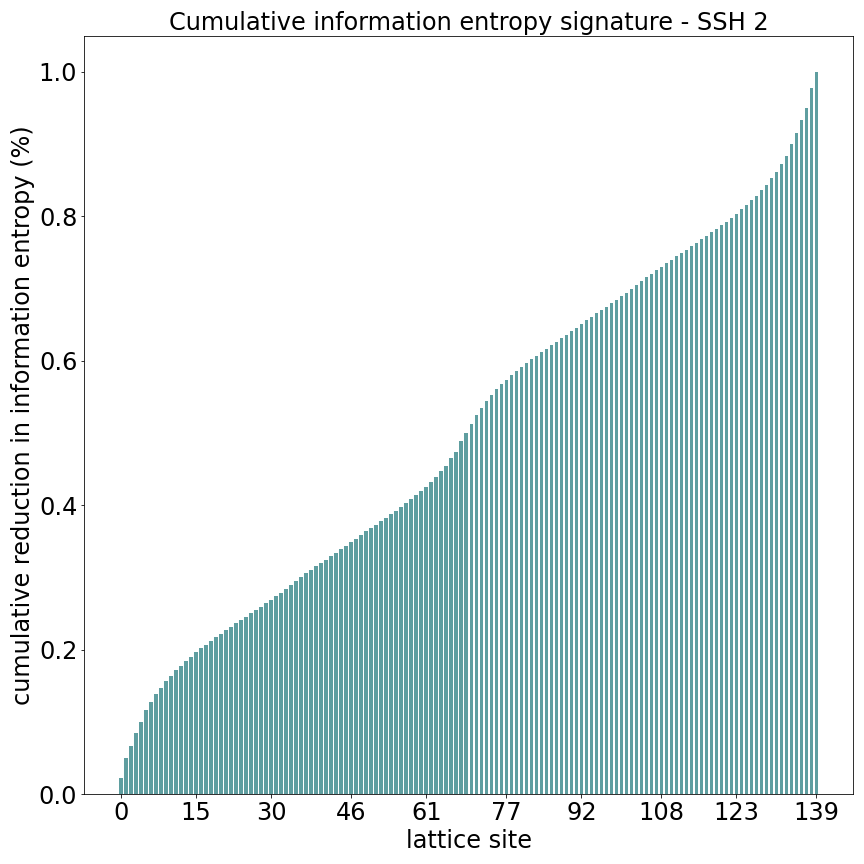}}
\subfigure[]{\label{cumulative_entropy_ssh2_180}\includegraphics[width=.32\textwidth]{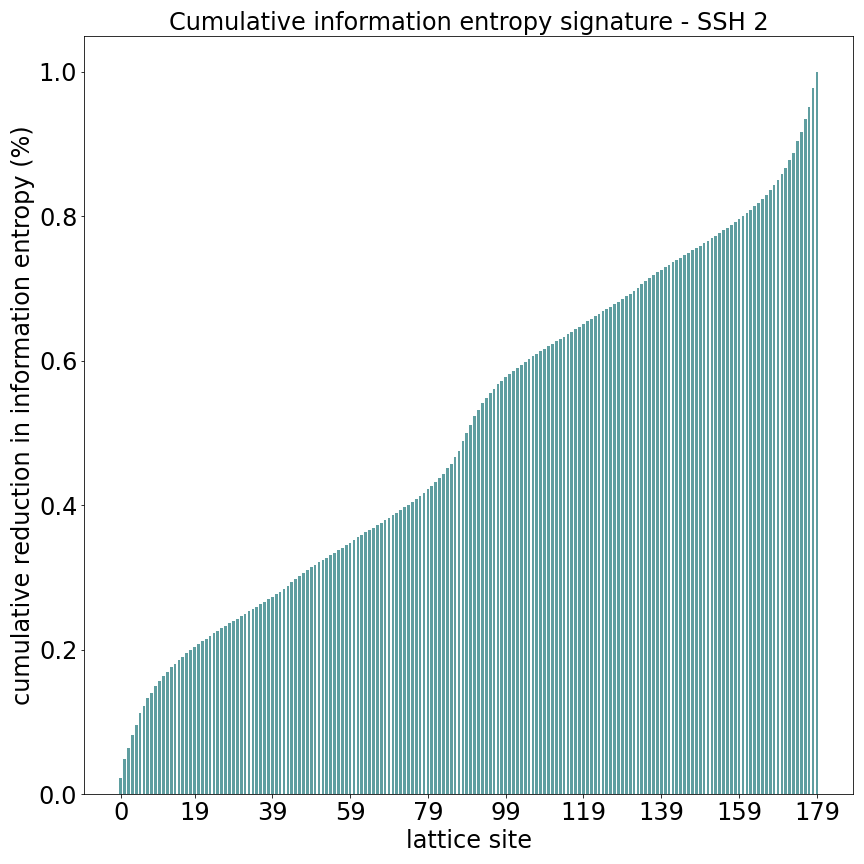}}
\subfigure[]{\label{cumulative_entropy_ssh2_220}\includegraphics[width=.32\textwidth]{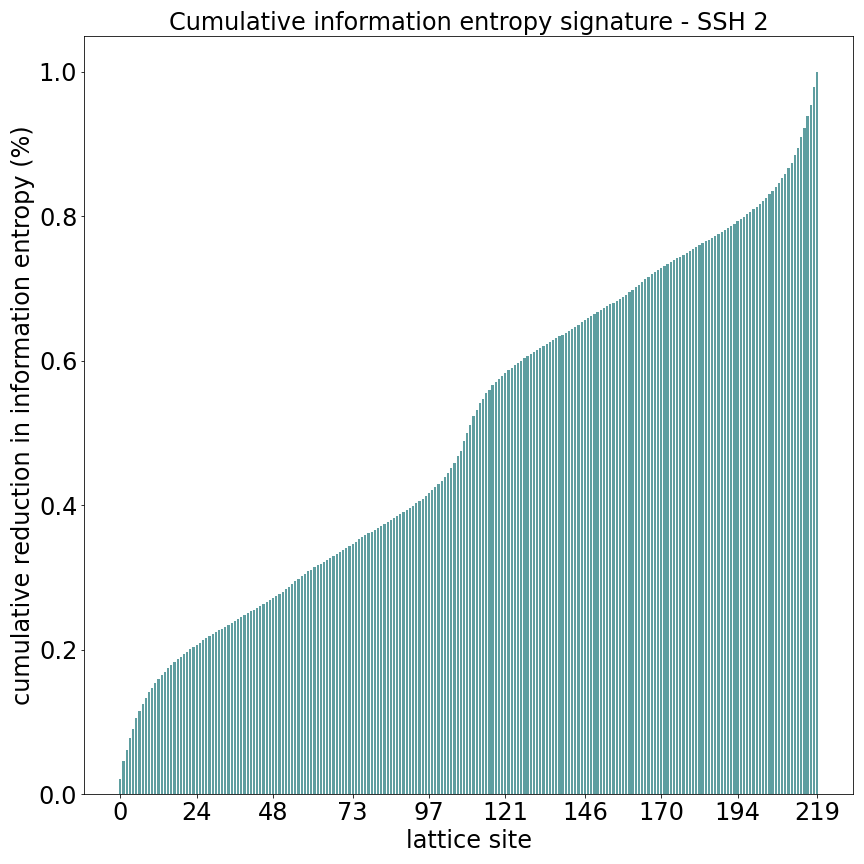}}
\caption{Cumulative entropy distributions for higher lattice sizes in experiment 2. (a) Cumulative entropy distributions for 70 unit cells. (b) Cumulative entropy distributions for 90 unit cells. (c) Cumulative entropy distributions for 110 unit cells.}
\label{cdf2}
\end{figure}

\end{document}